\documentclass[%
 reprint,
 amsmath,amssymb,
 aps,
]{revtex4-2}

\usepackage[dvipsnames]{xcolor}

\usepackage{enumitem}
\usepackage{lineno} 
\usepackage{graphicx}% Include figure files
\usepackage{dcolumn}% Align table columns on decimal point
\usepackage{bm}% bold math$
\usepackage{hyperref}% add hypertext capabilities
\usepackage{booktabs} % For \toprule, \midrule, \bottomrule
\usepackage{subcaption} % for subfigure environment
\begin{document}

\preprint{APS/123-QED}

%\title{Probing causal mechanisms in large-scale climate dynamics\\ with reduced-order neural models}
\title{Probing forced responses and causality in data-driven climate emulators:\\ conceptual limitations and the role of reduced-order models}
%\title{Probing causal mechanisms in climate dynamics: limitations of data-driven methods and the role of reduced-order models}

\author{Fabrizio Falasca}
\email{fabri.falasca@nyu.edu}
\affiliation{%
Courant Institute of Mathematical Sciences \\ New York University, New York, NY, USA 
}%

\date{\today}% It is always \today, today,

\begin{abstract}
A central challenge in climate science and applied mathematics is developing data-driven models of multiscale systems that capture both stationary statistics and responses to external perturbations. Current neural climate emulators aim to resolve the atmosphere–ocean system in all its complexity but often struggle to reproduce forced responses, limiting their use in causal studies such as Green’s function experiments. To explore the origin of these limitations, we first examine a simplified dynamical system that retains key features of climate variability. We interpret the results through linear response theory, providing a rigorous framework to evaluate neural models beyond stationary statistics and probe causal mechanisms. We argue that the ability of emulators of multiscale systems to reproduce perturbed statistics depends critically on (i) the choice of an appropriate coarse-grained representation and (ii) careful parameterizations of unresolved processes. These insights highlight reduced-order models, tailored to \textit{specific} goals, processes, and scales, as valuable alternatives to \textit{general-purpose} emulators. We next consider a real-world application, developing a neural model to investigate the joint variability of the surface temperature field and radiative fluxes. The model infers a multiplicative noise process directly from data, largely reproduces the system's probability distribution, and enables causal studies through forced responses. We discuss its limitations and outline directions for future work. These results expose key challenges in data-driven modeling of multiscale physical systems and underscore the value of coarse-grained, stochastic approaches, with response theory as a principled framework to guide model design and enhance causal understanding.\\

\textbf{This is a post-peer-review, pre-copyedit version of an article published in Physical Review Research. The authenticated version is available online at:} \href{https://doi.org/10.1103/2f4r-k8lr}{https://doi.org/10.1103/2f4r-k8lr}
\end{abstract}
\maketitle

\section{Introduction} \label{sec:intro}

A key challenge in current applied mathematics is to build data-driven models from partial observations of multiscale physical systems. One of the most complex examples is the climate system: a forced, dissipative, and chaotic system whose dynamics spans at least 15 orders of magnitude across spatial and temporal scales \cite{GhilLucarini}. This problem has been extensively studied over the past three decades, mainly in the realm of reduced-order models \cite{LucariniChekroun}. Seminal contributions include the work of Penland and collaborators \cite{Penland89,Penland95,PENLAND1996534}, the strategy of Kravtsov and colleagues \cite{KRAVTSOV,KONDRASHOV201533,STROUNINE2010145,Agarwal}, and the contributions of Majda et al. \cite{majdaVsGhil,majdaConstraints}. Recent proposals include Markov-chain and generative approaches \cite{Andre1,Andre2,GIORGINI2024134393,GiorginiScore,CycloLudoAndre}, as well as autoencoder-based, latent-space formulations \cite{PierreLatent}. Most prior work has focused on evaluating reduced-order models on equilibrium statistics and forecasting, with limited attention to responses to \textit{external} perturbations. Notable exceptions include the contributions of Majda and Qi \cite{majdaSIAM,QiMajdaChaos}. Capturing such responses is key for building models that not only mimic the system's dynamics but also faithfully capture causal relations across degrees of freedom \cite{Baldovin,FabCausal}.\\

Recently, autoregressive models based on deep learning architectures have been proposed as potential \textit{emulators} of complex turbulent systems such as the weather \cite{Pathak2022,gencast,Lam2023,Bi2023,fourcastnet3} and climate \cite{Kochkov2023,Watt-Meyer2023,Watt-Meyer2024,Lucie,OceanNet,Bire,Samudra,Camulator,lucie3D,SamudraACE}. Unlike traditional reduced-order models, these neural architectures aim to emulate the full hierarchy of spatial and temporal scales present in the system. For example, \cite{Samudra} and \cite{Camulator} emulate the full ocean and atmosphere components of state-of-the-art climate models. Such emulators often reproduce stationary statistics and generate forecasts across a range of horizons. They can potentially be used for tuning and sensitivity analysis \cite{M2Lines}, and to improve the study of rare events \cite{Lancelin}, offering a promising direction for fast and efficient tools in climate modeling. However, several limitations remain: for example, long rollouts can suffer from error accumulation that can lead to instabilities \cite{Chris,PedramStability}; rare events are not captured if absent from the training set, limiting their applicability for rare-event analysis in a changing climate \cite{PedramPNAS}; key atmospheric modes, such as the Quasi-Biennial Oscillation and the Southern Annular Mode, are not well represented \cite{Baxter}. A limitation particularly relevant for climate science, as opposed to weather forecasting, is that high skill in reproducing stationary statistics does not ensure accurate responses in perturbation experiments, thereby limiting the use of such models for causal studies of climate dynamics. This has two consequences: (i) neural emulators often fail to generalize in out-of-distribution climate change scenarios, and (ii) they may not capture the correct causal relations across variables, two issues that are likely related. The work of Van Loon et al. \cite{Senne} illustrates this limitation in the context of the ``pattern effect'' problem \cite{blochjohnsonetal2024,Dong2019AttributingPacific}, i.e. the influence of sea surface temperature (SST) warming patterns on the Earth’s energy imbalance. Using the ACE2 emulator \cite{Watt-Meyer2024} trained on the ERA5 reanalysis \cite{era5}, they constructed a Green’s function by perturbing SSTs regionally and measuring the radiative response. The resulting Green’s function failed to capture the causal relation between SST and radiative fluxes in a quantitative way. This is just one example of limitations of neural emulators in capturing responses to perturbations with additional examples in \cite{Samudra,Camulator,ZhangACE}. Even if such issues are addressed in specific cases, as shown by Wu et al (2025) \cite{Wu} in the context of the pattern effect, they highlight a broader challenge: building general-purpose emulators that reliably capture forced responses remains inherently difficult and context-dependent. Such limitations are also likely to become more pronounced in coupled settings. Additionally, emulators' skill in capturing forced responses is often assessed using single trajectories. In climate applications, however, the relevant quantities are probability distributions, particularly the responses of the mean and variance. We will show that focusing on the skill in capturing variance responses can reveal important limitations otherwise hidden when examining only the mean response. The limitations examined here are especially relevant for emulators built from observational data and intended to model ``real-world'' dynamics. \\ 

% Specifically, Cecconi et al. \cite{FutureFromPast} demonstrated how the length of the trajectory data needed to sample the state space, scales exponentially with the dimensionality of the system, leading to serious limitations of data-driven modeling for high-dimensional systems. The success of data-driven modeling is then ascribed to the dynamics being constrained on low dimensional manifolds embedded in a high-dimensional space. Here, we build on the pedagogical perspective in \cite{FutureFromPast} by considering a simplified low-dimensional system from the outset and exposing limitations in the case of a partially observed state vector.

While the field of neural emulation is advancing rapidly and yielding impressive results, it is useful to step back and examine its fundamental limitations in capturing forced responses, both from a theoretical standpoint and in controlled settings. In the first part of this paper, we build on a series of previous works by Vulpiani and collaborators on the limitations of data-driven modeling \cite{FutureFromPast,Hosni,BaldovinEntropy}.  We consider a simplified stochastic dynamical system and ask the following question: \textit{Can neural autoregressive models, trained on (practically) infinitely long time series, accurately capture responses to general external perturbations?} In addressing this question, we focus on two relevant scenarios in data-driven-only modeling:
\begin{itemize}[noitemsep]
\item \textit{Unknown equations, fully observed state vector.} Data-driven modeling when the entire state vector is available. Here, we assume no knowledge of the underlying equations, but we have access to very long time series of the complete system state vector.\\
\item \textit{Unknown equations, partially observed state vector.} Data-driven modeling in the presence of partial observations. This reflects common constraints in modeling high-dimensional systems from data, where the full set of relevant variables is typically unobserved \cite{e20070509}, for example because of  unresolved fast and small-scale processes, or even unknown \cite{BaldovinEntropy}. In this more realistic setting, we again assume no knowledge of the underlying equations and only partial access to the system’s state vector.
\end{itemize}
To tackle such questions we proceed pedagogically by focusing on a multiscale ``triad model'' first proposed by Majda and collaborators in \cite{NormalForms,Majda2010}. The model mimics the abstract structure of dynamical cores of large-scale climate models with a linear operator and energy conserving quadratic nonlinearities. It provides a simplified yet relevant mathematical framework to unambiguously explore fundamental (rather than technical) limitations in capturing forced responses in climate emulators. We address the questions above by training neural stochastic emulators on a long integration of the triad model and test the models' performance on stationary and perturbed cases. In doing so, we face the following issue: quantifying how an emulator respond to general forcing is an ill-posed question, as there are infinite possible forcings. Nonetheless, the linear response can be computed by convolving the forcing with an impulse response operator \cite{MajdaBook,Marconi}. This operator can be inferred from the model and is a \textit{building block} that an emulator must capture to reproduce the response to general external perturbations. Therefore, linear response theory is here proposed as a rigorous framework to evaluate neural models beyond stationary statistics and probe causal mechanisms \cite{Baldovin}. In the case of a fully observed state vector we show that the inferred neural model can reproduce stationary and perturbed statistics of the system. In the case of partial observations, as in realistic settings, the performance of a data-driven model necessarily depends on (i) the choice of proper variables to model and (ii) suitable parameterizations of unobserved processes. We then note that the lack of clear scale separation in realistic applications further complicates the problem, leading to non-unique coarse-graining choices tailored to specific goals. These issues are illustrated through a real-world example in a later section. The field of reduced order modeling, introduced earlier, has long sought to confront these challenges by formulating effective Langevin equations capturing the coarse-grained, slow dynamics of the system \cite{PENLAND1996534,Gammaitoni,Dalmedico}, though typically without addressing responses to external perturbations. Therefore, we will argue that a promising perspective is to build on such established approaches, emphasizing coarse-grained dynamics and stochastic parameterizations, while using neural networks as powerful tools within this framework.\\

We emphasize that the focus on a simplified dynamical system is not a limitation in this context. In fact, the goal is not to demonstrate the success of a method, but to illustrate, by analogy, fundamental issues of data-driven modeling that are already evident in low dimensions and are likely to become more pronounced in higher-dimensional settings \cite{Lenny2002}. The first part of this paper is pedagogical in nature, aiming to bridge perspectives across researchers in statistical physics, climate science, and machine learning, while also providing practical methodological tools for the design and evaluation of data-driven models.\\ 

Building on these ideas, we conclude this study by considering a real-world example. Throughout this paper, we emphasize, when possible, the distinction between ``real-world examples/systems'' and merely high-dimensional dynamics. By real-world examples we refer to situations where (i) we observe only the temporal evolution of a subset of variables without access to the full state vector (i.e. a projection of the full dynamics) or (ii) when even in the presence of all variables the system encompasses a huge number of spatial and temporal scales. Both cases (i) and (ii) lead to the need to formulate \textit{effective} rather than \textit{primitive} equations for both practical and conceptual reasons \cite{Gammaitoni}. For an early instance of case (ii), see the work of Charney (1948) \cite{charney1949},  Charney et al. (1950) \cite{neumann} and the discussion in Dalmedico (2001) \cite{Dalmedico}. The example we consider is the “pattern effect” introduced above, which requires Green’s function experiments to quantify the response of top-of-atmosphere (TOA) radiative fluxes to perturbations in the sea surface temperature (SST) field. To this end, we develop a simplified stochastic neural model that emphasizes a large-scale representation of SST and global-mean net radiative flux, rather than attempting to resolve all spatial and temporal scales of variability as in current emulators. We discuss the coarse-graining choices relevant for model construction, as well as a practical and efficient way to fit multiplicative noise from data. The model largely reproduces stationary statistics, especially the most energetic SST mode, corresponding to the El Ni\~no Southern Oscillation \cite{ENSOcomplexity}, and it enables Green’s function–like experiments to infer direct causal relations between SST and TOA radiative flux. The discussion also addresses the model’s limitations and potential methodological developments.\\

The paper is organized as follows: in Section \ref{sec:theory} we introduce the theoretical ideas underpinning this work. In Section \ref{sec:analogy} we discuss key conceptual limitations in data-driven modeling of real-world dynamical systems. In Section \ref{sec:real-world} we present a real-world application. A summary of the main findings and proposals follow in Section \ref{sec:discussion} and Section \ref{sec:conclusion}.

\section{Theoretical Framework: Data-driven stochastic climate modeling and response theory} \label{sec:theory}

We begin by outlining the main theoretical concepts that underpin this work. First, we briefly introduce general ideas for reduced-order stochastic modeling relevant for climate dynamics. We then review key aspects of response theory, which offers a rigorous framework to move beyond the evaluation of stationary dynamics, probe causal mechanisms, and ultimately guide model design.

\subsection{Reduced-order stochastic climate models: theoretical and practical considerations} \label{sec:ROMs}
A general dynamical system can be represented in the form
\begin{equation}
d \mathbf{u}/dt = \mathbf{F} + \mathbf{L}\mathbf{u} + \mathbf{n}(\mathbf{u}),
\label{eq:climate_eq}
\end{equation}
%In the case of analysis of climate anomalies (after removing diurnal or yearly cycles) we can then focus on the case $\mathbf{F} = 0$.  
where $\mathbf{F}$ is an external forcing, and $\mathbf{L}$, $\mathbf{n}(\mathbf{u})$ represent linear and nonlinear operators, respectively. $\mathbf{u}$ is not necessarily finite dimensional, and in the context of coupled atmosphere-ocean variability, it could in principle represent the deviation of all fields, at all scales, from their climatology \cite{Penland95}. However, for obvious practical reasons, Eq. \eqref{eq:climate_eq} needs to be simulated on finite spatial and temporal resolutions. In contrast to numerical weather prediction, which demands high spatial and temporal resolution, low-frequency climate variability is governed by a few dominant large-scale modes \cite{Penland95,KONDRASHOV201533,MajdaBook,Majda2010,Dubrulle,FabCausal}. The goal of reduced-order stochastic climate models is to capture the evolution of these modes and their interactions. This modeling choice allows us to move away from a grid-based representation of the system, emphasizing resolution-independent modes as fundamental components of the system. The cumulative effect of unresolved fast variables, present at smaller spatial scales, is parametrized as stochastic terms. More formally, we decompose $\mathbf{u}$ into slow and fast components, $\mathbf{u} = (\mathbf{x}, \mathbf{y})$, leading to an \textit{effective} Langevin equation of the form:
\begin{equation}
d \mathbf{x} = \mathbf{f}_{eff}(\mathbf{x})dt + \mathbf{\Sigma}(\mathbf{x})d\mathbf{W},
\label{eq:eff_eq}
\end{equation}
where $\mathbf{f}_{eff}$ represents the deterministic ``drift'' dynamics, $\mathbf{x}$ are the slow variables and $\mathbf{W}$ is the standard Wiener process. The noise covariance matrix $\mathbf{\Sigma}$ can be independent of $\mathbf{x}$, corresponding to additive noise, or state-dependent $\mathbf{\Sigma}(\mathbf{x})$ corresponding to multiplicative noise. In the latter case, the
equation is here interpreted in the It\^{o} sense. Equation~\eqref{eq:eff_eq} is mathematically justified in the limit of infinite time-scale separation between fast and slow modes \cite{GhilLucarini}. In the most general (and often intractable) setting, the effective dynamics of the slow variables also involve memory effects induced by the fast degrees of freedom, leading to non-Markovian dynamics. Kondrashov et al. (2015) \cite{KONDRASHOV201533} addressed this problem in a practical and efficient way by introducing the framework of multilevel stochastic models. Nonetheless, appropriate coarse-graining procedures have been shown to effectively average out non-Markovian contributions in many climate applications, yielding useful Markovian approximations \cite{Penland89,Penland95}. We refer the reader to \cite{LucariniChekroun} for a review on the role of non-Markovian contributions in effective climate equations and to \cite{Wouters,Vissio2018} for specific examples.\\

Reduced-order stochastic climate models, as in Eq.~\eqref{eq:eff_eq}, provide an efficient alternative to state-of-the-art climate models for studying low-frequency climate variability. They are much more computationally efficient and allow us for mechanistic understanding of climate dynamics by isolating core processes and examining their qualitative behavior. This helps bridging the gap between simulations and understanding \cite{HeldGap}. This has fueled the development of low-dimensional stochastic models learned either from data \cite{Hasselmann,Penland89,Penland95,majdaSIAM,majdaVsGhil,KONDRASHOV201533,LucariniChekroun,nanChen} or from known equations \cite{MTV1,MTV2,MTV3,MTV4,NormalForms}, as well as inference-based techniques to infer interactions across modes of variability \cite{Kane,Donges,Ilias_1,FabCausal}.\\

In practice, given a phenomenon of interest, we typically begin with spatiotemporal data consisting of $N$ time series of length $T$. We then often proceed as follows:\\

\paragraph*{Coarse-grained dynamics.} In spatiotemporal dynamical systems like climate, coarse-graining involves: (i) choosing a limited set of climate fields (e.g. temperature, velocities etc.), (ii) retaining a few relevant modes or large-scale averages, (iii) restricting the range of temporal scales. Throughout the paper, we will often refer to the so-called \textit{proper} variables as the coarse-grained representation obtained via (i–iii). The relevant modes are often identified through Empirical Orthogonal Functions (EOFs) \cite{MajdaPerspective,STROUNINE2010145}, by retaining the first $n < N$ of such bases. The correspondent principal components (PCs), serve as the reduced time series for model training. Although EOFs are convenient, they may not be always optimal, and alternatives such as principal interaction patterns \cite{CROMMELIN}, community detection \cite{FabCausal} and autoencoders \cite{latentPierre,PierreLatent} may be more suitable depending on the application. The performance of the final model is sensitive to all coarse-graining choices (i–iii); see, e.g., \cite{HeldFDT,Pedram2}. In climate dynamics, there is no clear space- and time- scale separation. In this case, identifying the “appropriate” coarse-grained representation depends on the specific goals of the researcher and remains one of the most difficult challenges in data-driven modeling, as no general mathematical principle exists to guide such choices from data alone in high-dimensional systems \cite{FutureFromPast,Baldovin,Baldovin}.\\

\paragraph*{Inferring a stochastic model.} Given the coarse-grained data $\mathbf{x}$, the goal is to infer a stochastic Markovian model as in Eq.~\eqref{eq:eff_eq}. We summarize several traditional approaches that motivated our work, while acknowledging that many other methods exist beyond those discussed here. The simplest example is the Linear Inverse Model (LIM) proposed by Penland \cite{Penland89}, where the drift is linear and the noise is additive, both estimated directly from data. LIMs have shown forecasting skill for sea surface temperature anomalies comparable to climate models \cite{Penland95,PENLAND1996534,Penland_Fowler,Capotondi}. Recent developments extended the LIM framework to include colored noise \cite{coloredLIM} and nonlinearities \cite{LIM_nonlinear}. Kravtsov et al. (2005) and Kondrashov et al. (2015) \cite{KRAVTSOV,KONDRASHOV201533} generalized this framework through multilevel regression, incorporating quadratic drift terms and non-Markovian effects. These models perform well for stationary statistics \cite{STROUNINE2010145} and forecasting \cite{KONDRAENSO}. The work of Majda and Harlim extended the multilevel regression framework by proposing stability constraints \cite{majdaConstraints}. A theoretically grounded alternative is the approach of Majda, Timofeyev, and Vanden-Eijnden \cite{MTV1,MTV2,MTV3,MTV4}, dubbed MTV by the authors, which derives effective equations for the coarse-grained variables $\mathbf{x}$ rather than assuming them. MTV has been successful for simplified models, though less so for complex climate systems, likely due to choices of coarse-graining steps (i-iii) shown above rather than intrinsic limitations, see also discussion in  \cite{STROUNINE2010145}. We will expand on this approach in Section~\ref{sec:analogy} and Appendix~\ref{app:MTV-triad}.\\

Overall, these works demonstrated that robust reduced-order stochastic climate models can be built through relatively simple regression strategies. In this paper, we build on these insights and investigate how simple neural architectures can infer both the drift and state-dependent noise in Eq.~\eqref{eq:eff_eq} with minimal assumptions. Our neural models retain a decomposition into linear and nonlinear terms, $\mathbf{L}$ and $\mathbf{n}(\mathbf{x})$, as in the general representation of dynamical systems in Eq. \eqref{eq:climate_eq}.

\subsection{Response theory: a (very) brief overview} \label{sec:response_theory}

In this paper, we argue that data-driven models should be tested on their ability to capture both stationary statistics and responses to external perturbations. Assessing responses to arbitrary forcings is ill-posed, as the space of possible perturbations is essentially infinite. A way forward is to consider small impulse perturbations, which allow us to define a response operator and derive general conclusions about causality and forced responses within the linear response regime.\\

\paragraph{Impulse response operator.} Given a $n$-dimensional coarse-grained state vector $\mathbf{x}(t) = [x_1,x_2,...,x_n](t)$, we perturb a degree of freedom $x_j$ with a small, impulse perturbation $\delta x_j(0)$, $x_j(0) \rightarrow x_j(0) + \delta x_j(0)$ \cite{Risken,Baldovin}. We then consider a general observable $A(x_k(t))$ (i.e. a general functional of the system) and quantify its response, in terms of ensemble average, as:
\begin{equation}
\delta \langle A(x_k(t)) \rangle = \langle A(x_k(t)) \rangle_\text{p} - \langle A(x_k(t)) \rangle~,
\label{eq:response_observable}
\end{equation}
where $\langle \cdot \rangle$ refers to ensemble average and the subscript ``p'' refers to the perturbed system. In this paper we are going to focus on two observables: (i) $A(x_k(t)) = x_k(t)$ and (ii) $A(x_k(t)) = (x_k(t) - \mu_k(t))^2$, with $\bm{\mu}(t)$ representing the time-dependent mean of the distribution. The two observables allow us to capture time-dependent responses in the mean and variance of the perturbed distribution respectively. The impulse response operator $\mathbf{R}(t) \in \mathbb{R}^{n,n}$ is then defined as:
\begin{equation}
R_{k,j}(t) = \lim_{\delta x_{j}(0)\to0} \frac{\delta \langle A(x_k(t)) \rangle}{\delta x_{j}(0)}.
\label{eq:pulse_response}
\end{equation}
We refer to Appendix \ref{app:response_operator} for details on how to infer $R_{k,j}(t)$ in practice. The utility of the inferred impulse response operator is two-fold: 
\begin{itemize}
    \item \textit{Causal inference.} Understanding causal relations across degrees of freedoms in physical systems requires to study the effect of interventions on the system \cite{IsmaelNew}. The inferred response operator quantifies time-dependent cause-effect interactions across degrees of freedoms $x_j$ and $x_k$ \cite{Aurell,Lucarini2018,Baldovin,LucariniPRL,FabCausal}. We refer to Appendix \ref{app:response_operator} for details on the relation between the response operator and causality.
    \item \textit{Response to generic time-dependent forcing.} The response operator $\mathbf{R}(t)$ can be used to predict the response of the system to small, but general, time-dependent external forcing $\mathbf{f}(t) \in \mathbb{R}^{n}$ as:
    \begin{equation}
    \delta \langle A(x_k(t)) \rangle = \sum_j \int_0^t R_{k,j}(\tau) f_j(t-\tau)  d\tau.
    \label{eq:convolution}
    \end{equation}
    Importantly, while we assume linearity in the system's response, the formulas above are valid for both linear and nonlinear systems.
\end{itemize}
A data-driven model capable of accurately capturing the impulse response operator will also capture the time-dependent causal interactions across degrees of freedom. Response theory can then be used to assess the skills of emulators by going beyond purely statistical metrics such as r-squared, assessing the causal sensitivity of the system to external perturbation \cite{FabCausal}. Thus, this approach opens new, rigorous avenues for evaluating and studying data-driven models. Moreover, recovering the impulse response operator ensures a faithful approximation of the system’s linear response to small general forcings. For this reason, we view the impulse response operator as a fundamental \textit{building block} that a data-driven model must reproduce to capture the responses to general perturbations. Importantly, the response operator can be inferred from data alone through the Fluctuation-Dissipation Theorem \cite{Marconi}, as discussed in Section \ref{sec:discussion} and in Appendix \ref{app:response_operator}.

\section{Limitations of data-driven modeling by analogy: the case of partial observations} \label{sec:analogy}

A fundamental limitation of purely data-driven modeling of real-world, multiscale climate systems is that many degrees of freedom remain unobserved. Even when the governing equations are known, as in meteorology, climate fields are only sampled at finite temporal and spatial resolutions, leading to unresolved fast and small-scale processes. The challenge is even greater in settings where governing equations are unknown and provide no guidance for emulator design. In both cases, data-driven approaches can reproduce the variability of the retained variables, including their probability distributions, autocorrelation functions, and short-term forecasts. However, the response of probability distributions to perturbations, particularly the mean and variance in realistic climate applications, can uncover structural deficiencies stemming from unresolved dynamics that would otherwise remain hidden from stationary statistics alone. The goal of this section is to illustrate such limitations within a controlled mathematical setting, where they can be made explicit and unambiguous.

\subsection{A triad model as a simplified prototype for coarse-grained, stochastic climate modeling}

We focus on a simplified yet relevant dynamical system that retains essential features of high-dimensional geophysical flows, making the conclusions of this section meaningful for climate applications. The model was introduced by Majda and collaborators \cite{NormalForms} and later used in \cite{Majda2010} in the context of climate dynamics. In what follows, we present the model and refer to Appendix \ref{app:MTV-triad} for an in-depth discussion on its relevance as a building block for complex turbulent flows. The triad model considered in this study takes the following form: 
\begin{equation}
\begin{split}
d x_1 &= (L_{11}x_1 + L_{12}x_2 + L_{13}x_3 + I x_1 x_2)dt \\
d x_2 &= (-L_{12}x_1 -I x_1^2 -\gamma_2 \epsilon^{-1} x_2) dt + \sigma_2 \epsilon^{-1/2} dW_2 \\
d x_3 &= (-L_{13}x_1 -\gamma_3 \epsilon^{-1} x_3) dt + \sigma_3 \epsilon^{-1/2} dW_3\\
\end{split}
\label{eq:triad_system} 
\end{equation}
This idealized system mimics the functional structure of turbulent geophysical flows, featuring linear couplings $L_{11}, L_{12}, L_{13}$ and energy-conserving quadratic nonlinearities. The parameter $I$ controls the strength of the nonlinearity: when $I = 0$, the model reduces to purely linear dynamics and Gaussian statistics. The quadratic interactions typically arise when geophysical turbulent flows are expressed in terms of orthogonal bases such as EOFs \cite{majdaSIAM}. The parameter $\epsilon$ sets the time-scale separation between the slow mode $x_1$ and the fast modes $(x_2, x_3)$, allowing exploration of multiscale dynamics in a controlled setting. The terms $W_2$ and $W_3$ represent independent Wiener processes. The model generates smooth, unimodal distributions with relatively small (but significant) deviations from Gaussianity, and it exhibits fast decay of correlations. Both features are relevant in the study of low-frequency climate variability \cite{MajdaStructuralStability} and found in both general circulation models and observational data, see e.g. \cite{FranzkeMajda,Judith,Sura}.

\subsection{The case of unknown equations but a fully observed state vector} \label{sec:full_state_vector}

We consider the relatively simple case of a low-dimensional dynamical system with only one characteristic time scale (i.e. $\epsilon = 1$). The parameters chosen for this first test are: $L_{11} = -2$, $L_{12} = 1$, $L_{13} = 1$, $\gamma_2 = 0.6$, $\gamma_3 = 0.4$, $\sigma_2 = 1.2$, $\sigma_3 = 0.8$, $I = 1$, $\epsilon = 1$. We now integrate the system in \eqref{eq:triad_system} with an Euler–Maruyama scheme for a period of length $T = 10^7$ with time step $dt = 0.01$ (i.e. $10^5$ Model Time Units). Given the system's decorrelation time scale (shown later), the resulting time series can be effectively treated as infinitely long. Given this long trajectory of the full state vector $\mathbf{x}(t) = [x_1,x_2,x_3](t)$, we learn data-driven models directly from the time series. We construct both linear and nonlinear stochastic models. We specify the functional forms of these models and assess their ability to reproduce both the stationary and perturbed statistics of the original triad system.

\subsubsection{Functional form of the data-driven models considered} \label{sec:data_driven_models_full_state_vector}

We consider two data-driven models. To keep the main text focused, we leave the technical details of the fitting procedures, along with details of the neural networks, in Appendix \ref{app:fitting}.
\begin{itemize}
    \item \textit{Linear model.} We consider the following linear model:
    \begin{equation}
    d\mathbf{x} = \mathbf{L}\mathbf{x}dt + \mathbf{\Sigma}d\mathbf{W},
    \label{eq:LIM-triad}
    \end{equation}
    $\mathbf{L}$ is a linear operator, $\mathbf{\Sigma}\mathbf{\Sigma}^\text{T}$ is the noise covariance matrix and $\mathbf{W}$ is the standard Wiener process.
    \item \textit{Nonlinear model.} The nonlinear model takes the following form:
    \begin{equation}
    d\mathbf{x} = (\mathbf{L}\mathbf{x} + \mathbf{n}(\mathbf{x}))dt + \mathbf{\Sigma}d\mathbf{W}.
    \label{eq:ANN-triad}
    \end{equation}
    $\mathbf{L}$ is a linear operator and $ \mathbf{n}(\mathbf{x})$ is parametrized by a Multilayer Perceptron (MLP). Crucially, the term $(\mathbf{L} + \mathbf{n}(\mathbf{x}))$ is fitted jointly. In what follows we are going to often refer to this model as ``neural model'' or ``neural emulator''.
\end{itemize}
The two models above are trained on a long simulation of the triad system in Eq. \eqref{eq:triad_system}.\\

Note that this is the only Section in this work where we compare linear and nonlinear autoregressive models. The linear method is considered for two reasons. First, we show that linear inverse modeling (LIM), as originally proposed in \cite{Penland89}, serves as a strong baseline for neural autoregressive models, successfully capturing both the mean and variance of the system’s stationary distribution, the autocorrelation functions, and the ensemble mean response to external perturbations. Second, and consistent with theoretical expectations \cite{Majda2010}, we demonstrate that linear models are fundamentally limited in capturing changes in the full probability distribution: for instance, they cannot represent any response in the variance to external forcings. This limitation restricts their usefulness for studying forced changes in distributions beyond the mean.

\subsubsection{Stationary statistics}

\begin{figure*}[tbhp]
\centering
\includegraphics[width=1\textwidth]{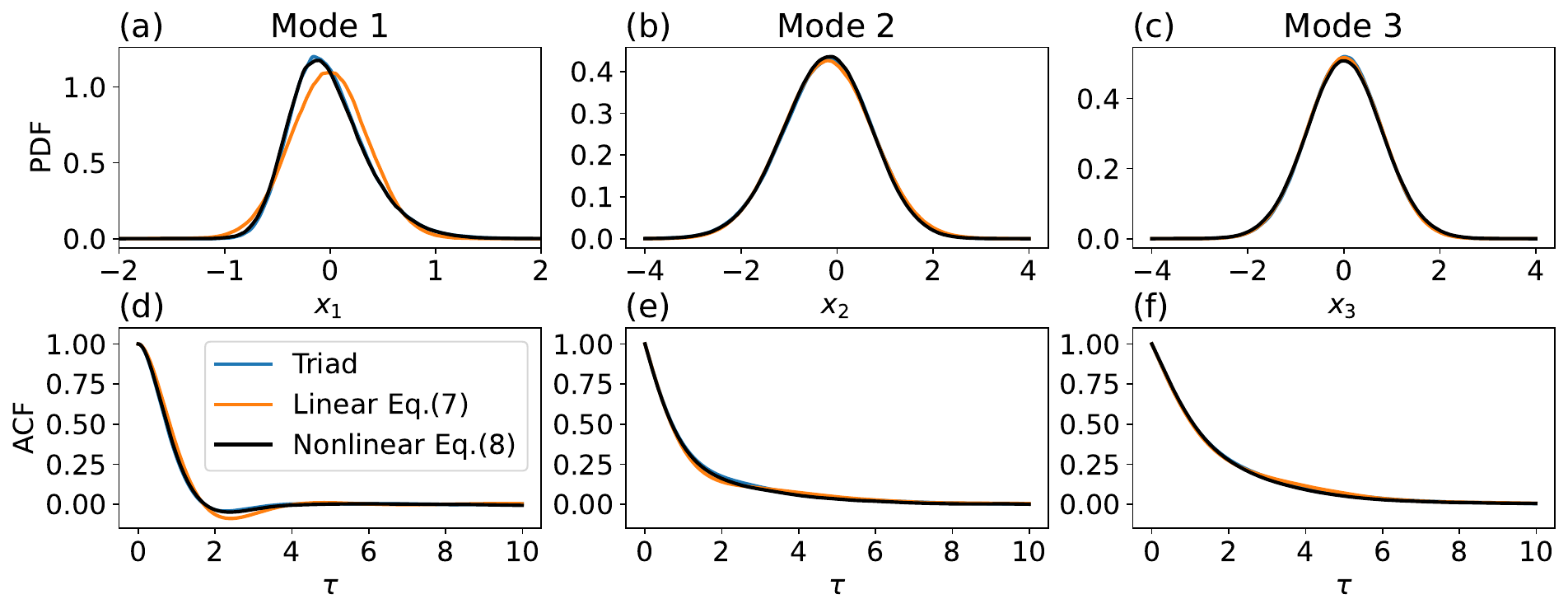}
\caption{First row: stationary distributions of the variables $x_1(t), x_2(t)$ and $x_3(t)$ as modeled by the original triad system in Eq. \eqref{eq:triad_system}, the linear emulator in Eq. \eqref{eq:LIM-triad} and the neural emulator in Eq. \eqref{eq:ANN-triad}. Stationary distributions are obtained after running the three systems for $10^5$ model time units. Second row: same as the first row but for the autocorrelation functions.}
\label{fig:triad_PDF_and_ACF}
\end{figure*}

We integrate the two data-driven models in Eq. \eqref{eq:LIM-triad} and Eq. \eqref{eq:ANN-triad} for $T = 10^7$, yielding time series of the same length as those obtained from the original triad system. We then compare the modeled and original probability density functions (PDFs) and autocorrelation functions (ACFs), as shown in Figure~\ref{fig:triad_PDF_and_ACF}. The neural model described in Eq.~\eqref{eq:ANN-triad} successfully captures the stationary distribution of the triad system. In particular, it outperforms the linear model by accurately reproducing the non-Gaussian features of $x_1$. Both models also reproduce the autocorrelation functions well, as seen in the second row of Figure~\ref{fig:triad_PDF_and_ACF}. In summary, the neural emulator is capable of capturing both the PDF and autocorrelations of the full state vector $\mathbf{x}(t)$. At the same time, the linear model serves as a strong baseline for evaluating more sophisticated data-driven approaches.

\subsubsection{Response of probability distributions to \textit{external} perturbations}

\begin{figure*}[tbhp]
\centering
\includegraphics[width=1\textwidth]{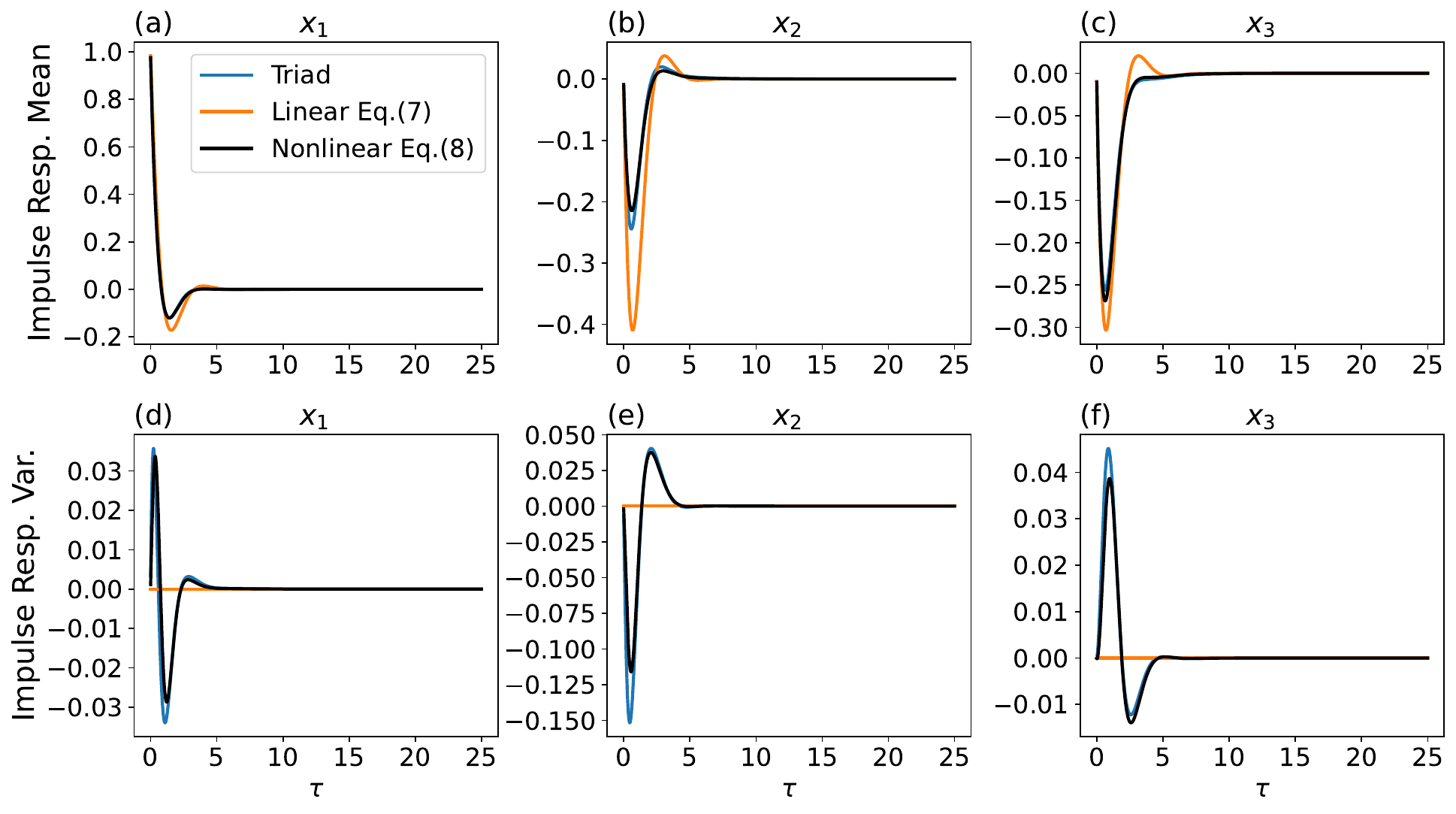}
\caption{Impulse response functions as modeled by the original triad system in Eq. \eqref{eq:triad_system}, the linear emulator in Eq. \eqref{eq:LIM-triad} and the neural emulator in Eq. \eqref{eq:ANN-triad}. First row: response of the ensemble mean to \textit{impulse} perturbations imposed on $x_1(0)$ at time $t = 0$. Second row: response of the ensemble variance to \textit{impulse} perturbations imposed on $x_1(0)$ at time $t = 0$. Impulse response functions are computed using an ensemble size of $N_e = 10^5$ members.}
\label{fig:triad_impulse}
\end{figure*}

\paragraph{Impulse response functions.} We now examine the impulse response of the original triad model (Eq.~\eqref{eq:triad_system}), the linear model (Eq.~\eqref{eq:LIM-triad}), and the nonlinear neural model (Eq.~\eqref{eq:ANN-triad}). Specifically, we analyze the time-dependent response of observables $A(x_k(t))$, with $k = 1,2,3$, to an impulse perturbation imposed on $x_1(0)$ at time $t = 0$. We choose as observables the mean and variance of the distribution. The results, shown in Figure~\ref{fig:triad_impulse}, reveal that the nonlinear neural model captures the mean response of the distribution quantitatively, while the linear model captures it only qualitatively. The most significant differences arise in the response of the variance: the linear model yields exactly zero variance response, as expected \cite{Majda2010}. In contrast, the neural model captures the variance response with relatively high fidelity. We therefore consider the neural model presented here as a skillful model both in terms of reproducing the stationary distribution as well as the response operator.\\

\begin{figure*}[tbhp]
\centering
\includegraphics[width=1\textwidth]{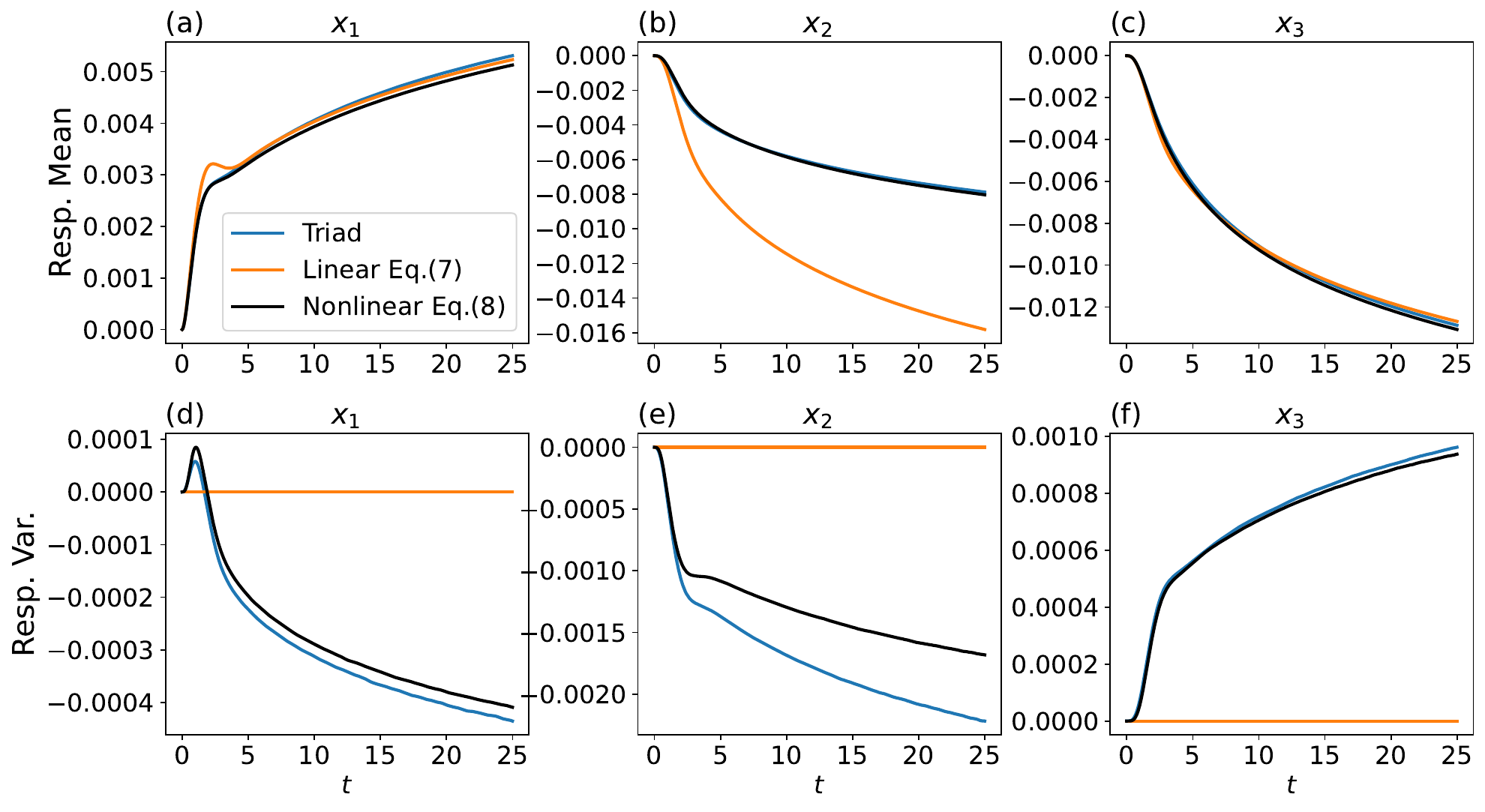}
\caption{Response to a small logarithmic forcing $F = (0.01 log(1+t),0,0)$ imposed on the right-hand-side in Eq. \eqref{eq:triad_system}, the linear emulator in Eq. \eqref{eq:LIM-triad} and the neural emulator in Eq. \eqref{eq:ANN-triad}. First row: response of the ensemble mean to external forcing. Second row: response of the ensemble variance to external forcing. Responses are computed using an ensemble of $N_e = 10^5$ members.}
\label{fig:triad_log-forced}
\end{figure*}

\paragraph{Responses to a specific logarithmic forcing.} we now consider a more targeted test and introduce a logarithmic forcing term to the right-hand side of both the original and data-driven triad model. The forcing is applied only to $x_1$ and takes the form $(0.01 log(1+t),0,0)$ per time step. Each model is integrated for 2500 time steps (25 Model Time Units), and we compute the difference between the forced and unforced simulations in terms of ensemble mean and variance responses. Results are presented in Figure~\ref{fig:triad_log-forced}. The neural model (Eq. \eqref{eq:ANN-triad}) accurately reproduces the forced responses across all three components. The linear model (Eq. \eqref{eq:LIM-triad}) performs qualitatively well but exhibits significant deviations in the ensemble mean response of the second mode $x_2$ as shown in Figure~\ref{fig:triad_log-forced}(b). This is consistent with the earlier Section: the largest error in the linear model’s impulse response also occurred in $x_2$ (see Figure \ref{fig:triad_impulse}(b)). This supports the interpretation that accurately capturing the impulse response operator is critical for a model’s reliability in predicting responses to general, time-dependent forcings. As expected, the linear model is not able to capture the response in ensemble variance. The neural model shows skillful response, capturing the time-dependent changes in ensemble variance relatively well, with the largest error in $x_2$. In general, considering that the changes in variance are very small, we conclude that the neural emulator is skillful in responding to the imposed logarithmic forcing.

\subsubsection{Main takeaways}

Given a (practically) infinitely long trajectory of a stochastic dynamical system, and assuming the ideal condition where the full state vector is observed, it is possible to construct a neural stochastic model that is stable, reproduces the stationary statistics, and responds accurately to small but general external perturbations. The analysis does not rule out that other systems may struggle to capture responses even with full observations. These issues, while important, are considered here as technical, rather than fundamental, and  point to the need for more sophisticated architectures, improved algorithms, or the inclusion of additional physical constraints within existing frameworks.\\ 

While the results shown in this Section are encouraging, they remain far from real-world scenarios. Emulators of climate models would require full access to all climate fields at all temporal and spatial scales, which is often impractical due to the enormous data demands. However, if that requirement were met, then the main challenge would reduce to the need for very long time series for model training \cite{FutureFromPast}. More fundamentally, emulators built from observational data and aimed at capturing ``real-world'' dynamics typically have access only to partial observations of a complex, interconnected system.

\subsection{The case of unknown equations and a partially observed state vector} \label{sec:partial-obs}

We now return to the triad model defined in Eq. \eqref{eq:triad_system} and consider the more realistic scenario of model inference under partial observations and multiple timescales. The parameters chosen for this test are: $L_{11} = -2$, $L_{12} = 0.2$, $L_{13} = 0.1$, $\gamma_2 = 0.6$, $\gamma_3 = 0.4$, $\sigma_2 = 1.2$, $\sigma_3 = 0.8$, $I = 1$, $\epsilon = 0.01$. In this setting the variable $x_1$ clearly is the slow mode, forcing the much faster modes $x_2$ and $x_3$. We again integrate the system in Eq. \eqref{eq:triad_system} with an Euler–Maruyama scheme for a period of length $T = 10^7$ with time step $dt = 0.01$ (i.e. $10^5$ Model Time Units).\\

Unlike the case considered in the previous section, where the entire state vector $\mathbf{x}(t) = [x_1,x_2,x_3](t)$ was available, we now consider a more challenging scenario: only one component of the system is observed (or available or chosen), either $x_1$, $x_2$, or $x_3$. The modeling task is thus reduced to constructing a scalar stochastic model for the observed component alone. We require the scalar stochastic model to satisfy two criteria:
\begin{enumerate}[label=\roman*.]
\item It must reproduce the stationary statistics of the observed component;
\item It must correctly capture the mean and variance response to perturbations as in the original triad system.
\end{enumerate}
Both criteria are nontrivial, with the second posing the greater difficulty. Together they mirror core challenges in data-driven modeling built from partial observations. Satisfying these criteria leads to two central challenges:

\begin{itemize}
    \item \textit{Choice of variable to model.} The reduced-order model depends critically on the choice of the modeled variable. Not all components can be selected to satisfy criteria (i) and (ii) discussed above. The only viable option is to model the slow mode $x_1$, since the cumulative effect of the much faster $x_2$ and $x_3$ can be represented stochastically. Identifying such a \textit{proper} variable is a necessary prerequisite for constructing a meaningful model under partial observations \cite{MajdaBook,FutureFromPast,Lacorata,BaldovinPlos}. In realistic spatiotemporal climate dynamics, such choice translates into selecting appropriate climate fields together with a narrow range of spatiotemporal scales.  We will give a practical example in Section IV. 
    \item \textit{The need for suitable parametrizations.} Having identified the proper variable $x_1$ to model, the next challenge is to appropriately parametrize the cumulative effects of the unresolved components $x_2$ or $x_3$ onto the slow mode $x_1$. This leads to the classic problem of stochastic parametrization \cite{ArnoldHM,Dawson,LucariniChekroun,GhilLucarini,MajdaPerspective,StochBAMS}. As we will show, the success of the resulting model, particularly in capturing forced responses in the variance, hinges critically on the quality of this parametrization.
\end{itemize}

%A brief remark: in partially observed settings, one might be tempted to appeal to Takens’ embedding theorem \cite{Takens}. This is not a practical option: the theorem does not extend to stochastic systems, and it becomes infeasible in high-dimensional settings \cite{Cecconi,Baldovin,Lucente}.

Two brief remarks. (1) In partially observed settings, one might be tempted to appeal to Takens’ embedding theorem \cite{Takens}. This is often not a practical option in real-world systems: the theorem does not extend to stochastic systems, and it becomes infeasible in high-dimensional settings \cite{Cecconi,Baldovin,Lucente}. (2) For the specific case of the triad model in Eq.~\eqref{eq:triad_system}, Majda and collaborators showed that an analytical reduced-order equation for the slow variable $x_1$ can be rigorously derived in the limit $\epsilon \to 0$ \cite{NormalForms,Majda2010}; see discussion in Appendix \ref{app:MTV-triad}. Our aim here is not to test such analytical strategies, which require full knowledge of the governing equations and are often limited in real-world settings; see footnote \footnote{Even moderately more complex systems already expose the practical limitations of rigorous stochastic mode reduction procedures; see Strounine et al.\ (2008) in the context of quasi-geostrophic modeling \cite{STROUNINE2010145}.}. Rather, we use the triad model as a simple conceptual tool to highlight key challenges of data-driven modeling of multiscale systems. In realistic settings, both the governing equations and the full set of relevant variables are typically unknown \cite{Cecconi}: even before specifying a model structure, identifying the slow variables is nontrivial, non-unique and highly problem-dependent.

In what follows, we assume that the proper slow variable, $x_1$, has been correctly identified. Given this, we examine how the choice of stochastic parametrization and temporal coarse-graining influences the ability of neural models to emulate both stationary statistics and responses to perturbations. This discussion serves as a precursor to the more challenging real-world application in Section~\ref{sec:real-world}, where identifying the relevant variables requires a series of nontrivial coarse-graining decisions.

\subsubsection{Functional form of the reduced order models considered} \label{sec:data_driven_models_partial_state_vector}

Given the correct choice of variable $x_1$ to model we now proceed in proposing two data-driven scalar models. Technical details of the fitting procedures are left to Appendix \ref{app:fitting}. We consider two neural models:
\begin{itemize}
    \item \textit{Neural model with additive noise.} We consider the following neural model fitted using an artificial neural network, i.e. a Multilayer Perceptron (MLP). The model takes the following form:
    \begin{equation}
    dx = (L x + n(x)) dt + \sigma dW,
    \label{eq:NN-scalar-additive}
    \end{equation}
    $L$ is a linear operator (here just a scalar), $n(x)$ is a nonlinear operator and $\sigma^2$ is the noise variance. $W$ is the standard Wiener process. 
    \item \textit{Neural model with multiplicative noise.} We consider a neural model with multiplicative:
    \begin{equation}
    dx = (L x + n(x)) dt + \sigma(x) dW.
    \label{eq:NN-scalar-multiplicative}
    \end{equation}
    We use the same drift $(L x + n(x))$ fitted for the additive model in Eq. \eqref{eq:NN-scalar-additive}, so that differences can only arise from the stochastic parametrization. The multiplicative (state dependent) noise $\sigma(x)$ is fitted with an additional neural network, see Appendix \ref{app:fitting}.
\end{itemize}
To train the two models above we consider a long simulation of the triad system in Eq. \eqref{eq:triad_system} using the set of parameters reported in Section \ref{sec:partial-obs}. We then extract only the time series of $x_1$ and train the models above on  this long, scalar time series $x_1(t)$.

\subsubsection{Stationary and perturbed statistics of the scalar model} \label{sec:scalar_model}

We integrate the two data-driven scalar models in Eq. \eqref{eq:NN-scalar-additive} and \eqref{eq:NN-scalar-multiplicative} for $T = 10^7$ time steps, yielding time series of the same length as the time series of $x_1(t)$ used for training. To evaluate the models' ability to capture both stationary and perturbed statistics, we examine the stationary probability distribution and the impulse response operator in ensemble mean and variance.\\

\begin{figure*}[tbhp]
\centering
\includegraphics[width=1\textwidth]{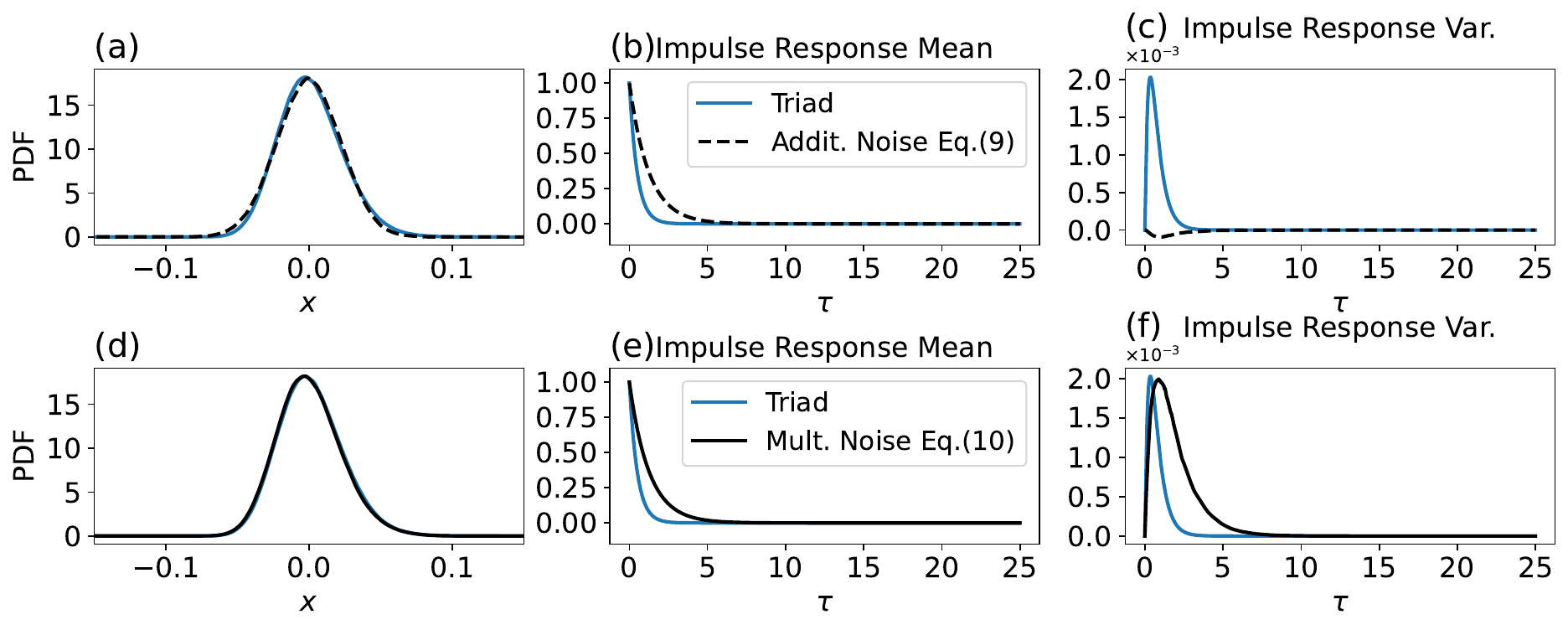}
\caption{First row: stationary probability distribution (panel (a)); ensemble mean response to an impulse perturbation (panel (b)); ensemble variance response to an impulse perturbation (panel (c)) for the variable $x_1$ in the triad system in Eq. \eqref{eq:triad_system} and the variable $x$ in the neural scalar emulator with additive noise in Eq. \eqref{eq:NN-scalar-additive}. Second row: same as the first row but we are now considering the neural scalar emulator with multiplicative noise in Eq. \eqref{eq:NN-scalar-multiplicative}. Impulse response functions are computed using an ensemble of $N_e = 10^6$ members.}
\label{fig:triad_impulse_reduced}
\end{figure*}

The neural model with additive noise (Figure \ref{fig:triad_impulse_reduced}(a)) fails to accurately capture the stationary statistics in a quantitative way. Moreover, the ensemble mean response to an impulse perturbation decays too slowly, and notably, the model fails to reproduce significant responses in the ensemble variance. In contrast, incorporating multiplicative noise (Eq. \eqref{eq:NN-scalar-multiplicative}) leads to a substantially improved representation of the stationary distribution (Figure \ref{fig:triad_impulse_reduced}(d)) and the model now captures a qualitatively correct signal in the ensemble variance, albeit with a similarly slow decay.\\ 

These results indicate that the inclusion of multiplicative noise improves the reduced-order model's ability to represent both stationary statistics and variance responses to perturbations. Moreover, as discussed in Appendix \ref{app:MTV-triad}, the use of multiplicative noise in reduced-order modeling is not an ad-hoc addition but is theoretically justified and explicitly predicted by rigorous mode reduction strategies \cite{MTV1}. We emphasize two key points:
\begin{enumerate}[label=\roman*.]
\item The analysis in Figure \ref{fig:triad_impulse_reduced} highlights the importance of evaluating a model’s ability in predicting responses in higher-order moments, such as the variance. In the simple example discussed here, focusing solely on the ensemble mean response would lead to the incorrect conclusion that the two models respond to perturbations in the same way (see panels (b) and (e) in Figure \ref{fig:triad_impulse_reduced}).
\item Matching the stationary distribution quantitatively does not guarantee equally accurate responses to perturbations, as shown in Figure \ref{fig:triad_impulse_reduced}(d,e,f).
\end{enumerate}

The presence of slowly decaying responses in the ensemble mean and variance to perturbations is indicative of a system that is no longer Markovian. The reduced-order stochastic scalar model is trained solely on the time series of $x_1(t)$. However, the time series $x_1(t)$ is generated by the full triad system in Eq.~\eqref{eq:triad_system} and implicitly reflects the influence of the unobserved variables $x_2$ and $x_3$ on it. The variability of the unobserved variables $x_2$ and $x_3$ manifest as memory effects in the reduced order scalar model for $x_1$. One strategy for addressing such memory effects is the multilevel regression approach first introduced by Kratsov, Kondrashov and Ghil (2005) \cite{KRAVTSOV} and later generalized by Kondrashov, Checkroun and Ghil (2015) \cite{KONDRASHOV201533}. The multilevel regression strategy augments the reduced-order model with a hierarchy of linear equations solving for the residuals from the deterministic dynamics and it can be thought as a practical implementation of the Mori–Zwanzig formalism in statistical mechanics \cite{KONDRASHOV201533,LucariniChekroun}. An alternative, pragmatic approach for handling non-Markovian effects is to coarse-grain the data in the temporal dimension, for instance, by averaging over a time window. This suppresses fast fluctuations within that window leading to an effective ``Markovianization'' of the dynamics. Due to its simplicity and practicality, this method is commonly used in the literature (see e.g. \cite{Penland95}) and is also adopted in the present study.\\

\begin{figure*}[tbhp]
\centering
\includegraphics[width=1\textwidth]{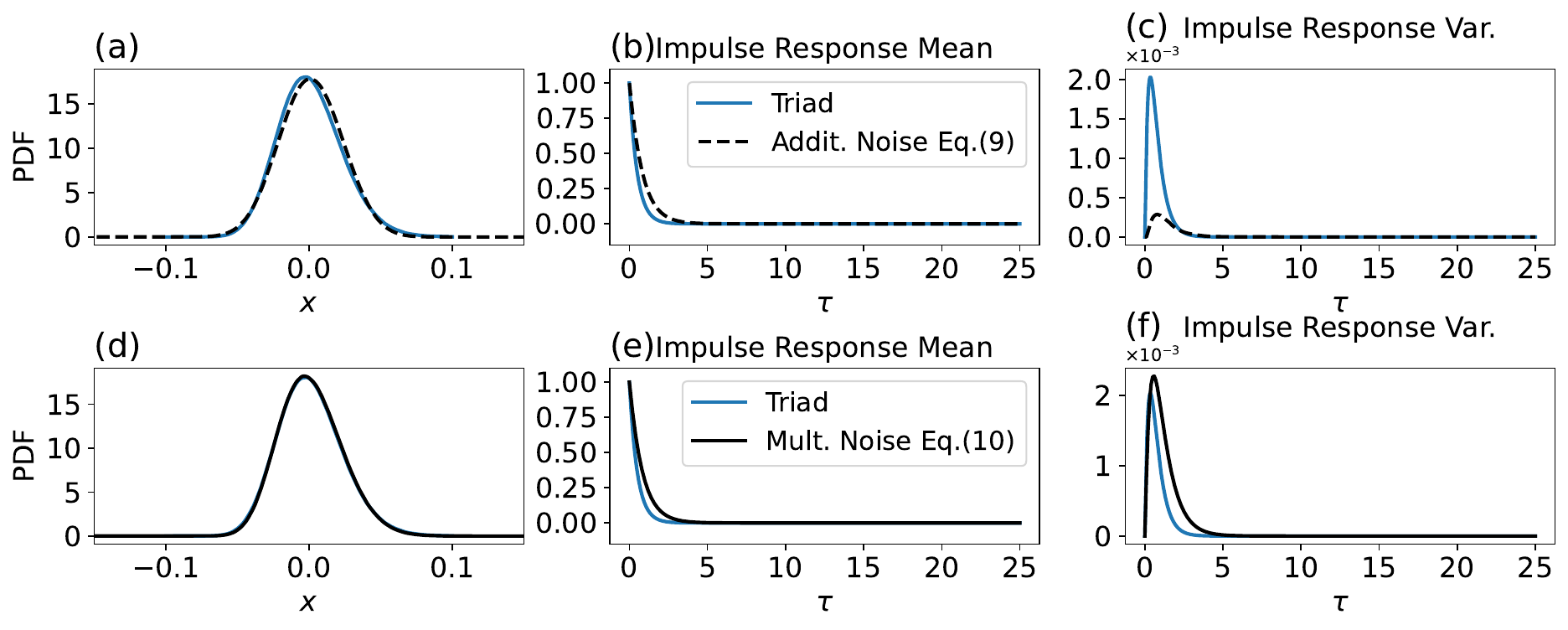}
\caption{Same as in Figure \ref{fig:triad_impulse_reduced} however here the scalar emulators in Eq. \eqref{eq:NN-scalar-additive} and Eq. \eqref{eq:NN-scalar-multiplicative} have been trained on a long trajectory of $x_1(t)$ after coarse-graining by averaging every 10 time-steps. No additional preprocessing was performed for the original triad system in Eq. \eqref{eq:triad_system}, i.e. the blue curves in this Figure and in Figure \eqref{fig:triad_impulse_reduced} are the same.}
\label{fig:triad_impulse_reduced_Coarse_Grained}
\end{figure*}

We consider again the time series $x_1(t)$ from the long simulation of the triad system in Eq.~\eqref{eq:triad_system}. As an additional step, we average $x_1(t)$ over a short temporal window of 10 time steps (equivalent to $0.1$ Model Time Units). This averaging smooths out fast variability and effectively defines a coarse-grained system that is ``more Markovian''. We then retrain the neural models with additive and multiplicative noise on the coarse-grained data. Results are shown in Figure~\ref{fig:triad_impulse_reduced_Coarse_Grained}. Now, the model with multiplicative noise not only reproduces the stationary distribution but also closely matches the variance response, capturing its peak magnitude and deviating only slightly in its temporal decay. One could, in principle, tune the averaging window to match the autocorrelation of the emulated $x(t)$ with that of the original $x_1(t)$, thereby ensuring closer agreement in the decay of responses. Furthermore, training on multiple time steps should also result in a more accurate model. We view these as methodological refinements for future work, and leave it outside the scope of this study.

\subsubsection{Main takeaways}

Even with access to very long time series, modeling multiscale real-world dynamical systems from data alone remains fundamentally constrained by observing only a subset of the full state vector. The most severe limitation arises from the choice of variables: (i) the model’s skill depends critically on which variables are observed and included. Once these variables are fixed, (ii) the influence of unobserved degrees of freedom must be adequately parametrized to capture responses in probability distributions. Although these issues can be illustrated with low-dimensional examples, they become much more severe in realistic spatiotemporal climate systems, where there is little or no scale separation and variable selection involves nontrivial, non-unique decisions about which fields to include, and which spatial and temporal scales to retain — that is, how to construct an appropriate coarse-grained representation of the system.\\ 

These considerations highlight intrinsic limitations of general-purpose emulators of coupled climate systems in capturing responses in mean and variance of the distribution to \textit{external} perturbations. In contrast, reduced-order neural models are often tailored to specific problems, allowing one to leverage prior literature, insights, and results to guide variables and coarse-graining choices. Such prior knowledge can be essential for constructing data-driven models and for clearly delineating their domains of validity.

\section{A real-world example: radiative responses to temperature perturbations in a climate model} \label{sec:real-world}

Building on the insights discussed above, we present a stochastic reduced-order model tailored to a specific real-world, high-dimensional test-case. We introduce the scientific problem at hand and the proposed emulator to tackle this specific question. We then highlight the coarse-graining choices defining the proper variables to model. Finally, we showcase the skill of this emulator in studying causal mechanisms in large-scale climate dynamics, as well as its limitations and future work.

\subsection{Scientific problem at hand} \label{sec:scientific-problem}

\paragraph{The ``pattern effect'' problem.} Radiative forcing (e.g., from changes in $\text{CO}_2$) drives surface warming, which alters sea surface temperatures (SSTs) and associated radiative feedbacks. Recent studies show that even for the same global-mean warming, different SST warming patterns can yield different feedbacks. For example, eastern Pacific warming tends to amplify warming through positive feedbacks, while western Pacific warming has a stabilizing effect \cite{blochjohnsonetal2024,Dong2019AttributingPacific}. This sensitivity of Earth’s energy imbalance to spatial SST patterns is known as the “pattern effect.” The pattern effect describes the fast response of top-of-atmosphere (TOA) radiative fluxes to spatial temperature variations and can be formalized in energy balance models via a Taylor expansion around the steady-state temperature \cite{Meyssignac}.\\

A common approach to study the pattern effect is to relate SST patterns to TOA radiative fluxes using Green’s function experiments \cite{barsugli2002global}, where the radiative response to localized SST perturbations is quantified with atmosphere-only models \cite{zhouetal2017,Zhang2023UsingCM4,Dong2020IntermodelModels,Dong2019AttributingPacific,alessi2023surface}. These experiments are valuable for testing model sensitivities but are computationally costly and typically require large SST perturbations (1–4 K) to obtain a measurable response within short simulations. Such large perturbations can trigger nonlinear behavior, undermining the linearity assumptions of the Green’s function framework \cite{williams2023circus}. This motivates the need for fast, interpretable data-driven models that can simulate responses of large ensembles at low cost and with small perturbation amplitudes. Furthermore, since climate models are known to be biased, data-driven emulators offer, in principle, a complementary way to learn climate dynamics directly from observations.\\

\paragraph{Possible emulator strategies.} Two broad strategies exist for building emulators. One is to construct a general-purpose emulator of the climate system and then apply it to specific problems. However, as shown by Van Loon et al. (2025) \cite{Senne} in the context of the ACE2 emulator and the pattern effect, this strategy may not always succeed. A second approach for large-scale climate dynamics is to design an emulator tailored to a given problem, e.g. \cite{PENLAND1996534}. While this sacrifices generality, it can also offer key advantages by clarifying the choice of variables and scales to model, grounding the design on prior literature, and isolating only the processes most relevant to the problem.

\subsection{A reduced-order stochastic model of surface temperature and radiative fluxes}\label{sec:mult-noise-highD}

\paragraph*{Goals and model formulation.} Our goal is to develop a skillful emulator of large-scale sea surface temperature (SST) and radiative fluxes that captures both stationary variability and responses to perturbations, enabling the study of their causal interactions. In this work, we focus on ensemble-mean responses, which allows for direct comparison with previous studies. We adopt the following functional form for the neural model:
\begin{equation}
d\mathbf{x} = (\mathbf{L}\mathbf{x} + \mathbf{n}(\mathbf{x}))dt + \mathbf{\Sigma}(\mathbf{x})d\mathbf{W},
\label{eq:ANN-SST-TOA}
\end{equation}
where $\mathbf{W}$ is the standard Wiener process. In Appendix \ref{app:fitting}, we propose a practical algorithm to fit the multivariate multiplicative noise $\mathbf{\Sigma}(\mathbf{x})$.\ Differences with the case of additive noise mainly arise in the reconstruction of the stationary density as further discussed in Appendix \ref{app:add-vs-mult-noise}.\\

As outlined throughout the paper, a central challenge in data-driven modeling lies in selecting preprocessing steps to define the coarse-grained dynamics $\mathbf{x}$ to be modeled. These choices, guided by the study’s objectives, are critical for identifying the proper variables, since the climate system involves not only SST and radiative fluxes but also many other interacting variables. This is thus a practical instance of a partially observed state vector. Below, we describe the dataset and preprocessing procedure, which largely follows \cite{FDTPatternEffect}, with further details provided in that work.

\subsubsection{Data, preprocessing and the choice of proper variables} \label{sec:data}

\paragraph{Data.} We consider outputs from the state-of-the-art coupled climate model GFDL-CM4 \citep{cm4}, which provides a 600-year pre-industrial control run with constant $\text{CO}_2$. The ocean component is MOM6 \citep{OM4} with 0.25$^\circ$ horizontal resolution and 75 vertical layers, and the atmosphere is AM4 \citep{AM4a,AM4b} with ~100 km resolution and 33 vertical layers. The model in Eq. \eqref{eq:ANN-SST-TOA} is trained on this control run. We consider two climate fields: sea surface temperature (SST) and the global mean net radiative flux at the top of the atmosphere (TOA). The SST field is restricted to latitudes $[-45^\circ, 45^\circ]$, following earlier studies that highlight the importance of radiative feedbacks in the tropics and subtropics \cite{Dong2019AttributingPacific,Zhang2023UsingCM4}.\\

\paragraph{A possible choice of proper variables.} We coarse-grain the GFDL-CM4 control run in space and time, averaging daily data to monthly means and removing the seasonal cycle to focus on anomalies rather than periodic signals. Considering 1 month averages in TOA fluxes is not arbitrary: the immediate response of TOA fluxes to perturbations in SST is fast (e.g., $\sim20$ days in tropical large-scale convective overturning motion in the GFDL-CM4 model) \cite{FDTPatternEffect}. This averaging filters out higher-frequency processes that mediate surface–TOA coupling, leading to a ``more Markovian'' system . Given so, \textit{direct} causal links between TOA fluxes and SST patterns can be detected as one time step responses; see Appendix \ref{app:response_operator}. We then high-pass filter the data with a cut-off of $10^{-1}$ years, removing (i) low-frequency (multidecadal) oscillations, which are poorly sampled even in 600-year runs, and (ii) slow drifts from incomplete ocean equilibration. This choice sacrifices the ability to capture multidecadal variability but ensures robust sampling of subdecadal statistics, yielding a larger effective sample size. Implicitly, the choice assumes that the response of the net radiative flux to SST perturbations occurs within a 10-year window. Next, we project the SST field $\mathbf{X}(t)\in \mathbb{R}^{N}$ onto an empirical orthogonal functions (EOFs) basis:
\begin{equation}
\mathbf{X}(t) = \sum_{i=1}^n x_i(t)\mathbf{e}_i + \sum_{j=n+1}^N y_j(t)\mathbf{e}_j,
\label{eq:EOF_decomposition}
\end{equation}
and retain the first $n=20$ modes. The combined steps of monthly averaging, high-pass filtering and projection over EOF bases define a coarse-grained representation of the SST variability. This choice follows earlier work showing the skill of linear Markov models for SST built on similar preprocessing \cite{Penland95,PENLAND1996534}. For radiative fluxes, we focus on the global mean TOA net flux, as the TOA flux exhibits little spatial structure and we are mainly interested in its average value. The state vector is then defined as $\mathbf{x}(t)\in\mathbb{R}^{n+1}$, consisting of the $n$ SST principal components concatenated with the global mean TOA flux.\\

Crucially, these choices yield an imperfect model that is skillful in some respects—such as capturing probability distributions and ensemble-mean responses to temperature perturbations—but inadequate in others, including autocorrelation structure and responses in ensemble variance. We discuss these successes and limitations in the following sections.\\

\paragraph{Forced simulations} We will test the model’s ability to reconstruct changes in global-mean TOA radiative flux given prescribed SST trajectories from two 150-years long forced runs: $1$pctCO$_2$ and $4\times$CO$_2$. The $1$pctCO$_2$ experiment imposes a 1\% $\text{CO}_2$ increase per year starting from pre-industrial conditions, while the $4\times$CO$_2$ experiment abruptly quadruples $\text{CO}_2$ at $t=0$. A complication is that TOA fluxes respond directly to rising $\text{CO}_2$, whereas our reduced-order model only uses SST and fluxes and does not explicitly include $\text{CO}_2$. To address this, we remove the portion of change of the net TOA radiative flux that is attributable solely to rising $\text{CO}_2$. This allow us to isolate and study the component of radiative feedback that is driven by changes in SST alone. This correction exploits the logarithmic dependence of global-mean forcing on $\text{CO}_2$ concentration \citep{Pierrehumbert,Romps}. As is standard, we subtract $8 \text{Wm}^{-2}$ in the $4\times$CO$_2$ run \cite{Zhao22investigation}, and apply a time-dependent correction $\alpha \log (C_t/C_0)$ in the $1$pctCO$_2$ run, where $C_t$ and $C_0$ denote $\text{CO}_2$ concentrations in the forced and control simulations.

\subsubsection{Stationary statistics and forced responses}

\paragraph{Stationary statistics.} Given the $n+1$ dimensional state vector $\mathbf{x}(t)\in \mathbb{R}^{n+1}$ defined above, we train the emulator in Eq.~\eqref{eq:ANN-SST-TOA} on the long control run of GFDL-CM4. The trained model is then used to simulate a trajectory of equal length (i.e. 600 years at monthly time step). The stationary probability distributions of the most energetic SST mode, corresponding to the El Ni\~no Southern Oscillation (ENSO) \cite{ENSOcomplexity}, and the net radiative TOA flux are well reproduced Figures~\ref{fig:SST-TOA}(a,b). Statistics and comparisons with the case of additive noise are reported in Appendix \ref{app:add-vs-mult-noise}. The main discrepancies arise in the autocorrelation functions, shown for the first SST mode and the TOA flux in Figure~\ref{fig:SST-TOA}(c,d). This limitation is further discussed in Section ``Limitations'' below.\\

\paragraph{Forced responses.} The pattern-effect problem (Section~\ref{sec:scientific-problem}) concerns identifying causal links between SST warming patterns and TOA flux changes. Before using our emulator for Green’s function–like studies to investigate the pattern-effect, we first assess its skill as a reconstruction tool. Specifically, we reconstruct the global-mean TOA flux from two prescribed, forced SST trajectories, i.e. from the $1$pctCO$2$ and $4\times$CO$2$ simulations. Forced SSTs are projected onto the first $n=20$ EOFs from the control run, yielding coarse-grained trajectories $\mathbf{x}_{\text{SST}}^{f}(t)$, $f$ denotes “forced.” Given $\mathbf{x}_{\text{SST}}^{f}(t)$, the emulator returns the TOA variable $x_{\text{TOA}}(t+1)$ at the next time step, initialized with $x_{\text{TOA}}(0)=0$. Results (Figure~\ref{fig:SST-TOA}(e,f)) show that the emulator largely reproduces TOA flux changes in both scenarios. This reconstruction is possible because of the focus on one-month averages, reducing the time-dependent TOA response, driven by an atmospheric reorganization at sub-monthly scales, into a one-step prediction. For the 1pctCO$_2$ experiment, the linear trend in net TOA flux is $-0.043 W m^{-2} yr^{-1}$ in GFDL-CM4 and $-0.034 W m^{-2} yr^{-1}$ in the emulator. We then compute correlations between the reconstructed and simulated radiative flux at the TOA, after removing a linear trend in 1pctCO$_2$ and an initial 50-year transient in 4xCO$_2$. At monthly resolution, correlations are $r = 0.37$ (1pctCO2) and $r = 0.43$ (4xCO2). If yearly averages are considered (as in the current literature, see e.g. \cite{blochjohnsonetal2024}) these increase to $r = 0.70$ (1pctCO2) and $r = 0.74$ (4xCO2). All correlations are significant at the $5$ sigma level under a null hypothesis of zero correlation, tested via bootstrap resampling with replacement.

\begin{figure*}[tbhp]
\centering
\includegraphics[width=1\textwidth]{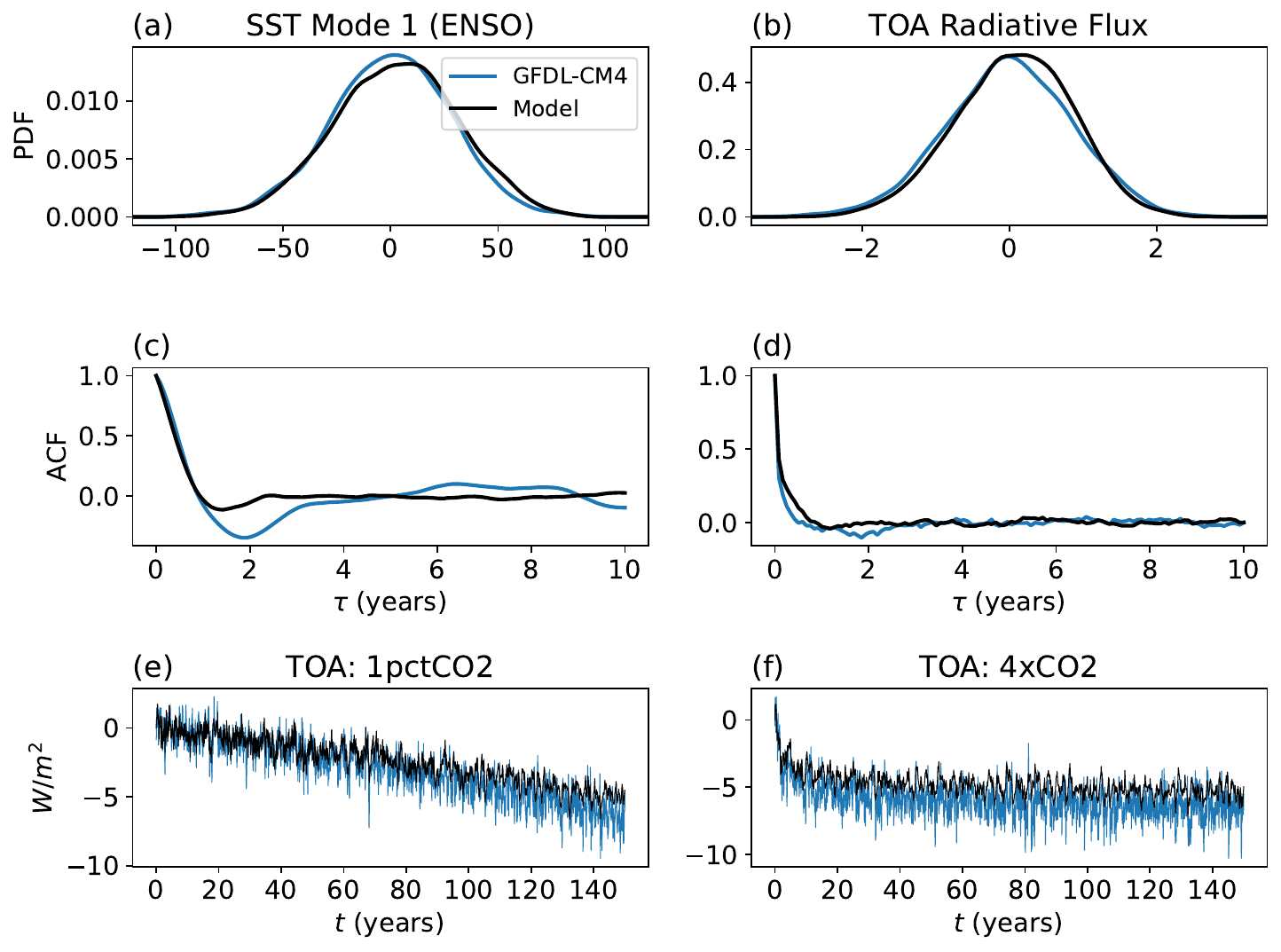}
\caption{Panel (a): stationary distributions obtained for the 1st principal component of the sea surface temperature (SST) field as in the original GFDL-CM4 climate model and as emulated by the neural emulator with multiplicative noise in Eq. \eqref{eq:ANN-SST-TOA}. Units: nondimensional. Panel (b): same as panel (a) but for the global mean radiative flux at the top of the atmosphere (TOA). Units: $W/m^2$. The stationary distributions have been obtained after simulating the neural emulator for 600 years. Second row: same as the first row but for the autocorrelation functions. Third row: reconstruction of the net radiative flux at the TOA by the neural emulator in Eq. \eqref{eq:ANN-SST-TOA} after constraining the SST degrees of freedom to follow a precomputed trajectory in two forced experiments in the GFDL-CM4 model. The forced experiments considered are the $1$pctCO$_2$ and $4\times$CO$_2$, see Section \ref{sec:data} for details. Note: the net flux at the TOA variable is negative upward (i.e., leaving the planet).}
\label{fig:SST-TOA}
\end{figure*}

\subsubsection{Interpretability and causal understanding through ensemble mean responses}

In the previous section, we showed that the reduced-order neural emulator can reconstruct the net radiative flux at the top of the atmosphere (TOA) from prescribed changes in the sea surface temperature (SST) field. This amounts to the mapping $\mathbf{x}_{\text{SST}}^{\text{f}}(t) \rightarrow x_{\text{TOA}}(t+1)$, where $\mathbf{x}^{\text{f}}_{\text{SST}}(t)\in\mathbb{R}^n$, $x_{\text{TOA}}(t)\in\mathbb{R}$, and $x_{\text{TOA}}(0)=0$. We now ask the following question: \textit{What are the SST patterns that contribute the most to this skillfull prediction?} Addressing this question requires moving beyond prediction tasks and identify the causal influence of individual SST components on radiative fluxes. We frame the problem in terms of linear response theory, focusing on the direct link $x_{j,\text{SST}}(t) \rightarrow x_{\text{TOA}}(t+1)$, where $x_{j,\text{SST}}(t)$ denotes the $j$-th SST component. We refer the reader to Appendix \ref{app:response_operator} for details on the causal interpretation of the response operator.\\

We apply an impulse to the $j$-th SST component at $t = 0$ and compute the ensemble-mean response of the net TOA flux at $t = 1$, denoted $\delta\langle x_{\text{TOA}}(1)\rangle$ (see Appendix~\ref{app:response_operator} for details). This quantity, referred to as $R_{\text{TOA},j}(1)$, represents the one-step ensemble-mean response and is estimated using $10^6$ initial conditions to ensure robust statistics. By focusing on the one-step response, we isolate fast atmospheric adjustments while neglecting the slower oceanic feedbacks (see Section 7 in \cite{FDTPatternEffect} for details). This setup is thus conceptually closer to atmospheric-only Green’s function experiments, where the ocean acts as a prescribed boundary condition rather than a dynamically coupled component of the system.\\

The vector $\mathbf{R}_{\text{TOA}}(1)\in \mathbb{R}^{n}$ is projected back to the full $N$-dimensional SST field using the EOFs, $\mathbf{e} \in \mathbb{R}^{n,N}$:
\begin{equation}
\begin{split}
\mathbf{S} = \mathbf{R_{\text{TOA}}}(1) \cdot \mathbf{e} = \sum_j^n R_{\text{TOA},j}(1) ~ \mathbf{e}_j.
\end{split}
\label{eq:response_1_time_step-projection} 
\end{equation}
At each ocean grid point $i$, the sensitivity map $\mathbf{S} \in \mathbb{R}^N$ approximates the sensitivity $\frac{\partial x_{\text{TOA}}}{\partial x_{\text{SST},i}}$, where $x_{\text{TOA}}$ is the global mean radiative flux at the TOA and $x_{\text{SST},i}$ is the SST at location $i$.\\ 

\begin{figure}[tbhp]
\centering
\includegraphics[width=1\linewidth]{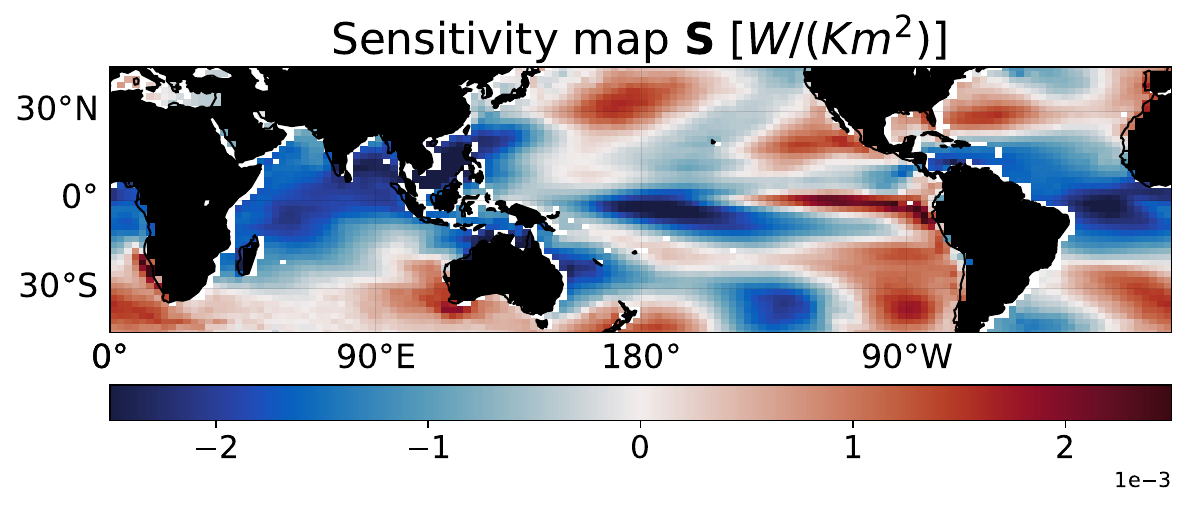}
\caption{Sensitivity maps obtained using the neural emulator in Eq. \eqref{eq:ANN-SST-TOA}. At each SST grid point $i$, we show the one–time-step response in the ensemble mean ($\mathbf{S}^{\mu}$) of the global-mean net radiative flux at the TOA following an impulse perturbation in SST at that location. The computation is carried out in a 21-dimensional reduced space spanned by empirical orthogonal functions, and the maps represent the projection of these low-dimensional results back to the full grid.}
\label{fig:Sensitivity_map}
\end{figure}

The sensitivity map $\mathbf{S}$ is shown in Figure \ref{fig:Sensitivity_map}. The obtained sensitivity map $\mathbf{S}$ is in qualitative agreement with previous results in response theory (Figure 2(a) in \cite{FDTPatternEffect}), with other data-driven approaches (see Figure 7(b) in \cite{kang2023recent} and Figure 2 in \cite{senneNN}), and with the one obtained by directly perturbing the SST boundary condition of atmosphere-only climate models (see third row of Figure 2 in \cite{blochjohnsonetal2024}, see Figure 5(a) in \cite{Dong2019AttributingPacific}). In the case of atmosphere-only climate models, the maps exhibit fewer spatial features due to the coarse resolution of the perturbation patches (spanning tens of degrees) in numerical models experiments. Focusing on smaller perturbation patches (approaching grid-scale perturbations) yields more finely resolved spatial features in climate-model experiments (see Figure 3 in \cite{Leif}), a point sometimes invoked to motivate higher-resolution modeling for forced responses. Our results, however, show that similarly detailed sensitivity patterns can be obtained with a reduced-order model using only 20 principal components. This suggests that high spatial resolution is not required for this problem; what matters is representing the system in an appropriate reduced (coarse-grained) representation. For additional statistical reconstruction approaches to the pattern effect that yield structures similar to those in Figure \ref{fig:Sensitivity_map}, see the recent analysis by Fredericks et al. (2025) \cite{Leif}.\\

The dominant feature of $\mathbf{S}$ is a dipole in the tropical Pacific: negative sensitivity in the west and positive sensitivity in the east \citep{Dong2019AttributingPacific}. This implies that eastern Pacific warming produces a positive radiative feedback, while western Pacific warming produces the opposite. The mechanisms are well established \citep{Pascale,williams2023circus}: in the east, high climatological low-level cloud cover decreases under warming, reducing reflected shortwave radiation and yielding positive sensitivity. In the west, deep convection transports surface warming vertically and then horizontally via gravity waves, strengthening the inversion over the eastern Pacific and other stratocumulus regions, increasing low-level cloud fraction and reflected shortwave radiation, thus producing a negative feedback. This analysis confirms that the accurate reconstructions in the forced regime (Figure~\ref{fig:SST-TOA}(e,f)) arise from physically consistent mechanisms.\\

\subsubsection{Limitations and a few ways forward}

Our coarse-grained representation enabled the development of a data-driven model to study the direct causal interaction $x_{\text{SST}} \rightarrow x_{\text{TOA}}$ through perturbation experiments, while also capturing the stationary distribution well. However, these coarse-graining choices also introduce some drawbacks. We discuss these limitations below, further delineating the model’s domain of validity and outlining possible ways forward:
\begin{itemize}
    \item The model does not fully capture the decay of autocorrelations. For example, the first SST mode in Figure~\ref{fig:SST-TOA}(a) suggests residual non-Markovian behavior. This is not an issue for our focus on \textit{direct} causal links (responses at the shortest time step), but it would matter for responses over longer time horizons. Remedies include: (i) expanding the state vector by including slower oceanic variables and more EOF components, (ii) incorporating memory terms in the stochastic parameterization \cite{KONDRASHOV201533}, or (iii) further coarse-graining the temporal resolution (e.g., 3-month averaging). Such choices will depend on the intended scientific application.
    \item We found that responses in ensemble variance are sensitive to retraining the multiplicative noise term, so we do not present these results. The fitting algorithm in Appendix~\ref{app:fitting} requires far more data than available here (only 7200 samples for training and validation), to achieve high accuracy. In this data-scarce setting, exploring simple analytical forms for the multiplicative noise may be beneficial and it is left for future work. Importantly, data-driven approaches, such as the Fluctuation–Dissipation Theorem (FDT) \cite{MajdaBook,Marconi}, can provide an approximate benchmark for responses. This allows calibration of models with different stochastic terms, particularly to test variance responses where direct benchmarks are often unavailable. This strategy aligns with recent proposals such as \cite{majdaSIAM} and will be explored in future work.
    \item While our emulators are empirically stable, stability is not guaranteed, which may matter if tested against large perturbations. Future work could impose additional constraints on the neural architectures, similar to previous approaches for physically constrained regression models as in \cite{majdaConstraints,KONDRASHOV201533}.
\end{itemize}

\subsubsection{Main takeaways}

In this Section, we developed a stochastic neural emulator that captures the joint evolution of the sea surface temperature field and the global mean net radiative flux at the top of the atmosphere in the GFDL-CM4 climate model. Addressing a specific, well-posed problem, proved crucial in guiding the construction of the model: it enabled us to build on previous literature, restrict attention to just two variables and a limited range of spatial and temporal scales. These simplifications led to a stable and computationally efficient emulator that responds to external forcing and can be used to investigate causal relationships via perturbation experiments. Such experiments focused on very large ensembles ($10^6$ trajectories), currently impossible to perform with climate models.

\section{Discussion} \label{sec:discussion}

In summary, the first part of this work discussed important limitations of purely data-driven models of complex climate systems. We highlighted the potential of combining traditional reduced-order modeling approaches, including state-dependent (multiplicative) stochastic parameterizations, with recent advances in neural networks to study large-scale, low-frequency climate dynamics. We emphasized the need to evaluate such models against both stationary and perturbed statistics. In this context we discussed how response theory provides rigorous guidance to move beyond stationary evaluation and probe causal mechanisms. These ideas were then illustrated through a climate-dynamics problem focusing on inferring causal relations between surface warming and changes in net radiative fluxes. Given careful variable selection, reduced-order neural stochastic models can capture both stationary and perturbed statistics in a computationally efficient way. Below, we expand on these points and outline directions for future work.

\subsection{Limits of data-driven-only climate emulators and the role of stochastic approaches}

In the first part of this paper, we distinguished two possible scenarios in data-driven modeling:
\begin{enumerate}[label=\roman*.]
\item \textit{Unknown equations, fully observed state vector.} In this idealized setting, it is possible to infer a model from data alone that reproduces the system’s stationary statistics and responds accurately to small external perturbations.
\item \textit{Unknown equations, partially observed state vector.} This more realistic case presents fundamental difficulties, even with access to high-quality data and powerful neural model architectures. The most critical limitation is (a) identifying an appropriate coarse-grained representation; once this is determined, (b) the cumulative effect of unobserved degrees of freedom needs to be accurately parametrized. Learning such parametrizations from data remains a central challenge, and we showed that multiplicative noise can be advantageous over additive noise, though it requires a larger dataset for reliable training.
\end{enumerate}
These results highlight intrinsic limitations of purely data-driven modeling of high-dimensional, multiscale climate systems, where the full set of relevant variables is unobserved or even unknown. This is especially true for emulators built to capture real-world dynamics from observational data, where observations are inherently partial. In this context, the ``correct'' choice of variables, scales, and stochastic parameterizations depends ultimately on the specific scientific goals \cite{Smith2001}. This perspective helps explain the continued success of Linear Inverse Models (LIMs) proposed by Penland, C. (1989) \cite{Penland89}, which rely on careful selection of climate fields and preprocessing choices to capture targeted aspects of climate variability; see, for example, \cite{Penland95,PENLAND1996534,Penland2025} among others.\\

We emphasize that the analysis presented here applies to data-driven models of turbulent geophysical systems that must capture both stationary statistics and responses to external perturbations, as in (causal) Green’s function experiments \cite{blochjohnsonetal2024}. For Earth system models, these issues mainly concern autoregressive climate emulators, particularly coupled atmosphere–ocean systems, and should not be confused with short-term weather forecasting using neural networks (e.g., \cite{AIFS,fourcastnet3}), where long-term stability and forced responses are less relevant. Our study also does not address statistics-based emulators for climate change projections, which have proven effective for investigating responses of key climate variables to CO$_2$ forcing \cite{Bjorn,Gosha,AndreEmulation}.\\

\paragraph*{Future work.} The discussion above suggests two possible directions for future research in autoregressive emulators of climate. First, a stochastic formulation of neural autoregressive climate emulators is preferable to a purely deterministic one. Probabilistic formulations align more closely with the inherently stochastic and uncertain nature of climate dynamics \cite{PalmerQJRMS}. Recent advances in weather emulation already point in this direction such as FourecastNet3 \cite{fourcastnet3}, GenCast \cite{fourcastnet3} and the novel work of Alet et al. (2025) \cite{Alet}. New interesting directions also include probabilistic emulation in the framework of stochastic interpolants \cite{ProbabilisticSI,ProbabilisticSIPedram,Nikolaj}. Second, the ability of climate emulators in capturing forced responses in the full probability distribution, akin to Green’s function experiments, remains largely unexplored, especially for responses in higher-order moments. A promising way forward is to evaluate novel architectures on their ability to capture the mean and variance response to external perturbations in a hierarchy of simplified models relevant for large scale climate variability, such as the 2-scale Lorenz–96 system \cite{Lorenz96}. These controlled testbeds enable rigorous quantification of forced responses and could inform the design of future architectures for data-driven climate modeling.

\subsection{Response theory as a theoretical framework for model evaluation}

We argued that the impulse response operator is a building block that a model should capture in order to reproduce causal interactions across degrees of freedom and to accurately respond to external perturbations (Section \ref{sec:response_theory} and Appendix \ref{app:response_operator}). This perspective enables us to move beyond evaluation metrics based solely on stationary statistics and instead consider the effect of external interventions. Assessing impulse responses is straightforward in the case of data-driven emulators built from simple numerical models, i.e. models that allow us to simulate thousands of ensemble members, as in the first part of this study. In such cases, one can actively perturb the reference model to generate a ground truth for the response operator. However, for models constructed from observational data, such direct perturbations are not possible. In this context, one can leverage the Fluctuation-Dissipation Theorem (FDT) to predict the response operator from data alone \cite{MajdaStructuralStability,Baldovin}. Traditionally, the quasi-Gaussian approximation have yield reliable estimates of impulse responses in the ensemble mean \cite{GRITSUN,Majda2010}. Recent advances in score-based generative models have shown the potential to extend these predictions to responses in higher-order moments \cite{LudovicoFabriAndre}.\\

\paragraph*{Future work.} In the absence of direct perturbation experiments, FDT provides a rigorous bridge between stationary and forced dynamics \cite{Marconi,MajdaBook}. With sufficiently long integrations of current climate emulators, one can readily focus on representative large-scale modes and estimate their impulse responses from the stationary record, thereby extending evaluation beyond stationary statistics at minimal computational cost. A recent example illustrating this approach in the climate context can be found in \cite{FabCausal}. Importantly, the analysis reveals the need to test emulators in their ability to respond to higher-order moments. \cite{LudovicoFabriAndre} have shown that coupling the FDT with generative models provide a flexible framework to estimate responses in the full probability distribution. Moreover, the response operator at the shortest time step encodes \textit{direct} causal links, i.e. the input–output relationships across variables in a reduced-order emulator (see Appendix \ref{app:response_operator}). Future work will leverage this information to suppress spurious, non-causal input-output dependencies during training, similarly to the recent work of Chen and Liu (2024) \cite{nanChen}, leading  to causally constrained, sparse reduced-order models.

\subsection{A realistic application: radiative responses to temperature perturbations}

Building on the theoretical insights introduced in the first part, the second part of the paper presented a stochastic neural emulator for the coarse-grained dynamics of sea surface temperature (SST) and net radiative flux at the top-of-atmosphere (TOA). The model was trained on a long unforced simulations of a coupled climate model and designed with a specific physical task in mind: capturing radiative feedbacks to SST anomalies. The emulator focuses on the evolution of 20 principal components of the SST field and on the scalar global mean, net radiative flux; unobserved degrees of freedom are parametrized as multiplicative noise through an additional neural network. The emulator largely reproduces the stationary statistics of the system and allows us to reconstruct TOA radiative fluxes in two forced simulations. We inferred the direct causal links between temperature modes and TOA fluxes through a perturbation approach, akin to Green's function-like experiments with climate models \cite{blochjohnsonetal2024}, targeting the so-called ``pattern effect'' \cite{Dong2019AttributingPacific}. Such responses are then projected back onto the EOFs to obtain maps of sensitivity in mean and variance change of the radiative flux at the TOA to SST warming perturbations. We discussed similarities of the sensitivity map for the ensemble mean with recent studies, indicating that the emulator’s skill in reconstructing TOA fluxes under forcing arises from physically meaningful mechanisms.We also identified limitations. In particular, the variance response is sensitive to the training of the multiplicative-noise component, restricting the emulator’s reliability in perturbation experiments beyond those targeting the ensemble mean. We outline possible paths forward to resolve this issue. These findings highlight the potential of reduced-order neural emulators to perform thousands of perturbation experiments, eventually shifting the focus of climate feedback analysis from mean responses to the responses of probability distributions.\\

\section{Conclusion} \label{sec:conclusion}

Modeling the climate system from data alone is constrained by observing only a subset of variables of a  multiscale, high-dimensional dynamical system. This raises the fundamental challenge of identifying the proper variables (both fields and spatiotemporal scales), a fundamental problem in physics emphasized in classical work \cite{Onsager} and revisited in recent studies \cite{Lucente}. Such choices are inherently non-unique, especially because the climate system lacks clear space- and time-scale separation, and are therefore largely determined by the study’s scientific goal rather than by the pursuit of an unattainable perfect model \cite{Smith2001}. In contrast to numerical weather prediction, which demands high spatial and temporal resolution, low-frequency climate variability is dominated by a few recurrent spatiotemporal patterns, or modes of variability, and their interactions. Thus, at least for certain tasks, the multiscale structure of the climate system can be an advantage, allowing us to focus on the coarse-grained dynamics while treating unresolved, fast processes stochastically. Traditional reduced-order models stand out from general-purpose emulators by targeting specific climate problems from the outset, guided by prior literature, physical insight, and a clear understanding of the relevant variables and scales at play.  Such data-driven modeling approach has a long history in climate science \cite{LucariniChekroun}, though most efforts have emphasized reproducing stationary statistics or short-term forecasting \cite{Penland89,Penland95,KONDRAENSO} rather than forced responses. A promising perspective is to build on such established reduced-order approaches, including those developed by Majda and Qi \cite{majdaSIAM,QiMajdaChaos}, while leveraging flexible neural networks architectures to discover low-dimensional latent representations \cite{latentPierre,PierreLatent} and fit deterministic and stochastic components. Furthermore, tools from response theory such as the FDT \cite{FabCausal,LudovicoFabriAndre,ValerioData2} can be used to guide model development and eventually lead to reduced-order models that efficiently capture responses in the mean and variance to external perturbations, advancing causal inference and attribution analyses in climate science.

\appendix

\section{Response operator: practical estimation and causal interpretation} \label{app:response_operator}

\subsection{Inferring the response operator in practice}

Given a model $d\mathbf{x} = \mathbf{f}(\mathbf{x})dt + \mathbf{\Sigma}(\mathbf{x})d\mathbf{W}$ we build the impulse response operator as follows:

\begin{enumerate}
    \item We simulate a very long trajectory of the model, starting from a random initial condition and remove an initial transient. We then sample $N_e$ random points from the simulation to define an ensemble of $N_e$ initial conditions on the model’s attractor. $N_e$ should be very large (i.e. $N_e \gg 1$) in order to sample the whole attractor and approximate averages over the invariant distribution.
    \item For each one of the $N_e$ initial conditions, we impose an impulse perturbation $\Delta_j$ to the degree of freedom $x_j$ at time $t = 0$. The amplitude of the perturbation  should be theoretically infinitesimally small, ensuring linearity of the response even in nonlinear systems. In practice we do as follows: we consider the long time series of $x_j(t)$ from the long control integration above and define $\Delta_j = 10^{-2} \sigma_j$, where $\sigma_j$ is the standard deviation of time series $x_j(t)$.
    
    %For each one of the $N_e$ initial conditions, we impose an impulse perturbation $x_j(0) \rightarrow x_j(0) + \Delta_j \delta(t)$ of size $\Delta_j$ on the $j$-th degree of freedom at time $t = 0$. $\delta(t)$ is the Dirac delta function. The amplitude $\Delta_j$ should be theoretically infinitesimally small, ensuring linearity of the response even in nonlinear systems. In practice we do as follows: we consider the long time series of $x_j(t)$ from the long control integration above and define $\Delta_j = 10^{-2} \sigma_j$, where $\sigma_j$ is the standard deviation of time series $x_j(t)$. 
    \item Therefore, for a given initial condition $\mathbf{x}(0)$, we simulate two trajectories: a control trajectory without perturbation and a perturbed one, where an impulse perturbation $\Delta_j$ has been imposed on the $j$-th degree of freedom at time $t = 0$. This procedure is repeated in parallel for all initial conditions, resulting in an ensemble of $N_e$ paired of control and perturbed trajectories.
    \item At each time $t$ we then estimate the time dependent ensemble average $\langle A(x_k(t)) \rangle$ of an observable $A(x_k(t))$ for both the perturbed and unperturbed run. We refer to the perturbed and control ensemble average respectively as $\langle A(x_k(t)) \rangle_{\text{p}}$ and $\langle A(x_k(t)) \rangle$. The observables considered in this study are going to be: (i) $A(x_k(t)) = x_k(t)$ and (ii) $A(x_k(t)) = (x_k(t) - \mu_k(t))^2$, with $\bm{\mu}(t)$ representing the time-dependent mean of the distribution, therefore quantifying the response in ensemble mean and variance.
    \item We define the impulse response operator for observable $A(x_k(t))$ as
    \begin{equation*}
        R_{k,j}(t) = \frac{\langle A(x_k(t)) \rangle_{\text{p}} - \langle A(x_k(t)) \rangle}{\Delta_j}.
    \end{equation*}
\end{enumerate}

\subsection{Causal interpretation of the response operator}

Causality in natural systems is inferred through external interventions: an agent perturbs the system and the response is measured \cite{IsmaelNew,Aurell}. Climate modeling follows the same principle through Green’s function experiments, where one variable is perturbed and the response of another is evaluated \cite{blochjohnsonetal2024,LucariniPRL}. Linear response theory formalizes this principle, providing a rigorous framework for causal analysis both in models and directly from data \cite{Baldovin,FabCausal,LudovicoFabriAndre}.\\

Because responses are inherently time-dependent, causal links through responses can be distinguished in three forms, all encoded in the time-dependent response operator: (i) direct links, (ii) indirect links, and (iii) cumulative links (or causal sensitivity). In what follows, we use the triad system in Eq. \eqref{eq:triad_system} as a simple example to illustrate these ideas in a clear and unambiguous way. Finally, we summarize recent approaches for estimating responses directly from data and refer to Baldovin et al. (2020) \cite{Baldovin} for details on the theoretical foundations of this approach.\\

\begin{figure}[h!]
\centering
% Panel (a)
% Panel (b)
\begin{subfigure}[b]{0.1\textwidth}
    \centering
    % Use align instead of equation for compactness
    \begin{align*}
    \frac{d x_1}{dt} &= f_1(x_1,x_2,x_3) \\
    \frac{d x_2}{dt} &= f_2(x_1,x_2) \\
    \frac{d x_3}{dt} &= f_3(x_1,x_3)
    \end{align*}
    \caption{Functional form of the triad system in Eq. \eqref{eq:triad_system}.}
    \label{fig:triad_graph_b}
\end{subfigure}
\begin{subfigure}[b]{0.3\textwidth}
    \centering
    \includegraphics[width=\linewidth]{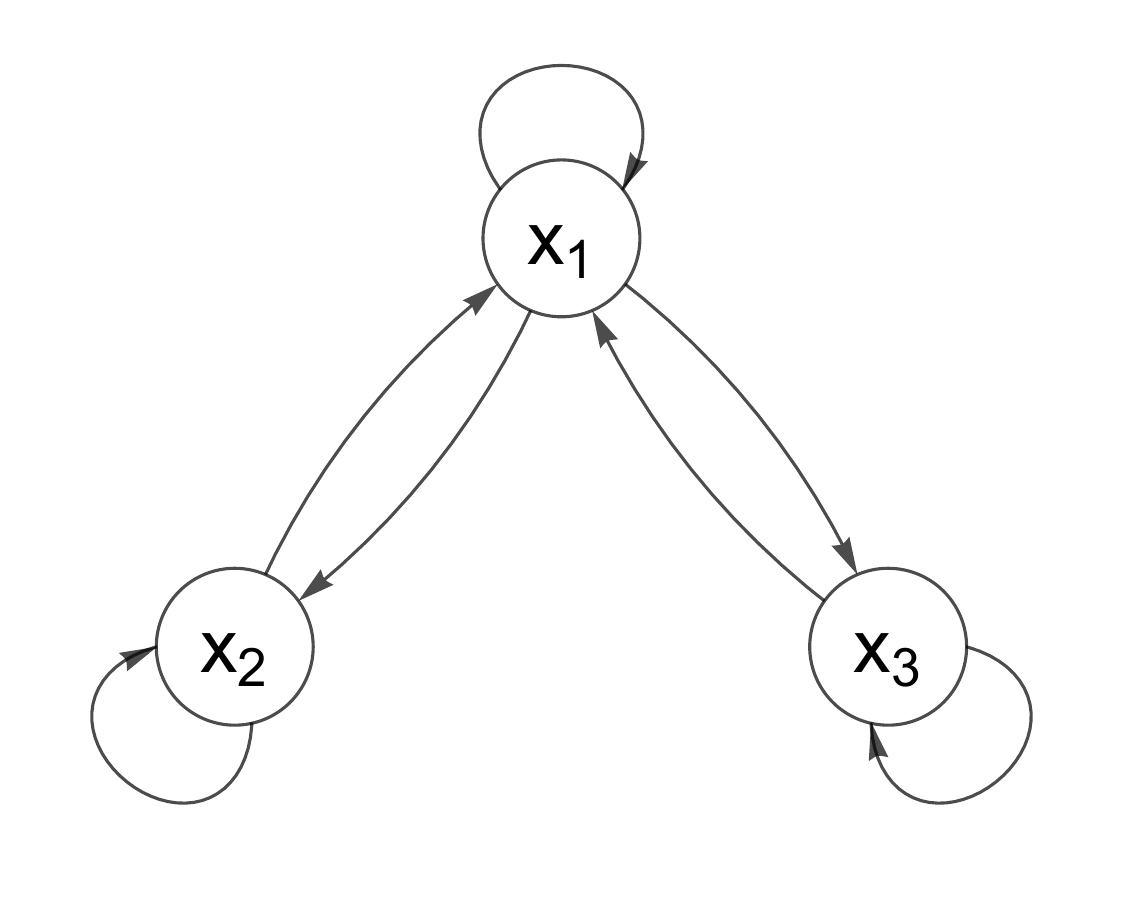}
    \caption{Causal graph of the triad system.}
    \label{fig:triad_graph_a}
\end{subfigure}
\hfill
\caption{Panel (a) Functional form of the triad system in Eq. \eqref{eq:triad_system}. Panel (b) Graph structure of the triad system. Note from panel (b) that causality in dynamical systems can involve cycles, in contrast with the acyclic structures often assumed in applications outside physics, see e.g. \cite{Kusner}.}
\label{fig:causal_triad_combined}
\end{figure}

Consider a high-dimensional, stochastic dynamical system of the kind:
\begin{equation}
\begin{split}
\frac{d x_i}{dt} = f_i(x_1,x_2,...,x_N) + \text{stochastic forcing};\\ ~ \text{with} ~ i = 1,2,...,N ~.
\end{split}
\label{eq:general_system} 
\end{equation}
In this setting, the existence of a \textit{direct} causal link $x_j \rightarrow x_k$ is determined by whether $x_j$ appears in the arguments of $f_k$. More formally, the \textit{absence} of a direct causal link $x_j \rightarrow x_k, j \neq k$ is expressed by: 
\[
\langle \frac{\partial f_k(\mathbf{x})}{\partial x_j} \rangle = 0.
\]
We adopt the view that causal estimates in real-world, high-dimensional stochastic systems are most naturally, and most practically, formulated as ensemble averages, making the ensemble-averaged Jacobian the appropriate structural object for direct causal links. As a concrete example, we return to the triad system in Eq. \eqref{eq:triad_system}, for which the underlying direct causal links are known exactly. This example provides a straightforward, intuitive illustration of how perturbation experiments reveal causal interactions. Figure \ref{fig:causal_triad_combined} reports both (a) its functional form and (b) its causal graph, which encode the same information.
Causal relations can then be probed by adding independent impulse perturbations to each degree of freedom and record the system's response. Responses are computed up to a long time horizon $\tau_{\max}$, with discrete times $t = 0, dt, 2dt, \dots, \tau_{max}$. Since we are interested in causal relations, it suffices to consider the identity observable $A(x_k(t)) = x_k(t)$, i.e. the response of the ensemble mean \cite{Baldovin}. We now clarify the three different types of causal links mentioned at the start of this Section.\\

\paragraph*{Direct causal links.} The direct causal links are encoded in the response operator $\mathbf{R}(dt)$ evaluated at the shortest time step $t = dt$.  The intuition follows directly from the graph representation in Figure \ref{fig:causal_triad_combined}(b) and is rather straightforward.
\begin{itemize}
    \item Any perturbation applied to $x_2$ will affect, on average, $x_2$ itself and $x_1$ after one time step, but not $x_3$. Hence, we infer $\langle \partial f_3(\mathbf{x}) / \partial x_2 \rangle= 0$.
    \item A perturbation applied to $x_3$ will affect, on average, $x_3$ and $x_1$, but not $x_2$, so $\langle \partial f_2(\mathbf{x}) / \partial x_3 \rangle= 0$. 
    \item By contrast, a perturbation to $x_1$ propagates to all three variables within a single step, indicating direct causal links from $x_1$ to both $x_2$ and $x_3$.
\end{itemize}

\paragraph*{Indirect causal links.} Causal relations in physical systems extend beyond direct interactions. A variable $x_j$ may affect another variable $x_k$ indirectly through an intermediate variable $x_i$, forming a path $x_j \rightarrow x_i \rightarrow x_k$. Such indirect links are central in climate dynamics \cite{RungeCausalEffects} and in complex systems more broadly \cite{Ay}. Within the response framework, these paths are revealed by the time-dependent operator $R_{k,j}(t)$: the first non-zero response identifies the shortest causal path in the graph connecting $x_j$ to $x_k$. Crucially, if $R_{k,j}(t) = 0$  for all time $t$, then no external intervention on $x_j$ can ever influence the behavior of $x_k$.\\

\paragraph*{Cumulative causal links or causal sensitivity.} Finally, Baldovin et al. \cite{Baldovin} introduced a cumulative (in time) measure of causation between two variables, defined as $D_{k,j} = \int_{0}^{\tau_{max}} R_{k,j}(t)dt$. Such cumulative link, can be also referred to as ``causal sensitivity''. Such a measure is especially relevant in practical experiments with high-dimensional models where interactions unfold across multiple spatial and temporal scales. A concrete example using climate models, is provided by the approach of Bloch-Johnson et al. (2024) \cite{blochjohnsonetal2024}, where causal links are inferred between patterns of surface warming and the net radiative flux at the top of the atmosphere. Their method reconstructs responses to perturbations by: (i) applying surface warming perturbations, (ii) integrating an atmospheric-only model with and without the perturbation, and (iii) computing the average (in time) difference between the two runs after removing an initial transient. In this case, the causal link emerges precisely as a cumulative (in time) atmospheric response to the surface perturbations.\\

We emphasize the crucial role of ensemble averages in establishing the general validity of response theory. Evaluating responses along single trajectories of nonlinear systems is strongly affected by chaos, restricting the applicability of these ideas to short times and very small perturbations. In contrast, focusing on observables, i.e. quantities averaged over the invariant density, ensures the robustness of the theory and largely removes the dependence on chaos. For a detailed discussion of this point, we refer the reader to \cite{FALCIONI1995481}. Future efforts to estimate responses in climate science, which often rely on perturbing single trajectories, would therefore benefit from focusing on ensemble-averaged observables.\\

\paragraph*{Response operator from data alone.} Given the assumption of Markovianity it is possible to infer the system's response operator from a long stationary trajectory using the Fluctuation-Dissipation Theorem (FDT) \cite{Baldovin}. The FDT formula provides a link between stationary and perturbed dynamics by predicting forced responses from stationary statistics alone \cite{Falcioni}.  Specifically:
\begin{equation}
R_{k,j}(t) = - \Big\langle A(x_{k}(t)) \frac{\partial \ln \rho(\mathbf{x})}{\partial x_{j}} \Big|_{\mathbf{x}(0)} \Big\rangle, 
\label{eq:response_general_FDT} 
\end{equation}
where $A(x_{k}(t))$ is a general observable, $\rho(\mathbf{x})$ is the stationary distribution. $\langle \cdot \rangle$ refers to ensemble average, approximated through temporal averages. $\frac{\partial \ln \rho(\mathbf{x})}{\partial x_{j}} \Big|_{\mathbf{x}(0)}$ represents the so-called ``score function'' evaluated at the initial state, i.e. on the attractor. Therefore, in principle the response operator can be computed from data if the score $\nabla \ln \rho(\mathbf{x})$ is available. In practice, the score is often inaccessible, and most applications rely on the quasi-Gaussian (qG) approximation, by replacing $\rho(\mathbf{x})$ with a Gaussian distribution $\rho^{G}(\mathbf{x})$ with same mean and covariances \cite{Leith,MajdaBook}. For responses in the ensemble mean the qG approximation achieves high skill even for nonlinear systems \cite{GRITSUN,Baldovin,LudovicoFabriAndre}. Recent work \cite{LudovicoFabriAndre} has demonstrated that generative modeling techniques can be used to efficiently estimate the score directly from data, leading to high skill in computation of higher-order responses in strongly nonlinear regimes. Other interesting new proposals have been recently formulated in the context of Koopman theory and can be used for future exploration of forced responses from data alone \cite{ValerioData1,ValerioData2,FDTKoopman}.\\

It is worth emphasizing that the central limitation discussed in the main text, i.e. the need to choose the proper variables for the system to be Markovian, remains relevant for the inference of response operators via FDT as well (and for any causal inference methodology); see the discussion in \cite{Baldovin}.

\section{Relevance of the the triad system for geophysical turbulence} \label{app:MTV-triad}

Here, we motivate the choice of the triad model analyzed in Section \ref{sec:analogy} as a building block for geophysical turbulent flows. We further present and test the associated analytical stochastic model for the slow variable $x_1$ (as discussed in Section \ref{sec:data_driven_models_partial_state_vector}). For detailed mathematical derivations and additional examples, we refer to the original work by Majda and collaborators \cite{NormalForms,MTV1,MTV2,MajdaBook,majdaSIAM}.

\subsection{Link to geophysical turbulence}

Projecting many geophysical turbulent flows, such as quasigeostrophic models, onto a complete set of orthogonal basis functions (spectral or empirical modes such as EOFs), give rise to the following generic form for the modes amplitudes:
\begin{equation}
d \mathbf{u}/dt = \mathbf{F} + \mathbf{L}\mathbf{u} + \mathbf{B}(\mathbf{u},\mathbf{u}),
\label{eq:abstract_form}
\end{equation}
where $\mathbf{F}$ is an external forcing, $\mathbf{L}$ is a linear operator and $\mathbf{B}(\mathbf{u},\mathbf{u})$ is a nonlinear, quadratic operator. The linear operator $\mathbf{L} = \mathbf{L}' + \mathbf{D}$, is formed by a skew-symmetric operator $\mathbf{L}'$ and by a diagonal dissipation matrix $\mathbf{D}$ with negative real entries. The quadratic nonlinearity should be energy-conserving, satisfying $\mathbf{u} \cdot \mathbf{B}(\mathbf{u},\mathbf{u}) = 0$ \cite{Lorenz,MTV1,MajdaBook}. Eq. \eqref{eq:abstract_form} is the typical abstract form of dry cores of atmospheric models \cite{MajdaPerspective}. Galerkin truncations of turbulent systems often reduce to this structure, as in the Lorenz ’63 and Charney–DeVore models \cite{Lorenz,Charney,CROMMELIN}. $\mathbf{u} \in \mathbb{R}^N$ represents the full turbulent geophysical system, with $N$ infinite in principle and finite but very large in applications ($N \gg 1$). The state $\mathbf{u}$ is then decomposed into slow and fast components, $\mathbf{u} = (\mathbf{x}, \mathbf{y})$, i.e. $\mathbf{x}$ represents the first $n$ components of $\mathbf{u}$ and $\mathbf{y}$ the next $N-n$.\\ 

We consider the simple case of $N = 3$: one large-scale resolved mode $x_1$ and two fast modes $\mathbf{y} = (x_2,x_3)$. We then project the three modes onto the dynamics in Eq. \eqref{eq:abstract_form}. This gives rise to three equations, defined by a forcing term, linear and quadratic self- and cross-interaction across slow and fast time scales. The triad model in Eq. \eqref{eq:triad_system} is then obtained by approximating the quadratic self-interactions of the fast modes as a damping and stochastic forcing, as: $-\gamma_i \epsilon^{-1} x_i + \sigma_i \epsilon^{-1/2} \xi_i$, with $i = 2,3$. $\xi_i$ is a standard Gaussian white noise process. $\epsilon$ sets the time-scale separation between the slow and fast modes. The key assumption is that the detailed nonlinear self-interactions among fast modes are not important if the goal is to capture the coarse-grained slow dynamics. This approximation transforms the original deterministic equations into a stochastic system. The specific version used in Eq. \eqref{eq:triad_system} is further simplified by setting to zero a few interactions, such as the linear cross- and self-interactions of the fast modes as well as the external forcing terms.\\

Therefore, the relevance of the triad system (Eq. \eqref{eq:triad_system}) to this study is that it provides a minimal multiscale setting with linear and quadratic nonlinear interaction of the kind that naturally arise when geophysical turbulent flows are expressed in terms of orthogonal bases such as EOFs. Furthermore, the model is characterized by fast decays of correlations, and generates PDFs that are regular and quasi-Gaussian, as often seen in the analysis of large-scale modes of climate variability \cite{MajdaStructuralStability}. For additional details see \cite{NormalForms} and Section 4 of \cite{majdaSIAM}.\\

\subsection{Analytical stochastic model for\\ the slow variable $x_1$}

Starting from the triad model in Eq.~\eqref{eq:triad_system}, Majda and collaborators \cite{MTV1} showed that an analytical reduced-order equation for the slow variable $x_1$ can be rigorously derived in the limit $\epsilon \to 0$ using standard projection techniques \cite{MajdaBook,NormalForms,Majda2010}. The mode reduction predicts the functional form of the dynamics governing the resolved, slow variable \cite{MTV1,MTV2,MTV3,MTV4}. The resulting stochastic equation is:
\begin{equation}
dx = (\Tilde{F} + ax + bx^2 - cx^3)\,dt + (A - Bx)\,dW_2 + \sigma\,dW_3,
\label{eq:triad_MTV}
\end{equation}
where the coefficients are expressed in terms of parameters of the triad system in Eq. \eqref{eq:triad_system} as: $a = L_{11} + \epsilon (\frac{I^2\sigma_2^2}{2\gamma^2_2} - \frac{L_{12}^2}{\gamma_2} - \frac{L_{13}^2}{\gamma_3})$; $b = -2\epsilon\frac{L_{12}I}{\gamma_2}$; $c = \epsilon\frac{I^2}{\gamma_2}$; $B = -I \epsilon^{1/2}\frac{\sigma_2}{\gamma_2}$; $L = -B \frac{L_{12}}{I}$; $\Tilde{F} = -\frac{AB}{2}$; $\sigma = \epsilon^{1/2}L_{13}\frac{\sigma_3}{\gamma_3}$. $W_2$ and $W_3$ are independent Wiener processes. Interestingly, the equation for the coarse-grained variable $x_1$ displays a structure richer than that of the full triad system in Eq. \eqref{eq:triad_system}. In addition to the original linear and quadratic interactions, the reduced equation includes new terms: an effective forcing $\Tilde{F}$, a cubic nonlinearity, and both additive and multiplicative noise. The additive noise results from the linear interactions across fast modes, while the multiplicative noise reflects the advection of the slow mode by the fast modes \cite{MajdaPerspective}. Thus, Eq. \eqref{eq:triad_MTV} demonstrates that multiplicative noise arises from theoretical considerations, rather than being introduced ad hoc, providing a firm basis for its use in reduced-order stochastic climate models. The scalar model above has been used to test the Fluctuation–Dissipation Theorem in climate dynamics \cite{MajdaStructuralStability,LudovicoFabriAndre} as well as for inverse modeling of the leading principal component of a quasi-geostrophic model and the observed variability associated with the North Atlantic Oscillation (NAO), demonstrating its relevance for theoretical studies in climate dynamics.\\

\begin{figure}[tbhp]
\centering
\includegraphics[width=1\linewidth]{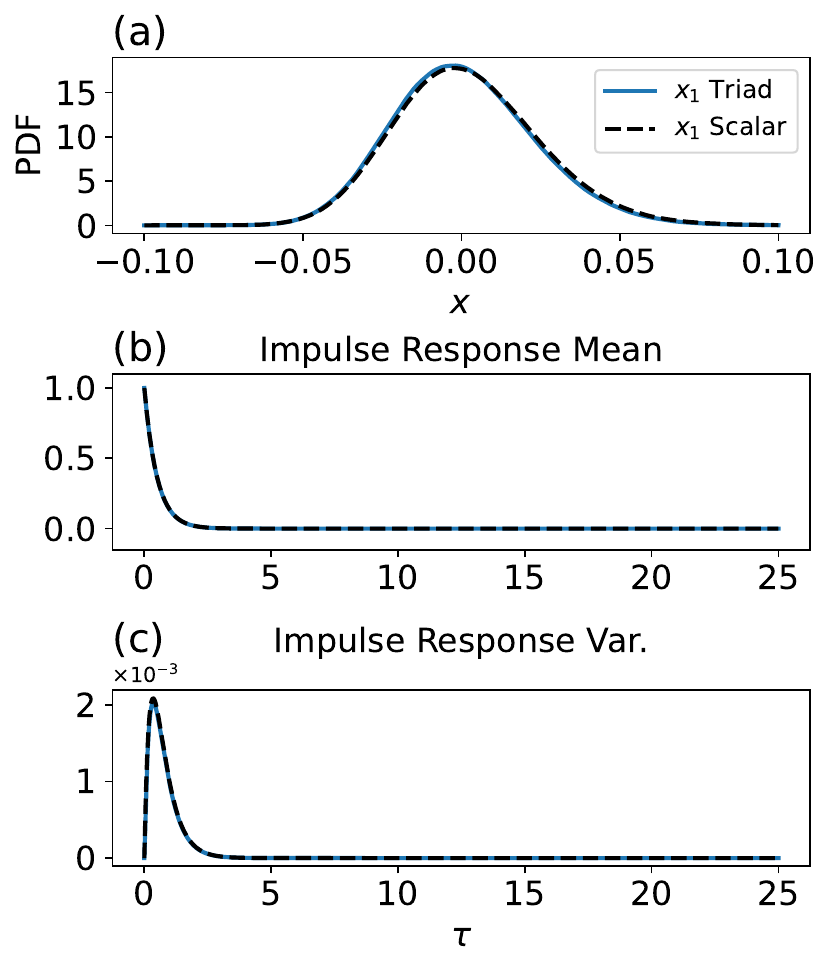}
\caption{In each panel: ``$x_1$ Triad'' represent statistics of $x_1$ in the original triad model in Eq. \eqref{eq:triad_system}; ``$x_1$ Scalar'' represent statistics of $x_1$ in the scalar reduced model in Eq. \eqref{eq:triad_MTV}. Panel (a): stationary probability distribution. Panel (b): response in ensemble mean. Panel (c): response in ensemble variance. Impulse response functions are computed using an ensemble of $N_e = 10^7$ members.}
\label{fig:MTV_response}
\end{figure}

We evaluate the model in Eq. \eqref{eq:triad_MTV} as done in Section \ref{sec:analogy}, so to reproduce the stationary statistics of the slow variable $x_1$ as well as its response to perturbations as in the original triad system in Eq. \eqref{eq:triad_system}. The results are reported in Figure \ref{fig:MTV_response}. The analysis shows that, given $\epsilon \ll 1$, it is indeed possible to build a Markovian model of the slow variable $x_1$ reproducing both stationary and perturbed statistics as in the original system.\\

Finally, as shown in this Section, rigorous model reduction frameworks, such as those developed by Majda and collaborators \cite{MTV2} and by Roberts \cite{Roberts2006,ROBERTS200812}, among others \cite{LucariniChekroun}, provide invaluable theoretical guidance for multiscale climate modeling. While these strategies were traditionally developed for systems with known equations of motion, they may also inspire future work aimed at incorporating such principles into data-driven-only emulators.\\

\section{Technical and practical details on the model formulations} \label{app:fitting}

\subsection{Multivariate data-driven stochastic models}

Consider a $N$-dimensional dynamical system described by a  stochastic differential equation $d \mathbf{x} = \mathbf{f}(\mathbf{x})dt + \mathbf{\Sigma}(\mathbf{x}) d \mathbf{W}$. This equation can be discretized by the forward Euler method as follows:
\begin{equation}
\begin{split}
\Delta \mathbf{x} = \mathbf{x}(t+\Delta t) - \mathbf{x}(t) = \mathbf{f}(\mathbf{x}(t))\Delta t + \mathbf{\Sigma}(\mathbf{x}) \sqrt{\Delta t} \bm{\xi}(t)~
\end{split}
\label{eq:functional_form_discrete} 
\end{equation}
where $\bm{\xi}(t)$ is a Gaussian white noise with zero mean, unit variance, and $\sqrt{\Delta t} \bm{\xi}(t)$ is the increment of the Wiener process $\Delta \mathbf{W} = \mathbf{W}(t+\Delta t) - \mathbf{W}(t)$. Without loss of generality, we rescale time units by the time sampling interval such that a non-dimensional time increment is $\Delta t = 1$ \cite{KRAVTSOV,KravtsovReview,KONDRASHOV201533,STROUNINE2010145}. The strategy is then to fit a rescaled version of Eq. \eqref{eq:functional_form_discrete}  from observational data alone. Such rescaled equation takes the following form:
\begin{equation}
\begin{split}
\Delta \mathbf{x} = \mathbf{x}(t+1) - \mathbf{x}(t) = \mathbf{f}(\mathbf{x}(t)) + \mathbf{\Sigma}(\mathbf{x}) \bm{\xi}(t).
\end{split}
\label{eq:functional_form_discrete_deltat1} 
\end{equation}
We proceed in two steps:
\begin{enumerate}[label=\roman*.]
\item \textit{Fitting the drift.} We compute $\Delta \mathbf{x} = \mathbf{x}(t+1) - \mathbf{x}(t)$ and fit the deterministic dynamics such that $\Delta \mathbf{x} \approx \mathbf{f}(\mathbf{x}(t))$. In the main paper, we considered two cases:
\begin{itemize}
    \item \textit{Linear drift.} In this case, given a trajectory $\mathbf{x}(t) \in \mathbb{R}^N$, we need to fit a matrix $\mathbf{L} \in \mathbb{R}^{N,N}$ such that $\Delta \mathbf{x} \approx \mathbf{L} \mathbf{x}(t)$ as close as possible. In practice, this is solved as a linear regression problem by minimizing the mean squared error. We use the linear regression module from the scikit-learn library \cite{scikit-learn}.
    \item \textit{Nonlinear drift.} We consider the following form: $\Delta \mathbf{x} \approx \mathbf{f}(\mathbf{x}(t)) = \mathbf{L} \mathbf{x}(t) + \mathbf{n}(\mathbf{x}(t))$, where $\mathbf{L} \in \mathbb{R}^{N,N}$ is the linear operator and $\mathbf{n}(\mathbf{x}(t))$ is the nonlinear operator parameterized by a multilayer perceptron (MLP). Both operators are fitted jointly by minimizing the mean squared error using a version of the gradient descent algorithm (Adam, \cite{kingma2014adam}) with weight decay. The MLP considered here always consisted of one hidden layer. The number of neurons in the hidden layer depended on the application: we considered 30 neurons in the experiments with the triad model in Section \ref{sec:data_driven_models_full_state_vector} and 1000 neurons in the real-world application in Section \ref{sec:real-world}. We played with more hidden layers but found a comparable performance, and sometimes a larger network also led to instabilities. The activation function used is the swish function (SiLU), which we found yields smoother responses to external perturbations. In contrast, the commonly used ReLU activation produced noisier responses. Furthermore, to ensure high accuracy in perturbation-response experiments, all data were processed using double-precision floating-point format.
\end{itemize}
\item \textit{Fitting the stochastic term.} Having fitted the deterministic dynamics $\mathbf{f}$, we compute the residuals $\mathbf{r}(t) \in \mathbb{R}^N$ as $\mathbf{r}(t) = \Delta \mathbf{x} - \mathbf{f}(\mathbf{x}(t))$. We then construct a suitable stochastic parametrization for this residual term. In the paper, we focus on two cases: additive and multiplicative noise.
\begin{itemize}
    \item \textit{Additive noise.} Given the residuals $\mathbf{r}(t)$, we proceed by computing their covariance matrix $\mathbf{C} \in \mathbb{R}^{N,N}$ as $\mathbf{C} = \langle \mathbf{r} \mathbf{r}^{\text{T}} \rangle$, where brackets denote time averaging. We then consider the Cholesky decomposition of $\mathbf{C}$ to derive a lower triangular matrix $\mathbf{\Sigma} \in \mathbb{R}^{N,N}$ such that $\mathbf{C} = \mathbf{\Sigma}\mathbf{\Sigma}^{\text{T}}$. The full fitted equation solved by the emulator can then be written as: 
    \begin{equation}
    \begin{split}
    \mathbf{x}(t+1) = \mathbf{x}(t) + \mathbf{f}(\mathbf{x}(t)) + \mathbf{\Sigma} \bm{\xi}(t)~
    \end{split}
    \label{eq:discrete_emulator_additive_noise} 
    \end{equation}
    and we remind the reader that $\bm{\xi}(t) \in \mathbb{R}^N$ is a Gaussian white noise with zero mean, unit variance.
    \item \textit{Multiplicative noise.} %The computed residuals $\mathbf{r} \in \mathbb{R}^{N,T}$ are a set of $N$ time series of length $T$. 
    Here, we consider state-dependent noise. To approximate it in a computationally efficient manner, we proceed as follows: 
    \begin{itemize}
        \item Given the residuals $\mathbf{r}(t)$ at each time $t$, we compute their outer product $\mathbf{C}(t) = \mathbf{r}(t) \mathbf{r}(t)^{\text{T}}$ which is a matrix of rank one. %This defines a state-dependent analog of the covariance matrix $\mathbf{C}(t) \in \mathbb{R}^{N,N}$ which has rank one.
        \item We train a neural network predicting a lower triangular matrix $\mathbf{\Sigma}(\mathbf{x}(t))$ with positive diagonal, which can be interpreted as a Cholesky factor, such that $\mathbf{C}(t) \approx \mathbf{\Sigma}(\mathbf{x}(t)) \mathbf{\Sigma}(\mathbf{x}(t))^\text{T}$ as close as possible. 
    \end{itemize}
    In practice, the loss function minimizes the mean squared error (MSE) between $\mathbf{C}(t)$ and $\mathbf{\Sigma}(\mathbf{x}(t)) \mathbf{\Sigma}(\mathbf{x}(t))^\text{T}$ at each batch. A minimizer of the MSE loss approximates the conditional mathematical expectation $\mathbb{E}[\mathbf{C}(t)|\mathbf{x}(t)$], that is the covariance matrix averaged over all residuals corresponding to the same state vector, as a function of this state vector, and can be seen as a generalization of a method presented in \cite{adler2018deep} to multiple dimensions. The positivity of the diagonal terms of the lower triangular matrix $\mathbf{\Sigma}(\mathbf{x}(t))$ is enforced through a softplus activation function. Note that this (high-dimensional) multiplicative noise parametrization is only applied in the real-world-like case, where data are scarce: to prevent overfitting, we use early stopping during training, halting after only 5 epochs.  The final fitted equation solved by the emulator is: 
    \begin{equation}
    \begin{split}
    \mathbf{x}(t+1) = \mathbf{x}(t) + \mathbf{f}(\mathbf{x}(t)) + \mathbf{\Sigma}(\mathbf{x}(t)) \bm{\xi}(t)~
    \end{split}
    \label{eq:discrete_emulator_multiplicative_noise} 
    \end{equation}
    where again $\bm{\xi}(t) \in \mathbb{R}^N$ is a Gaussian white noise with zero mean, unit variance.
\end{itemize}
\end{enumerate}

Furthermore, even in the real-world case of sea surface temperature fields and radiative fluxes, reducing the dimensionality to a few, $N = 21$, components led to a simplified computational problem where all neural networks could be trained on CPUs and on a personal computer.

\subsection{Scalar data-driven stochastic models}

Consider a trajectory of length $T$ of a $1$-dimensional dynamical system. As in the Section above, we assume that this scalar time series can be modeled as a discrete stochastic process with time-stepping $\Delta t = 1$, therefore rescaling our time-stepping by the sampling interval. The scalar equation we aim to fit is therefore:
\begin{equation}
\begin{split}
\Delta x = x(t+1) - x(t) = f(x(t)) + \sigma(x) \xi(t).
\end{split}
\label{eq:functional_form_discrete_scalar_deltat1} 
\end{equation}
The training of the deterministic (i.e. drift) dynamics follows exactly the same procedure of the multivariate case. Even the network considered are the same, with one hidden layer and 30 neurons. So we refer the reader to the Section above regarding the fitting of the deterministic dynamics and proceed in focusing on the stochastic case.\\

\textit{Fitting the stochastic term.}
\begin{itemize}
    \item \textit{Additive noise.} We compute the residual time series $r(t)=\Delta x - f(x(t))$ from the drift term. We then proceed in computing the standard deviation $\sigma$ of such residual time series. The full fitted equation solved by the emulator can then be written as: 
    \begin{equation}
    \begin{split}
    x(t+1) = x(t) + f(x(t)) + \sigma \xi(t)~
    \end{split}
    \label{eq:discrete_emulator_scalar_additive_noise} 
    \end{equation}
    and we remind the reader that $\xi(t)$ is a scalar Gaussian white noise with zero mean, unit variance.
    \item \textit{Multiplicative noise.} We again consider the residual time series $r(t)$ from the drift term. We then compute at each time step $t$ the square of the residual at that time step, i.e. $r(t)^2$. We then train a neural network to have $g(x(t)) \approx r(t)^2$ by minimizing the mean squared error \cite{adler2018deep}. The neural network is again a MLP with one hidden layer with 30 neurons. Differently, from the neural networks considered for the drift, here we added a softplus activation function after the last layer of the network to ensure that the variance estimates are nonnegative. The final fitted equation solved by the emulator is: 
    \begin{equation}
    \begin{split}
    x(t+1) = x(t) + f(x(t)) + \sqrt{g(x(t))}\xi(t).
    \end{split}
    \label{eq:discrete_emulator_scalar_multiplicative_noise} 
    \end{equation}
\end{itemize}

\section{Emulator of the SST field and net radiative flux: additive vs multiplicative noise}\label{app:add-vs-mult-noise}

In the main text, we considered the model defined in Eq.~\eqref{eq:ANN-SST-TOA} to emulate the joint variability of SST and net radiative flux at the TOA. Here we examine a simpler variant with additive noise:
\begin{equation}
d\mathbf{x} = (\mathbf{L}\mathbf{x} + \mathbf{n}(\mathbf{x}))dt + \mathbf{\Sigma}d\mathbf{W}.
\label{eq:ANN-SST-TOA-additive}
\end{equation}
The drift term $\mathbf{L}\mathbf{x} + \mathbf{n}(\mathbf{x})$ is identical to the one used in the multiplicative-noise model in the main text. We performed a 600-year-long emulation following the same protocol as in the main text.\\

The resulting autocorrelation functions, the reconstruction of net radiative flux, and the ensemble-mean response are all nearly indistinguishable from those obtained with multiplicative noise (Eq.~\eqref{eq:ANN-SST-TOA}), as shown in Figures~\ref{fig:SST-TOA}(c,d); \ref{fig:SST-TOA}(e,f); and \ref{fig:Sensitivity_map}. This demonstrates that the deterministic component (shared by both emulators) is responsible for setting these features.\\

\begin{figure*}[tbhp]
\centering
\includegraphics[width=1\textwidth]{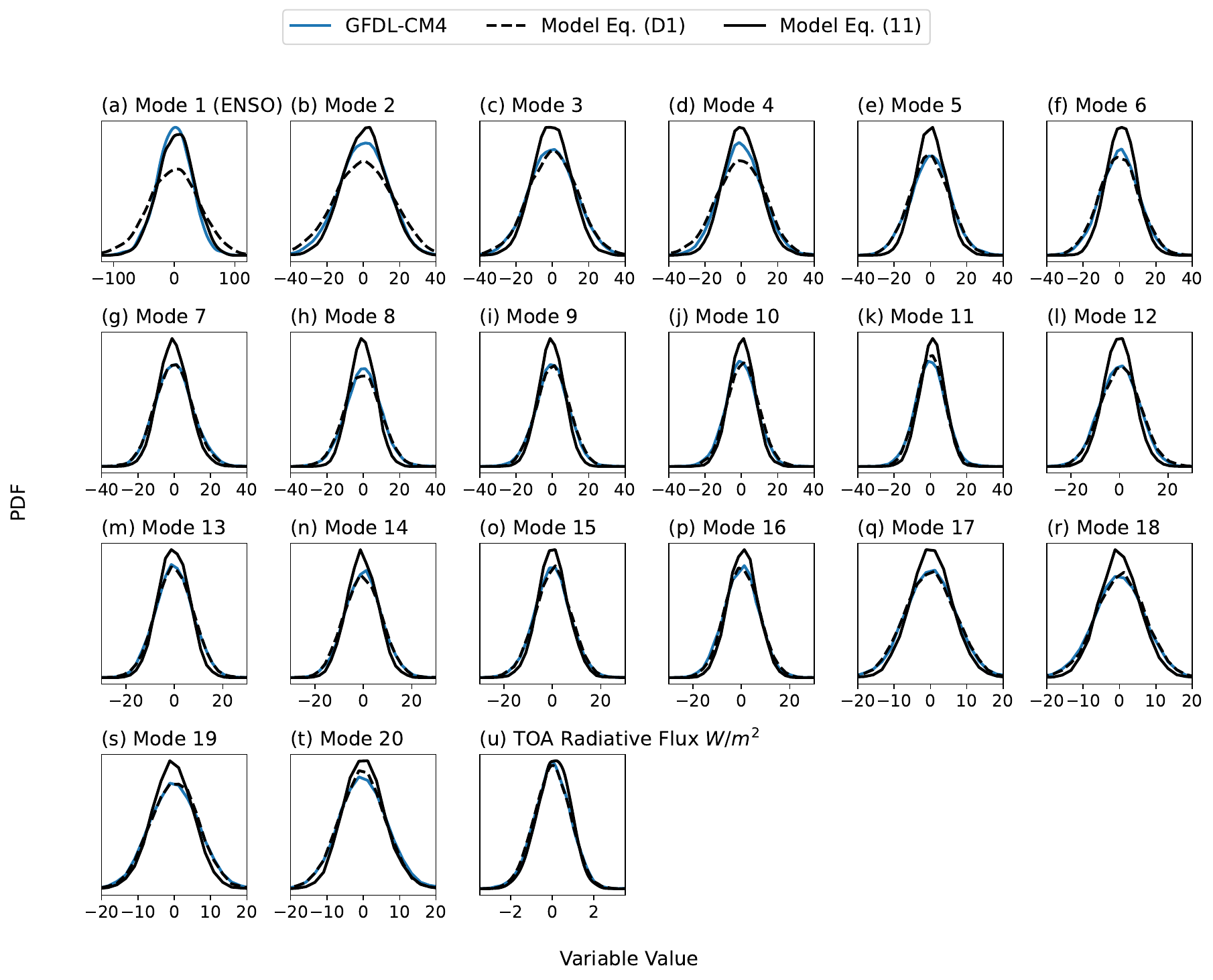}
\caption{Panel (a-t): stationary distributions obtained the first 20 principal components of the sea surface temperature (SST) field as in the original GFDL-CM4 climate model and emulated by the neural emulator with multiplicative (Eq. \eqref{eq:ANN-SST-TOA}) and additive (Eq. \eqref{eq:ANN-SST-TOA-additive}) noise. Units: nondimensional. Panel (u): same as panels (a-t) but for the global mean radiative flux at the top of the atmosphere (TOA). Units: $W/m^2$. The stationary distributions have been obtained after simulating the neural emulators for 600 years. Note that the system was trained on only 7200 time steps, which limits the accuracy of the learned multiplicative noise term.}
\label{fig:SST-TOA-Distributions}
\end{figure*}

The main difference arises in the reconstruction of probability distributions. Figure~\ref{fig:SST-TOA-Distributions} shows the PDFs for all 20 SST modes, as well as for the global-mean net radiative flux at the TOA. The PDF of the leading mode, associated with ENSO variability, is accurately reproduced by the emulator with multiplicative noise. Table~\ref{tab:ENSO-moments} reports the first four moments of the distributions of this first mode for the climate model and for the two emulators. The most pronounced discrepancy appears in the standard deviation, where the multiplicative-noise emulator shows substantially closer agreement. The multiplicative formulation also succeeds in capturing the PDF of the second mode.\\

Interestingly, for higher modes, each explaining a much smaller fraction of the variance, the emulator with additive noise performs slightly better. These differences, while visible, are modest if comparing the scale with the values of the first mode. We suspect that the very limited dataset (only 7200 samples in total for training and validation) is the main factor preventing the network from learning the multiplicative noise structure with higher accuracy. With a larger dataset, we expect the performance of the multiplicative-noise emulator to achieve more accurate reconstructions.

\begin{table}[h]
\centering
\caption{Mean, standard dev., skewness and kurtosis of SST Mode 1 (ENSO variability) for 600 years long runs.}
\label{tab:ENSO-moments}
\begin{tabular}{lllll}
\toprule
\textbf{Mode 1 (ENSO)} & \textbf{Mean} & \textbf{St.Dev.} & \textbf{Skewness} & \textbf{Kurtosis}\\
\midrule
CM4 Climate model       & 0.  & 28.27 &  -0.10 & 3.16\\
Eq. \eqref{eq:ANN-SST-TOA} (Multiplicative noise)   & 2.83  & 29.07 & -0.08 & 2.88 \\
Eq. \eqref{eq:ANN-SST-TOA-additive} (Additive noise)   & -0.65  & 42.31 & -0.13 & 3.33 \\
\bottomrule
\end{tabular}
\end{table}

\clearpage

\acknowledgments
Most of the ideas discussed in this work stem from collaborations and exchanges with colleagues and friends. I am grateful for the supportive and engaging environment of the Zanna Group at Courant, as well as for interactions within the broader M$^2$Lines project. I thank Laure Zanna and Carlos Fernandez-Granda for their feedback during early stages of the project. I am especially grateful to Pasha Perezhogin, Rory Basinski, Chris Pedersen, and Andre Souza for their thoughtful feedback and for many enriching discussions on data-driven modeling of the Climate system. I acknowledge an insightful email exchange with Prof. Tony Roberts, who drew my attention to recent advances on stochastic model reduction. I thank two anonymous reviewers for their very useful and constructive comments. This research received support through Schmidt Sciences, LLC, under the M$^2$LInES project. This work was supported in part through the NYU IT High Performance Computing resources, services, and staff expertise.

\bibliography{apssamp}% Produces the bibliography via BibTeX.

@PREAMBLE{
 "\providecommand{\noopsort}[1]{}" 
 # "\providecommand{\singleletter}[1]{#1}%" 
}

@article{gencast,
  title = {{Probabilistic weather forecasting with machine learning}},
  author = {Price, I. and Sanchez-Gonzalez, A. and Alet, F. and Andersson, T.R. and El-Kadi, A. and Masters, D. and Ewalds, T. and Stott, J. and Mohamed, S. and Battaglia, P. and Lam, R. and Willson, M.},
  journal = {Nature},
  volume = {637},
  pages = {84–90},
  year = {2025},
  doi = {https://doi.org/10.1038/s41586-024-08252-9}
}

@article{MajdaStructuralStability,
    title = {{High skill in low-frequency climate response through fluctuation dissipation theorems despite structural instability}},
    author = {Majda, A.J. and Abramov, R. and Gershgorin, B.},
    year = {2010},
    journal = {Proc. Natl. Acad. Sci.},
    number = {2},
    pages = {581–586},
    volume = {107},
    doi = {https://doi.org/10.1073/pnas.0912997107}
}

@article{NormalForms,
    title = {{Normal forms for reduced stochastic climate models}},
    author = {Majda, A.J. and Franzke, C. and Crommelin, D.},
    year = {2010},
    journal = {Proc. Natl. Acad. Sci.},
    number = {106},
    pages = {3649–3653},
    volume = {10},
    doi = {https://doi.org/10.1073/pnas.0900173106}
}

@article{FutureFromPast,
    title = {{Predicting the future from the past: An old problem from a modern perspective}},
    author = {Cecconi, F. and Cencini, M. and Falcioni, M. and Vulpiani, A.},
    year = {2012},
    journal = {Am. J. Phys.},
    number = {80},
    pages = {1001–1008},
    volume = {11},
    doi = {https://doi.org/10.1119/1.4746070}
}

@article{kang2023recent,
  title={Recent global climate feedback controlled by Southern Ocean cooling},
  author={Kang, Sarah M and Ceppi, Paulo and Yu, Yue and Kang, In-Sik},
  journal={Nature Geoscience},
  volume={16},
  number={9},
  pages={775--780},
  year={2023},
  publisher={Nature Publishing Group UK London}
}

@article{FranzkeMajda,
    title = {{Low-order stochastic mode reduction for a prototype atmospheric GCM}},
    author = {Franzke, C. and Majda, A.J.},
    year = {2006},
    journal = {Journal of the Atmospheric Sciences},
    number = {63},
    pages = {457–479},
    volume = {2},
    doi = {https://doi.org/10.1175/JAS3633.1}
}

@article{cm4,
  title = {{Structure and performance of GFDL's CM4.0 climate model}},
  author = {Held, I. M. and et al.},
  journal = {Journal of Advances in Modeling Earth Systems},
  volume = {11},
  pages = {3691–3727},
  year = {2019},
  doi = {doi.org/10.1029/ 2019MS001829}
}

@article{Andre1,
  title={Representing turbulent statistics with partitions of state space. Part 1. Theory and methodology.},
  author={Souza, A.N.},
  journal={J. Fluid Mech.},
  volume={997},
  number={A1},
  year={2024}
}

@article{Andre2,
  title={Representing turbulent statistics with partitions of state space. Part 2. The compressible Euler equations.},
  author={Souza, A.N.},
  journal={J. Fluid Mech.},
  volume={997},
  number={A2},
  year={2024}
}

@article{Ilias_1,
author = {I. Fountalis and C. Dovrolis and A. Bracco and B. Dilkina and S. Keilholz},
journal = {Appl. Netw. Sci.},
year = {2018},
title = {$\delta$-{MAPS} from spatio-temporal data to a weighted and lagged network between functional domain},
pages = {21},
volume = {3}
}

@article{Majda2010,
  title = {{Low-Frequency Climate Response and Fluctuation–Dissipation Theorems: Theory and Practice }},
  author = {Majda, A.J. and Gershgorin, B. and Yuan, Y.},
  journal = {Journal of the Atmospheric Sciences},
  volume = {67},
  number = {4},
  pages = {1186–1201 },
  year = {2010},
  doi = {https://doi.org/10.1175/2009JAS3264.1}
}

@article{Chris,
  title = {{Thermalizer: Stable autoregressive neural emulation of spatiotemporal chaos}},
  author = {Pedersen,C. and Zanna, L. and Bruna, J.},
  journal = {Arxiv},
  year = {2025},
  doi = {https://doi.org/10.48550/arXiv.2503.18731}
}

@Article{e20070509,
AUTHOR = {Chen, Nan and Majda, Andrew J.},
TITLE = {Conditional Gaussian Systems for Multiscale Nonlinear Stochastic Systems: Prediction, State Estimation and Uncertainty Quantification},
JOURNAL = {Entropy},
VOLUME = {20},
YEAR = {2018},
NUMBER = {7},
ARTICLE-NUMBER = {509},
URL = {https://www.mdpi.com/1099-4300/20/7/509},
PubMedID = {33265599},
ISSN = {1099-4300}
}

@article{LIM_nonlinear,
  title = {Nonlinear extensions of linear inverse models under memoryless or persistent random forcing},
  author = {Lien, Justin and Ando, Hiroyasu},
  journal = {Phys. Rev. Res.},
  volume = {7},
  issue = {3},
  pages = {L032038},
  numpages = {8},
  year = {2025},
  month = {Aug},
  publisher = {American Physical Society},
  doi = {10.1103/ds1j-fx3v},
  url = {https://link.aps.org/doi/10.1103/ds1j-fx3v}
}

@article{M2Lines,
  title = {{A Framework for Hybrid Physics-AI Coupled Ocean Models}},
  author = {Zanna, L. and et al.},
  journal = {Arxiv},
  year = {2025},
  doi = {https://arxiv.org/abs/2510.22676}
}

@article{Senne,
  title = {{Reanalysis-based Global Radiative Response to Sea Surface Temperature Patterns: Evaluating the Ai2 Climate Emulator}},
  author = {Van Loon, S. and Rugenstein, M. and Barnes, E. A.},
  journal = {Geophysical Research Letters},
  volume = {52},
  pages = {e2025GL115432},
  year = {2025},
  doi = {10.1029/2025GL115432},
  url = {https://doi.org/10.1029/2025GL115432}
}

@article{LucariniPRL,
  title = {Detecting and Attributing Change in Climate and Complex Systems: Foundations, Green's Functions, and Nonlinear Fingerprints},
  author = {Lucarini, Valerio and Chekroun, Micka\"el D.},
  journal = {Phys. Rev. Lett.},
  volume = {133},
  issue = {24},
  pages = {244201},
  numpages = {11},
  year = {2024},
  month = {Dec},
  publisher = {American Physical Society},
  doi = {10.1103/PhysRevLett.133.244201},
  url = {https://link.aps.org/doi/10.1103/PhysRevLett.133.244201}
}

@article{lucie3D,
  title = {{LUCIE-3D: A three-dimensional climate emulator for forced responses}},
  author = {Guan, H. and Arcomano, T. and Chattopadhyay, A. and Maulik, R.},
  journal = {Arxiv},
  year = {2025},
  doi = {https://doi.org/10.48550/arXiv.2509.02061}
}

@article{blochjohnsonetal2024,
author = {Bloch-Johnson, Jonah and Rugenstein, Maria A. A. and Alessi, Marc J. and Proistosescu, Cristian and Zhao, Ming and Zhang, Bosong and Williams, Andrew I. L. and Gregory, Jonathan M. and Cole, Jason and Dong, Yue and Duffy, Margaret L. and Kang, Sarah M. and Zhou, Chen},
title = {The Green's Function Model Intercomparison Project (GFMIP) Protocol},
journal = {Journal of Advances in Modeling Earth Systems},
volume = {16},
number = {2},
pages = {e2023MS003700},
doi = {https://doi.org/10.1029/2023MS003700},
url = {https://agupubs.onlinelibrary.wiley.com/doi/abs/10.1029/2023MS003700},
eprint = {https://agupubs.onlinelibrary.wiley.com/doi/pdf/10.1029/2023MS003700},
note = {e2023MS003700 2023MS003700},
year = {2024}
}

@article{GiorginiScore,
  title = {{Data-Driven Decomposition of Conservative and Non-Conservative Dynamics in Multiscale Systems}},
  author = {Giorgini, L.},
  journal = {Arxiv},
  year = {2025},
  doi = {https://doi.org/10.48550/arXiv.2505.01895}
}

@article{Kane,
  title = {{Bayesian structure learning for
climate model evaluation}},
  author = {O'Kane, T. J. and Harries, D. and Collier, M. A.},
  journal = {Journal of
Advances in Modeling Earth Systems},
  volume = {16},
  pages = {e2023MS004034},
  year = {2024},
  doi = {https://doi.org/10.1029/
2023MS004034}
}

@article{Lenny2002,
author = {Leonard A. Smith },
title = {What might we learn from climate forecasts?},
journal = {Proceedings of the National Academy of Sciences},
volume = {99},
pages = {2487-2492},
year = {2002},
doi = {10.1073/pnas.012580599},
URL = {https://www.pnas.org/doi/abs/10.1073/pnas.012580599},
eprint = {https://www.pnas.org/doi/pdf/10.1073/pnas.012580599}
}

@article{PedramPNAS,
author = {Sun, Y.Q. and Hassanzadeh, P. and Zand, M. and Abbot, D.S.},
title = {Can AI weather models predict out-of-distribution gray swan tropical cyclones?},
journal = {Proceedings of the National Academy of Sciences},
volume = {122},
number = {21},
pages = {e2420914122},
year = {2025},
doi = {10.1073/pnas.2420914122},
URL = {https://doi.org/10.1073/pnas.2420914122}
}

@article{PalmerQJRMS,
author = {Palmer, T.},
title = {A nonlinear dynamical perspective on model error: A proposal for non-local stochastic-dynamic parametrization in weather and climate prediction models},
journal = {Proceedings of the National Academy of Sciences},
volume = {127},
number = {572},
pages = {279-304},
year = {2001},
doi = {10.1002/qj.49712757202},
URL = { https://doi.org/10.1002/qj.49712757202}
}

@article{Judith,
author = {Berner, J. and Branstator, G.},
title = {Linear and Nonlinear Signatures in the Planetary Wave Dynamics of an AGCM: Probability Density Functions},
journal = {Journal of the Atmospheric Sciences},
volume = {64},
issue = {1},
pages = {117–136},
year = {2007},
doi = {https://doi.org/10.1175/JAS3822.1},
}

@article{FDTPatternEffect,
  title = {{A fluctuation-dissipation theorem perspective on radiative responses to temperature perturbations}},
  author = {Falasca, F. and Basinski, A. and Zanna, L. and Zhao, M.},
  journal = {Journal of Climate},
  volume = {38},
  issue = {14},
  pages = {3239–3260},
  year = {2025},
  doi = {https://doi.org/10.1175/JCLI-D-24-0479.1}
}

@article{senneNN,
author = {Van Loon, S. and Rugenstein, M. and Barnes, E. A.},
title = {Observation-based estimate of Earth;s effective radiative forcing},
journal = {Proceedings of the National Academy of Sciences},
volume = {122},
number = {24},
pages = {e2425445122},
year = {2025},
doi = {10.1073/pnas.2425445122},
URL = {https://www.pnas.org/doi/abs/10.1073/pnas.2425445122},
eprint = {https://www.pnas.org/doi/pdf/10.1073/pnas.2425445122}
}

@article{Camulator,
  title = {{CAMulator: Fast Emulation of the Community Atmosphere Model}},
  author = {Chapman, W.E. and Schreck, J.S. and Sha, Y. and Gagne II,D.J. and Kimpara,D. and Zanna, L. and Mayer,K.J. and Berner,J.},
  journal = {Arxiv},
  year = {2025},
  doi = {https://doi.org/10.48550/arXiv.2504.06007}
}

@article{LudovicoFabriAndre,
author = {Giorgini, L.T.  and Falasca, F.  and Souza, A.N.},
title = {Predicting forced responses of probability distributions via the fluctuation-dissipation theorem and generative modeling},
journal = {Proceedings of the National Academy of Sciences},
volume = {122},
number = {41},
pages = {e2509578122},
year = {2025},
doi = {10.1073/pnas.2509578122},
URL = {https://www.pnas.org/doi/abs/10.1073/pnas.2509578122},
eprint = {https://www.pnas.org/doi/pdf/10.1073/pnas.2509578122}
}

@article{Baxter,
  title = {{Benchmarking atmospheric circulation variability in an AI emulator, ACE2, and a hybrid model, NeuralGCM}},
  author = {Baxter, I. and Pahlavan, H. and Hassanzadeh, P. and Rucker, K. and Shaw, T.},
  journal = {Arxiv},
  year = {2025},
  doi = { 	
https://doi.org/10.48550/arXiv.2510.04466}
}

@article{Leif,
  title = {{Quantifying the radiative response to surface temperature variability: A critical comparison of current methods}},
  author = {Fredericks, L. and Rugenstein, M. and Thompson, D.W. and Van Loon, S. and Falasca, F. and Basinski-Ferris, R. and Ceppi, P. and Wu, Q. and Bloch-Johnson, J. and Alessi, M. and Kang, S. M.},
  journal = {Arxiv},
  year = {2025},
  doi = {https://doi.org/10.48550/arXiv.2511.00731}
}

@article{Lancelin,
  title = {{AI-boosted rare event sampling to characterize extreme weather}},
  author = {Lancelin, A. and et al.},
  journal = {Arxiv},
  year = {2025},
  doi = {https://doi.org/10.48550/arXiv.2510.27066}
}

@article{ValerioData1,
  title = {{Interpretable and Equation-Free Response Theory for Complex Systems}},
  author = {Lucarini, V.},
  journal = {Arxiv},
  year = {2025},
  doi = {https://doi.org/10.48550/arXiv.2502.07908}
}

@article{ValerioData2,
title = {A general framework for linking free and forced fluctuations via Koopmanism},
journal = {Chaos, Solitons \& Fractals},
volume = {202},
pages = {117540},
year = {2026},
issn = {0960-0779},
doi = {https://doi.org/10.1016/j.chaos.2025.117540},
url = {https://www.sciencedirect.com/science/article/pii/S096007792501553X}
}

@article{FDTKoopman,
  title = {Nonlinear correlations underlie linear response and causality},
  author = {Di Antonio, Gabriele and Vinci, Gianni Valerio},
  journal = {Phys. Rev. Res.},
  volume = {7},
  issue = {4},
  pages = {L042029},
  numpages = {11},
  year = {2025},
  month = {Oct},
  publisher = {American Physical Society},
  doi = {10.1103/vjvt-gdn1},
  url = {https://link.aps.org/doi/10.1103/vjvt-gdn1}
}

@article{Alet,
  title = {{Skillful joint probabilistic weather forecasting from marginals}},
  author = {Alet, F. and et al.},
  journal = {Arxiv},
  year = {2025},
  doi = {https://doi.org/10.48550/arXiv.2506.10772}
}

@ARTICLE{Pathak2022,
       author = {{Pathak}, Jaideep and {Subramanian}, Shashank and {Harrington}, Peter and {Raja}, Sanjeev and {Chattopadhyay}, Ashesh and {Mardani}, Morteza and {Kurth}, Thorsten and {Hall}, David and {Li}, Zongyi and {Azizzadenesheli}, Kamyar and {Hassanzadeh}, Pedram and {Kashinath}, Karthik and {Anandkumar}, Animashree},
        title = "{FourCastNet: A Global Data-driven High-resolution Weather Model using Adaptive Fourier Neural Operators}",
      journal = {arXiv e-prints},
         year = 2022,
        month = feb,
          doi = {10.48550/arXiv.2202.11214},
       eprint = {2202.11214},
 primaryClass = {physics.ao-ph},
}

@ARTICLE{fourcastnet3,
       author = {Bonev, B. and et al.},
        title = "{FourCastNet 3: A geometric approach to probabilistic machine-learning weather forecasting at scale}",
      journal = {arXiv e-prints},
         year = 2025,
        month = july,
          doi = {10.48550/arXiv.2507.12144},
        url = {https://doi.org/10.48550/arXiv.2507.12144}
}

@article{Lam2023,
author = {Remi Lam  and Alvaro Sanchez-Gonzalez  and Matthew Willson  and Peter Wirnsberger  and Meire Fortunato  and Ferran Alet  and Suman Ravuri  and Timo Ewalds  and Zach Eaton-Rosen  and Weihua Hu  and Alexander Merose  and Stephan Hoyer  and George Holland  and Oriol Vinyals  and Jacklynn Stott  and Alexander Pritzel  and Shakir Mohamed  and Peter Battaglia },
title = {Learning skillful medium-range global weather forecasting},
journal = {Science},
volume = {382},
number = {6677},
pages = {1416-1421},
year = {2023},
doi = {10.1126/science.adi2336},
URL = {https://www.science.org/doi/abs/10.1126/science.adi2336},
eprint = {https://www.science.org/doi/pdf/10.1126/science.adi2336}}

@article{Bi2023,
  title = {Accurate medium-range global weather forecasting with 3D neural networks},
  volume = {619},
  ISSN = {1476-4687},
  url = {http://dx.doi.org/10.1038/s41586-023-06185-3},
  DOI = {10.1038/s41586-023-06185-3},
  number = {7970},
  journal = {Nature},
  publisher = {Springer Science and Business Media LLC},
  author = {Bi,  Kaifeng and Xie,  Lingxi and Zhang,  Hengheng and Chen,  Xin and Gu,  Xiaotao and Tian,  Qi},
  year = {2023},
  month = jul,
  pages = {533–538}
}

@Inbook{Smith2001,
author="Smith, Leonard A.",
editor="Mees, Alistair I.",
title="Disentangling Uncertainty and Error: On the Predictability of Nonlinear Systems",
bookTitle="Nonlinear Dynamics and Statistics",
year="2001",
publisher="Birkh{\"a}user Boston",
address="Boston, MA",
pages="31--64",
isbn="978-1-4612-0177-9",
doi="10.1007/978-1-4612-0177-9_2",
url="https://doi.org/10.1007/978-1-4612-0177-9_2"
}

@article{Kochkov2023,
  title = {Neural general circulation models for weather and climate},
  volume = {632},
  ISSN = {1476-4687},
  url = {http://dx.doi.org/10.1038/s41586-024-07744-y},
  DOI = {10.1038/s41586-024-07744-y},
  number = {8027},
  journal = {Nature},
  publisher = {Springer Science and Business Media LLC},
  author = {Kochkov,  Dmitrii and Yuval,  Janni and Langmore,  Ian and Norgaard,  Peter and Smith,  Jamie and Mooers,  Griffin and Kl\"{o}wer,  Milan and Lottes,  James and Rasp,  Stephan and D\"{u}ben,  Peter and Hatfield,  Sam and Battaglia,  Peter and Sanchez-Gonzalez,  Alvaro and Willson,  Matthew and Brenner,  Michael P. and Hoyer,  Stephan},
  year = {2024},
  month = jul,
  pages = {1060–1066}
}

@ARTICLE{Watt-Meyer2023,
       author = {{Watt-Meyer}, Oliver and {Dresdner}, Gideon and {McGibbon}, Jeremy and {Clark}, Spencer K. and {Henn}, Brian and {Duncan}, James and {Brenowitz}, Noah D. and {Kashinath}, Karthik and {Pritchard}, Michael S. and {Bonev}, Boris and {Peters}, Matthew E. and {Bretherton}, Christopher S.},
        title = "{ACE: A fast, skillful learned global atmospheric model for climate prediction}",
      journal = {arXiv e-prints},
         year = 2023,
        month = oct,
          eid = {arXiv:2310.02074},
        pages = {arXiv:2310.02074},
          doi = {10.48550/arXiv.2310.02074},
archivePrefix = {arXiv},
       eprint = {2310.02074},
 primaryClass = {physics.ao-ph},
       adsurl = {https://ui.adsabs.harvard.edu/abs/2023arXiv231002074W}
}

@article{Meyssignac,
  title = {{Revisiting the Global Energy Budget Dynamics with a Multivariate Earth Energy Balance Model to Account for the Warming Pattern Effect }},
  author = {Meyssignac, B. and Guillaume-Castel, R. and Roca, R.},
  journal = {Journal of Climate},
  volume = {36},
  pages = {8113–8126 },
  year = {2023},
  doi = {https://doi.org/10.1175/JCLI-D-22-0765.1}
}

@article{Penland2025,
      author = {Sardeshmukh, P. and Penland, C. and Compo, G.},
      title = "Learning ENSO Dynamics from Data",
      journal = "Journal of Climate",
      year = "2025",
      publisher = "American Meteorological Society",
      address = "Boston MA, USA",
      doi = "10.1175/JCLI-D-25-0053.1",
      url = "https://journals.ametsoc.org/view/journals/clim/aop/JCLI-D-25-0053.1/JCLI-D-25-0053.1.xml"
}

@article{Zhang2023UsingCM4,
    title = {{Using a Green’s Function Approach to Diagnose the Pattern Effect in GFDL AM4 and CM4}},
    year = {2023},
    journal = {Journal of Climate},
    author = {Zhang, Bosong and Zhao, Ming and Tan, Zhihong},
    number = {4},
    month = {1},
    pages = {1105--1124},
    volume = {36},
    publisher = {American Meteorological Society},
    url = {https://journals.ametsoc.org/view/journals/clim/36/4/JCLI-D-22-0024.1.xml},
    doi = {10.1175/JCLI-D-22-0024.1},
    issn = {0894-8755}
}

@article{ZhangACE,
  title = {{The Equilibrium Response of Atmospheric Machine-Learning Models to Uniform Sea Surface Temperature Warming}},
  author = {Zhang, B. and Merlis, T.M.},
  journal = {Arxiv},
  year = {2025},
  doi = {https://doi.org/10.48550/arXiv.2510.02415}
}

@article{Dong2020IntermodelModels,
    title = {{Intermodel Spread in the Pattern Effect and Its Contribution to Climate Sensitivity in CMIP5 and CMIP6 Models}},
    year = {2020},
    journal = {Journal of Climate},
    author = {Dong, Yue and Armour, Kyle C. and Zelinka, Mark D. and Proistosescu, Cristian and Battisti, David S. and Zhou, Chen and Andrews, Timothy},
    number = {18},
    month = {9},
    pages = {7755--7775},
    volume = {33},
    publisher = {American Meteorological Society},
    url = {https://journals.ametsoc.org/view/journals/clim/33/18/jcliD191011.xml},
    doi = {10.1175/JCLI-D-19-1011.1},
    issn = {0894-8755}
}

@article{alessi2023surface,
  title={Surface temperature pattern scenarios suggest higher warming rates than current projections},
  author={Alessi, Marc J and Rugenstein, Maria AA},
  journal={Geophysical Research Letters},
  volume={50},
  number={23},
  pages={e2023GL105795},
  year={2023},
  publisher={Wiley Online Library}
}

@article{barsugli2002global,
  title={Global atmospheric sensitivity to tropical SST anomalies throughout the Indo-Pacific basin},
  author={Barsugli, Joseph J and Sardeshmukh, Prashant D},
  journal={Journal of Climate},
  volume={15},
  number={23},
  pages={3427--3442},
  year={2002},
  publisher={American Meteorological Society}
}

@article{Romps,
  title={Why the forcing from carbon dioxide scales as the logarithm of its concentration},
  author={Romps, D. M. and Seeley, J. T. and Edman, J. P.},
  journal={Journal of Climate},
  volume={35},
  number = {13},
  pages={4027–4047},
  year={2022},
  doi = {https://doi.org/10.1175/jcli-d-21-0275.1}
}

@article{Zhao22investigation,
  author = {M. Zhao},
  title =	 {An Investigation of the Effective Climate Sensitivity in 
            {GFDL}’s New Climate Models {CM4.0} and {SPEAR}},
  journal = {Journal of Climate},
  year =	 {2022},
  volume = {35},
  pages =	{5637~5660}, 
 doi = {DOI: 10.1175/JCLI-D-21-0327}
}

@book{Pierrehumbert,
  title={{Principles of Planetary Climate}},
  author={Pierrehumbert, R. T.},
  publisher={Cambridge University Press},
  year={2010}
}

@article{williams2023circus,
  title={Circus tents, convective thresholds, and the non-linear climate response to tropical SSTs},
  author={Williams, Andrew IL and Jeevanjee, Nadir and Bloch-Johnson, Jonah},
  journal={Geophysical Research Letters},
  volume={50},
  number={6},
  pages={e2022GL101499},
  year={2023},
  publisher={Wiley Online Library}
}

@article{zhouetal2017,
author = {Zhou, Chen and Zelinka, Mark D. and Klein, Stephen A.},
title = {Analyzing the dependence of global cloud feedback on the spatial pattern of sea surface temperature change with a Green's function approach},
journal = {Journal of Advances in Modeling Earth Systems},
volume = {9},
number = {5},
pages = {2174-2189},
doi = {https://doi.org/10.1002/2017MS001096},
url = {https://agupubs.onlinelibrary.wiley.com/doi/abs/10.1002/2017MS001096},
eprint = {https://agupubs.onlinelibrary.wiley.com/doi/pdf/10.1002/2017MS001096},
year = {2017}
}

@ARTICLE{Watt-Meyer2024,
       author = {Watt-Meyer, O. and Henn, B. and McGibbon, J. and Clark, S. K. and Kwa, A. and Perkins, W. A. and Bretherton, C. S.},
        title = "{ACE2: Accurately learning subseasonal to decadal atmospheric variability and forced responses}",
      journal = {arXiv e-prints},
         year = {2024},
          doi = {https://arxiv.org/abs/2411.11268}
}

@article{Samudra,
  title = {{Samudra: An AI global ocean emulator for climate.}},
  author = {Dheeshjith, S. and Subel, A. and Adcroft, A. and Busecke, J. and Fernandez‐Granda, C. and Gupta,S. and Zanna, L.},
  journal = {Geophysical Research Letters},
  volume = {52},
  pages = {e2024GL114318},
  year = {2025},
  doi = {https://doi.org/10.1029/2024GL114318}
}

@article{Onsager,
  title = {{Fluctuations and irreversible processes}},
  author = {Onsager, L. and Machlup, S.},
  journal = {Phys. Rev.},
  volume = {91},
  pages = {1505},
  year = {1953},
  url = {https://journals.aps.org/pr/abstract/10.1103/PhysRev.91.1505}
}

@inbook{Takens,
 author = {Takens, F.},
 booktitle = {Lect. Notes in Mathematics},
 pages = {21–48},
 publisher = {Springer, Berlin,Heidelberg},
 title = {{Detecting strange attractors in turbulence, in Dynamical Systems and Turbulence}},
 editor = {Rand,D.A. and Young, L.S.},
 volume = {898},
 pages = {366},
 year = {1981}
}

@article{neumann,
  title = {{Numerical Integration of the Barotropic Vorticity Equation}},
  author = {Charney, J. G. and Fj\"ortoft, R. and Von Neumann, J.},
  journal = {Tellus},
  volume = {2},
  issue = {4},
  year = {1950},
  url = {https://doi.org/10.1111/j.2153-3490.1950.tb00336.x}
}

@article{charney1949,
  title = {{ON A PHYSICAL BASIS FOR NUMERICAL PREDICTION OF LARGE-SCALE MOTIONS IN THE ATMOSPHERE }},
  author = {Charney, J. G.},
  journal = {Journal of Meteorology},
  volume = {6},
  issue = {6},
  pages = {372–385 },
  year = {1949},
  url = {https://doi.org/10.1175/1520-0469(1949)006<0372:OAPBFN>2.0.CO;2}
}

@article{Dalmedico,
  title = {{History and Epistemology of Models: Meteorology (1946–1963) as a Case Study.}},
  author = {Dalmedico, A.},
  journal = {Archive for History of Exact Science},
  volume = {55},
  pages = {395–422},
  year = {2001},
  url = {https://doi.org/10.1007/s004070000032}
}

@article{Gammaitoni,
  title = {{Prediction and Inference: From Models and Data to Artificial Intelligence}},
  author = {Gammaitoni L. and Vulpiani, A.},
  journal = {Foundations of Physics},
  volume = {54},
  number = {67},
  year = {2024},
  doi = {https://doi.org/10.1007/s10701-024-00803-4}
}

@article{Hosni,
  title = {{Forecasting in Light of Big Data}},
  author = {Hosni, H. and Vulpiani, A.},
  journal = {Philos. Technol.},
  volume = {31},
  pages = {557–569},
  year = {2018},
  doi = {https://doi.org/10.1007/s13347-017-0265-3}
}

@article{MajdaPerspective,
  title = {{An applied mathematics perspective on stochastic modelling for climate}},
  author = {Majda, A.J. and Franzke, C. and Boualem, K.},
  journal = {Phil. Trans. R. Soc. A.},
  volume = {366},
  pages = {2427–2453},
  year = {2008},
  doi = {http://doi.org/10.1098/rsta.2008.0012}
}

@article{HeldGap,
  title = {{The Gap between Simulation and Understanding in Climate Modeling}},
  author = {Held, I.M.},
  journal = {Bulletin of the American Meteorological Society},
  pages = {1609–1614},
  year = {2005},
  doi = {https://doi.org/10.1175/BAMS-86-11-1609}
}

@article{GhilLucarini,
  title = {The physics of climate variability and climate change},
  author = {Ghil, Michael and Lucarini, Valerio},
  journal = {Rev. Mod. Phys.},
  volume = {92},
  issue = {3},
  pages = {035002},
  numpages = {77},
  year = {2020},
  month = {07},
  publisher = {American Physical Society},
  doi = {10.1103/RevModPhys.92.035002},
  url = {https://link.aps.org/doi/10.1103/RevModPhys.92.035002}
}

@article{Dawson,
  title = {Simulating weather regimes: impact of model resolution and stochastic parameterization},
  author = {Dawson, A. and Palmer, T.N.},
  journal = {Clim Dyn},
  volume = {44},
  pages = {2177-2193},
  year = {2015},
  url = {http://dx.doi.org/10.1007/s00382-014-2238-x}
}

@article{ArnoldHM,
  title = {Stochastic parametrizations and model uncertainty in the Lorenz’ 96 system},
  author = {Arnold, H.M. and Moroz, I.M. and Palmer, T.N.},
  journal = {Phil Trans R Soc},
  volume = {371},
  pages = {20110479},
  year = {2013},
  url = {http://dx.doi.org/10.1098/rsta.2011.0479}
}

@article{GIORGINI2024134393,
title = {Reduced Markovian models of dynamical systems},
journal = {Physica D: Nonlinear Phenomena},
volume = {470},
pages = {134393},
year = {2024},
issn = {0167-2789},
doi = {https://doi.org/10.1016/j.physd.2024.134393},
url = {https://www.sciencedirect.com/science/article/pii/S0167278924003439},
author = {Ludovico Theo Giorgini and Andre N. Souza and Peter J. Schmid}
}

@article{Lorenz96,
  author = {Lorenz, E.N.},
  title = {Predictability: a problem partly solved},
  year = {1995},
  journal = {Seminar on Predictability, 4-8 September 1995},
  volume = {1},
  pages = {1-18},
  month = {1995},
  publisher = {ECMWF}
}

@article{LucariniChekroun,
  title = {{Theoretical tools for understanding the climate crisis from Hasselmann’s programme and beyond}},
  author = {Lucarini, V. and Chekroun, M.D.},
  journal = {Nat Rev Phys (2023)},
  year = {2023},
  url = {https://doi.org/10.1038/s42254-023-00650-8}
}

@article{ENSOcomplexity,
author = {Timmermann, A. and et al.},
journal = {Nature},
year = {2018},
title = {{El Ni\~{n}o-Southern Oscillation complexity}},
pages = {535-545},
volume = {559}
}

@article{Donges,
author = {Donges, J. and Zou, Y. and Marwan, N. and Kurths, J.},
journal = {Eur. Phys. J. Spec. Top.},
year = {2009},
title = {{Complex networks in climate dynamics}},
volume = {174},
pages = {157–179},
url = {https://doi.org/10.1140/epjst/e2009-01098-2}
}

@article{Baldovin,
author = {Baldovin, M. and Cecconi, F. and Vulpiani, A.},
journal = {Physical Review Research},
year = {2020},
title = {{Understanding causation via correlations and linear response theory}},
volume = {2},
pages = {043436}
}

@article{Ay,
author = {Ay, N. and Polani, D.},
title = {Information flows in causal networks},
journal = {Adv.
Complex Syst.},
volume = {11},
number = {1},
pages = {17-41},
year = {2008}}

@article{Lucie,
  title = {{LUCIE: A Lightweight Uncoupled ClImate Emulator with long-term stability and physical consistency for O(1000)-member ensembles}},
  author = {Guan, H. and Arcomano, T. and Chattopaday, A. and Maulik, R.},
  journal = {arXiv},
  year = {2025},
  doi = {https://doi.org/10.48550/arXiv.2405.16297}
}

@article{OceanNet,
  title = {OceanNet: a principled neural operator-based digital twin for regional oceans},
  author = {Chattopadhyay, A. and Gray, M. and Wu, T. and Lowe, A.B. and He, R.},
  journal = {Sci. Rep.},
  year = {2024},
  volume = {14},
  pages = {21181},
  url = {https://doi.org/10.1038/s41598-024-72145-0}
}

@article{Dubrulle,
  title = {{How many modes are needed to predict climate bifurcations? Lessons from an experiment}},
  author = {Dubrulle, B. and Daviaud, F. and Faranda, D. and Mari\'e, L. and Saint-Michel, B.},
  journal = {Nonlin. Processes Geophys.},
  year = {2022},
  volume = {29},
  pages = {17–35},
  doi = {doi.org/10.5194/npg-29-17-2022}
}

@article{Marconi,
author = {Marconi, U.M.B. and Puglisi, A. and Rondoni, L. and Vulpiani, A.},
journal = {Phys. Rep.},
volume = {461},
number = {111},
year = {2008},
title = {Fluctuation-dissipation: Response theory in statistical
physics}
}

@article{Lucente,
  title = {Inference of time irreversibility from incomplete information: Linear systems and its pitfalls},
  author = {Lucente, D. and Baldassarri, A. and Puglisi, A. and Vulpiani, A. and Viale, M.},
  journal = {Phys. Rev. Res.},
  volume = {4},
  issue = {4},
  pages = {043103},
  numpages = {21},
  year = {2022},
  month = {Nov},
  publisher = {American Physical Society},
  doi = {10.1103/PhysRevResearch.4.043103},
  url = {https://link.aps.org/doi/10.1103/PhysRevResearch.4.043103}
}

@article{nanChen,
  title = {Minimum reduced-order models via causal inference},
  author = {Chen, N. and Liu, H.},
  journal = {Nonlinear Dyn},
  volume = {113},
  pages = {11327–11351},
  year = {2025},
  doi = {https://doi.org/10.1007/s11071-024-10824-3}
}

@article{FALCIONI,
title = {Correlation functions and relaxation properties in chaotic dynamics and statistical mechanics},
journal = {Physics Letters A},
volume = {144},
number = {6},
pages = {341-346},
year = {1990},
issn = {0375-9601},
doi = {https://doi.org/10.1016/0375-9601(90)90137-D},
url = {https://www.sciencedirect.com/science/article/pii/037596019090137D},
author = {Falcioni, M. and Isola, S. and Vulpiani, A.}
}

@article{CROMMELIN,
    title = {{Strategies for Model Reduction: Comparing Different Optimal Bases}},
    author = {Crommelin, D.T. and Majda, A.J.},
    year = {2004},
    journal = {Journal of the Atmospheric Sciences},
    pages = {2206–2217},
    volume = {61},
    doi = {https://doi.org/10.1175/1520-0469(2004)061<2206:SFMRCD>2.0.CO;2}
}

@article{Charney,
    title = {{Multiple Flow Equilibria in the Atmosphere and Blocking}},
    author = {Charney, J.G. and DeVore, J.G.},
    year = {1979},
    journal = {Journal of the Atmospheric Sciences},
    pages = {1205–1216},
    volume = {36},
    issue = {7},
    doi = {https://doi.org/10.1175/1520-0469(1979)036<1205:MFEITA>2.0.CO;2}
}

@article{Hasselmann,
author = {Hasselmann, K.},
journal = {Tellus},
year = {1976},
title = {Stochastic climate models Part I. Theory.},
volume = {28},
pages = {473-485},
doi = {https://doi.org/10.1111/j.2153-3490.1976.tb00696.x}
}

@article{Pascale,
author = {Guilyardi, E. and Braconnot, P. and Jin, F.-F. and Kim, Seon Tae and Kolasinski, M. and Li. T. and Musat, I.},
journal = {Journal of Climate},
year = {2009},
title = {{Atmosphere Feedbacks during ENSO in a Coupled GCM with a Modified Atmospheric Convection Scheme}},
volume={2},
pages = {5698–5718}
}

@inproceedings{Kusner,
 author = {Kusner, Matt J and Loftus, Joshua and Russell, Chris and Silva, Ricardo},
 booktitle = {Advances in Neural Information Processing Systems},
 editor = {I. Guyon and U. Von Luxburg and S. Bengio and H. Wallach and R. Fergus and S. Vishwanathan and R. Garnett},
 pages = {},
 publisher = {Curran Associates, Inc.},
 title = {Counterfactual Fairness},
 url = {https://proceedings.neurips.cc/paper/2017/file/a486cd07e4ac3d270571622f4f316ec5-Paper.pdf},
 volume = {30},
 year = {2017},
address = {Long Beach, CA, USA}
}

@article{Dong2019AttributingPacific,
    title = {{Attributing Historical and Future Evolution of Radiative Feedbacks to Regional Warming Patterns using a Green’s Function Approach: The Preeminence of the Western Pacific}},
    year = {2019},
    journal = {Journal of Climate},
    author = {Dong, Yue and Proistosescu, Cristian and Armour, Kyle C. and Battisti, David S.},
    number = {17},
    month = {9},
    pages = {5471--5491},
    volume = {32},
    publisher = {American Meteorological Society},
    url = {https://journals.ametsoc.org/view/journals/clim/32/17/jcli-d-18-0843.1.xml},
    doi = {10.1175/JCLI-D-18-0843.1},
    issn = {0894-8755}
}

@article{Capotondi,
author = {Zhao, Y. and Capotondi, A.},
journal = {npj Clim Atmos Sci},
year = {2024},
title = {{The role of the tropical Atlantic in tropical Pacific climate variability}},
volume={7},
number={140},
url={https://doi.org/10.1038/s41612-024-00677-3}
}

@article{Pedram2,
  title={The Linear Response Function of an Idealized Atmosphere. Part II: Implications for the Practical Use of the Fluctuation–Dissipation Theorem and the Role of Operator’s Nonnormality},
  author={Hassanzadeh, P. and Kuang, Z.},
  journal={Journal of The Atmospheric Science},
  pages={3441–3452},
  year={2016},
  doi = {https://doi.org/10.1175/JAS-D-16-0099.1}
}

@article{GRITSUN,
  title={Climate Response Using a Three-Dimensional Operator Based on the Fluctuation–Dissipation Theorem},
  author={Gritsun, A. and Branstator, G.},
  journal={Journal of The Atmospheric Science},
  pages={2558–2575},
  year={2007},
  doi = {https://doi.org/10.1175/JAS3943.1}
}

@article{Leith,
  title={Climate response and fluctuation dissipation},
  author={Leith, C. E.},
  journal={Journal of The Atmospheric Science},
  volume = {32},
  pages={2022–2026},
  year={1975}
}

@article{majdaSIAM,
  title={{Strategies for Reduced-Order Models for Predicting the Statistical Responses and Uncertainty Quantiﬁcation in Complex Turbulent Dynamical Systems}},
  author={Majda, A. and Qi, Di},
  journal={SIAM REVIEW },
  volume={60},
  issue = {3},
  year={2018},
  doi = {https://doi.org/10.1137/16M1104664}
}

@article{QiMajdaChaos,
  title={{Linear and nonlinear statistical response theories with prototype applications to sensitivity analysis and statistical control of complex turbulent dynamical systems}},
  author={Qi, Di and Majda, A. },
  journal={Chaos},
  volume={29},
  issue = {10},
  year={2019},
  doi = {https://doi.org/10.1063/1.5118690}
}

@article{Aurell,
  title = {Causal analysis, correlation- response, and dynamic cavity},
  author = {Aurell, E. and Del Ferraro, G.},
  journal = {J. of Phys.: Conference Series},
  year = {2016},
  volume = {699},
  pages = {012002},
  doi =  {doi:10.1088/1742-6596/699/1/012002}
}

@article{IsmaelNew,
  title = {{Reflections on the asymmetry of causation}},
  author = {Ismael, J.},
  journal = {Interface Focus},
  volume = {12},
  pages = {20220081},
  year = {2023},
  url = {http://doi.org/10.1098/rsfs.2022.0081}
}

@article{Lucarini2018,
  title = {{Revising and Extending the Linear Response Theory
for Statistical Mechanical Systems: Evaluating Observables as Predictors and Predictands}},
  author = {Lucarini, V.},
  journal = {J Stat Phys},
  volume = {173},
  pages = {1698–1721},
  year = {2018},
  url = {https://doi.org/10.1007/s10955-016-1506-z}
}

@article{HeldFDT,
  title = {{Applying the Fluctuation–Dissipation Theorem to a Two-Layer Model of Quasigeostrophic Turbulence }},
  author = {Lutsko, N. and Held, I.M. and Zurita-Gotor,P.},
  journal = {Journal of the Atmospheric Sciences},
  volume = {72},
  pages = {3161–3177},
  year = {2015},
  doi = {https://doi.org/10.1175/JAS-D-14-0356.1}
}

@article{Penland89,
  title={{Random Forcing and Forecasting Using Principal Oscillation Pattern Analysis}},
  author={Penland, C.},
  journal={Monthly Weather Review},
  volume={117},
  pages={2165-2185},
  year={1989}
}

@article{Penland95,
  title={{The optimal growth of tropical sea surface temperature anomalies}},
  author={Penland, C. and Sardeshmukh, P.D.},
  journal={Journal of Climate},
  volume={8},
  number = {8},
  pages={1999–2024},
  year={1995}
}

@article{PENLAND1996534,
title = {A stochastic model of IndoPacific sea surface temperature anomalies},
journal = {Physica D: Nonlinear Phenomena},
volume = {98},
number = {2},
pages = {534-558},
year = {1996},
note = {Nonlinear Phenomena in Ocean Dynamics},
issn = {0167-2789},
doi = {https://doi.org/10.1016/0167-2789(96)00124-8},
url = {https://www.sciencedirect.com/science/article/pii/0167278996001248},
author = {C'ecile Penland}
}

@Article{Penland_Fowler,
AUTHOR = {Penland, C\'ecile and Fowler, Megan D. and Jackson, Darren L. and Cifelli, Robert},
TITLE = { Forecasts of Opportunity for Northern California Soil Moisture},
JOURNAL = {Land},
VOLUME = {10},
YEAR = {2021},
NUMBER = {7},
ARTICLE-NUMBER = {713},
URL = {https://www.mdpi.com/2073-445X/10/7/713},
ISSN = {2073-445X},
DOI = {10.3390/land10070713}
}

@article{MTV1,
  title = {{Models for stochastic climate
prediction}},
  author = {Majda, A. J. and Timofeyev, I. and Vanden-Eijnden},
  journal = {Proc. Natl. Acad. Sci. USA},
  volume = {96},
  pages = {14687–14691},
  year = {1999},
}

@article{MTV2,
  title = {{A mathematical framework for
stochastic climate models}},
  author = {Majda, A. J. and Timofeyev, I. and Vanden-Eijnden},
  journal = {Proc. Natl. Acad. Sci. USA},
  volume = {54},
  pages = {891–974},
  year = {2001},
}

@article{StochBAMS,
  title = {{Stochastic Parameterization: Toward a New View of Weather and Climate Models }},
  author = {Berner, J. and et al.},
  journal = {Bulletin of the American Meteorological Society},
  volume = {98},
  issue = {3},
  pages = {565–588 },
  year = {2017},
  doi = {https://doi.org/10.1175/BAMS-D-15-00268.1}
}

@article{KRAVTSOV,
  title={{Multilevel Regression Modeling of Nonlinear Processes: Derivation and Applications to Climatic Variability }},
  author={Kravtsov, S. and Kondrashov, D. and Ghil, M.},
  journal={Journal of Climate},
  volume={18},
  pages={4404–4424 },
  year={2005}
}

@article{RungeCausalEffects,
  title={{Identifying causal gateways and mediators in complex spatio-temporal systems}},
  author={Runge, J. and Petoukhov, V. and Donges, J. and Hlinka, J. and Jajcay, N. and Vejmelka, M. and Hartman, D. and Marwan, N. and Paluš, M. and Kurths, J.},
  journal={Nat Commun},
  volume={6},
  pages={8502},
  year={2015},
  doi = {https://doi.org/10.1038/ncomms9502}
}

@article{era5,
  title={{The ERA5 global reanalysis}},
  author={Hersbach, H. and et al.},
  journal={Quarterly Journal
of the Royal Meteorological Society},
  volume={146},
  number = {730},
  pages={1999–2049},
  year={2020},
  doi = {https://doi.org/10.1002/qj.3803}
}

@article{FALCIONI1995481,
title = {The relevance of chaos for the linear response theory},
journal = {Physica A: Statistical Mechanics and its Applications},
volume = {215},
number = {4},
pages = {481-494},
year = {1995},
issn = {0378-4371},
doi = {https://doi.org/10.1016/0378-4371(94)00277-Z},
url = {https://www.sciencedirect.com/science/article/pii/037843719400277Z},
author = {M. Falcioni and A. Vulpiani}
}

@article{FabCausal,
  title = {Data-driven dimensionality reduction and causal inference for spatiotemporal climate fields},
  author = {Falasca, Fabrizio and Perezhogin, Pavel and Zanna, Laure},
  journal = {Phys. Rev. E},
  volume = {109},
  issue = {4},
  pages = {044202},
  numpages = {21},
  year = {2024},
  month = {Apr},
  publisher = {American Physical Society},
  doi = {10.1103/PhysRevE.109.044202},
  url = {https://link.aps.org/doi/10.1103/PhysRevE.109.044202}
}

@article{Agarwal,
  title={{A comparison of data-driven approaches to build low-dimensional ocean models}},
  author={Agarwal, N. and Kondrashov, D. and Dueben, P. and Ryzhov, E. and Berloff, P.},
  journal={Journal of Advances in Modeling Earth Systems},
  volume={13},
  pages={e2021MS002537},
  year={2005},
  url = {https:// doi.org/10.1029/2021MS002537}
}

@article{coloredLIM,
  title = {Colored linear inverse model: A data-driven method for studying dynamical systems with temporally correlated stochasticity},
  author = {Lien, Justin and Kuo, Yan-Ning and Ando, Hiroyasu and Kido, Shoichiro},
  journal = {Phys. Rev. Res.},
  volume = {7},
  issue = {2},
  pages = {023042},
  numpages = {16},
  year = {2025},
  month = {Apr},
  publisher = {American Physical Society},
  doi = {10.1103/PhysRevResearch.7.023042},
  url = {https://link.aps.org/doi/10.1103/PhysRevResearch.7.023042}
}

@article{majdaConstraints,
  title = {{Physics constrained nonlinear regression models for time series}},
  author = {Majda,A. and Harlim, J.},
  journal = {NONLINEARITY},
  volume = {26},
  pages = {201–217 },
  year = {2013},
  publisher = {IOP P UBLISHING},
  url = {http://dx.doi.org/10.1088/0951-7715/26/1/201}
}

@article{majdaVsGhil,
	author = {Majda, Andrew J. and Yuan, Yuan},
	title = {{Fundamental limitations of ad hoc linear and quadratic multi-level regression models for physical systems}},
	year = {2012},
	journal = {Discrete and Continuous Dynamical Systems - Series B},
	volume = {17},
	number = {4},
	doi = {10.3934/dcdsb.2012.17.1333}
}

@article{KONDRASHOV201533,
title = {Data-driven non-Markovian closure models},
journal = {Physica D: Nonlinear Phenomena},
volume = {297},
pages = {33-55},
year = {2015},
issn = {0167-2789},
doi = {https://doi.org/10.1016/j.physd.2014.12.005},
url = {https://www.sciencedirect.com/science/article/pii/S0167278914002413},
author = {Kondrashov, D. and Chekroun, M.D. and Ghil,M.}
}

@article{KONDRAENSO,
title = {A Hierarchy of Data-Based ENSO Models},
journal = {Journal of Climate},
volume = {18},
pages = {4425–4444},
year = {2005},
doi = {https://doi.org/10.1175/JCLI3567.1},
author = {Kondrashov, D. and Kratsov, S. and Robertson, A.W. and Ghil,M.}
}

@article{STROUNINE2010145,
title = {Reduced models of atmospheric low-frequency variability: Parameter estimation and comparative performance},
journal = {Physica D: Nonlinear Phenomena},
volume = {239},
number = {3},
pages = {145-166},
year = {2010},
issn = {0167-2789},
doi = {https://doi.org/10.1016/j.physd.2009.10.013},
url = {https://www.sciencedirect.com/science/article/pii/S0167278909003285},
author = {K. Strounine and S. Kravtsov and D. Kondrashov and M. Ghil}
}

@incollection{KravtsovReview,
    author = {Kravtsov, S. and M. Ghil and D. Kondrashov},
    title = "{Empirical Model Reduction and the Modeling Hierarchy in Climate Dynamics and the Geosciences}",
    booktitle = "{Stochastic Physics and Climate Modeling}",
    publisher = {Cambridge University Press},
    editor = {T. Palmer and P. Williams},
    year = {2009},
    pages = {35-72}
}

@article{Sura,
  title={{Multiplicative noise and non-Gaussianity: A paradigm for atmospheric regimes?}},
  author={Sura, P. and Newman, M. and Penland, C. and Sardeshmukh, P.},
  journal={Journal of Climate},
  volume={62},
  pages={1391–1409},
  year={2005}
}

@article{Lacorata,
  title={Fluctuation-Response Relation and modeling in systems with fast and slow dynamics},
  author={Lacorata, G. and Vulpiani, A.},
  journal={Nonlin. Processes Geophys.},
  volume={14},
  pages={681–694},
  year={2007},
  url = {https://doi.org/10.5194/npg-14-681-2007}
}

@book{MajdaBook,
  title={{Information Theory and Stochastics for Multiscale Nonlinear Systems}},
  author={Majda, A. J. and Abramov, R. V. and Grote, M. J.},
  publisher={CRM Monograph Series, American Mathematical Society},
  address={Providence, RI},
  year={2005}
}

@article{AM4a,
  title = {{The GFDL global atmosphere and
land model AM4.0/LM4.0: 1. Simulation characteristics with
prescribed SSTs.}},
  author = {Zhao, M. and Coauthors},
  journal = {Journal of Advances in Modeling Earth Systems},
  volume = {10},
  pages = {691–734},
  year = {2018},
  doi = {https://doi.org/10.1002/2017MS001208}
}

@article{AM4b,
  title = {{The GFDL global atmosphere and
land model am4.0/LM4.0: 2. Model description, sensitivity
studies, and tuning strategies.}},
  author = {Zhao, M. and Coauthors},
  journal = {Journal of Advances in Modeling Earth Systems},
  volume = {10},
  pages = {735–769},
  year = {2018},
  doi = {https://doi.org/10.1002/2017MS001208}
}

@article{BaldovinPlos,
  title = {{Langevin equations from experimental data: The case of rotational diffusion in granular media}},
  author = {Baldovin, M. and Puglisi, A. and Vulpiani, A.},
  journal = {PLoS ONE},
  volume = {14(2)},
  pages = {e0212135},
  year = {2019},
  doi = {https://doi.org/10.1371/journal.pone.0212135}
}

@article{MTV3,
  title = {{A priori test of a stochastic mode
reduction strategy}},
  author = {Majda, A. J. and Timofeyev, I. and Vanden-Eijnden},
  journal = {Physica D},
  volume = {170},
  pages = {206–252},
  year = {2002},
}

@article{MTV4,
  title = {{Systematic strategies for
stochastic mode reduction in climate}},
  author = {Majda, A. J. and Timofeyev, I. and Vanden-Eijnden},
  journal = {J. Atmos. Sci.},
  volume = {60},
  pages = {1705–1722},
  year = {2003},
}

@article{Wouters, 
title={Cluster-based reduced-order modelling of a mixing layer}, 
volume={03}, 
journal={Journal of Statistical Mechanics: Theory and
Experiment}, 
author={Wouters, J. and Lucarini, V.}, 
year={2012}, 
pages={P03003},
DOI={10.1088/1742-5468/2012/03/P03003}
}

@article{Vissio2018, 
title={A proof of concept for scale-adaptive parametrizations: the case of the Lorenz '96 model}, 
volume={144}, 
issue = {710},
journal={Quarterly Journal of the Royal Metereological Society}, 
author={Vissio, G. and Lucarini, V.}, 
year={2018 Part A}, 
pages={63-75},
DOI={ https://doi.org/10.1002/qj.3184}
}

@article {Bire,
      author = "Bire, S. and Lütjens, B. and Azizzadenesheli, K. and Anandkumar, A. and Hill, C.",
      title = {{Ocean emulation with Fourier neural operators: Double gyre}},
      journal = "Journal of Advances in Modeling Earth Systems",
      year = "2025",
      volume = "17",
      number = "e2023MS004137",
      url = "https://doi.org/10.1029/2023MS004137"
}

@article {Bjorn,
      author = "Lütjens, B. and Ferrari, R. and Watson‐Parris, D. and Selin, N. E.",
      title = {{The impact of internal variability on benchmarking deep learning climate emulators}},
      journal = "Journal of Advances in Modeling Earth Systems",
      year = "2025",
      volume = "17",
      number = "e2024MS004619",
      url = "https://doi.org/10.1029/2024MS004619"
}

@article {Gosha,
      author = "Geogdzhayev, G. and Souza, A. N. and Flierl, G. R. and Ferrari, R.",
      title = {{An EOF-Based Emulator of Means and Covariances of Monthly Climate Fields}},
      journal = "EGUsphere [preprint]",
      year = "2025",
      url = "https://doi.org/10.5194/egusphere-2025-3768"
}

@article{ROBERTS200812,
title = {Normal form transforms separate slow and fast modes in stochastic dynamical systems},
journal = {Physica A: Statistical Mechanics and its Applications},
volume = {387},
number = {1},
pages = {12-38},
year = {2008},
issn = {0378-4371},
doi = {https://doi.org/10.1016/j.physa.2007.08.023},
url = {https://www.sciencedirect.com/science/article/pii/S0378437107008655},
author = {Roberts, A.J.}
}

@article{Roberts2006, title={Resolving the Multitude of Microscale Interactions Accurately Models Stochastic Partial Differential Equations}, volume={9}, DOI={10.1112/S146115700000125X}, journal={LMS Journal of Computation and Mathematics}, author={Roberts, A. J.}, year={2006}, pages={193–221}}

@article{Cecconi,
author = {Cecconi, F. and Costantini, G. and Guardiani, C. and Baldovin, M. and Vulpiani, A.},
journal = {Physical Biology},
year = {2020},
title = {{Correlation, response and entropy approaches to allosteric behaviors: a critical comparison on the ubiquitin case}},
volume = {5},
pages = {056002}
}

@article{BaldovinEntropy,
  title = {The Role of Data in Model Building and Prediction: A Survey Through Examples},
  author = {Baldovin, M. and Cecconi, F. and Cencini, M. and Puglisi, A. and Vulpiani, A.},
  journal = {Phys. Rev. Res.},
  volume = {20},
  issue = {807},
  year = {2018},
  month = {Oct},
  publisher = {Entropy},
  doi = { doi:10.3390/e20100807}
}

@article{om4,
  title = {{The GFDL global ocean and sea ice model OM4.0: Model description and simulation features}},
  author = {Adcroft, A. and Anderson, W. and Balaji, V. and Blanton, C. and Bushuk, M. and Dufour, C. O. and et al.},
  journal = {Journal of Advances in Modeling Earth Systems},
  volume = {11},
  pages = {3167–3211},
  year = {2019},
  doi = {doi.org/ 10.1029/2019MS001726}
}

@article{AIFS,
  title = {{AIFS -- ECMWF's data-driven forecasting system}},
  author = {Lang, S. and et al.},
  journal = {Arxiv},
  year = {2025},
  doi = {https://doi.org/10.48550/arXiv.2406.01465}
}

@article{ProbabilisticSI,
  title = {{Probabilistic Forecasting with Stochastic Interpolants and Föllmer Processes}},
  author = {Chen, Y. and Goldstein, M. and Hua, M. and Albergo, M.S. and Boffi, N.M. and Vanden-Eijnden, E.},
  journal = {Arxiv},
  year = {2025},
  doi = {https://doi.org/10.48550/arXiv.2403.13724}
}

@article{ProbabilisticSIPedram,
  title = {{Reframing Generative Models for Physical Systems using Stochastic Interpolants}},
  author = {Zhou, A. and Wikner, A. and Lancelin, A. and Hassanzadeh, P. and Farimani, A. B.},
  journal = {Arxiv},
  year = {2025},
  doi = { 	
https://doi.org/10.48550/arXiv.2509.26282}
}

@article{PierreLatent,
  title = {{Incorporating Multivariate Consistency in ML-Based Weather Forecasting with Latent-space Constraints}},
  author = {Fan, H. and Xiao, Y. and Qu, Y. and Ling, F. and Fei, B. and Bai, L. and Gentine, P.},
  journal = {Arxiv},
  year = {2025},
  doi = {https://doi.org/10.48550/arXiv.2510.04006}
}

@article{AndreEmulation,
  title = {{Score-based generative emulation of impact-relevant Earth system model outputs}},
  author = {Bouabid, S. and Souza, A.N. and Ferrari, R.},
  journal = {Arxiv},
  year = {2025},
  doi = {https://doi.org/10.48550/arXiv.2510.04358}
}

@article{Nikolaj,
  title = {{Physics-aware generative models for turbulent fluid flows through energy-consistent stochastic interpolants}},
  author = {Mücke,N.T. and Sanderse, B.},
  journal = {Arxiv},
  year = {2025},
  doi = {https://doi.org/10.48550/arXiv.2504.05852}
}

@article{CycloLudoAndre,
  title = {{Reduced-Order Modeling of Cyclo-Stationary Time Series Using Score-Based Generative Methods}},
  author = {Giorgini, L.T. and Bischoff,T. and Souza, A.N.},
  journal = {Arxiv},
  year = {2025},
  doi = {https://doi.org/10.48550/arXiv.2508.19448}
}

@article{SamudraACE,
  title = {{SamudrACE: Fast and Accurate Coupled Climate Modeling with 3D Ocean and Atmosphere Emulators}},
  author = {Duncan, J.P.C. and et al.},
  journal = {Arxiv},
  year = {2025},
  doi = {https://doi.org/10.48550/arXiv.2509.12490}
}

@article{Wu,
  title = {{Applying the ACE2 emulator to SST green's functions for the E3SMv3 global atmosphere model}},
  author = {Wu, E. and Rebassoo, F. and Paul, P. and Proistosescu, C. and Nugent, J. and McCoy, D. and et al. },
  journal = {Journal of
Geophysical Research: Machine Learning
and Computation},
  year = {2025},
volume = {2},
pages = {e2025JH000774},
  doi = {https://doi.org/10.1029/2025JH000774}
}

@article{latentPierre,
  title = {{Physically Consistent Global Atmospheric Data Assimilation with Machine Learning in a Latent Space}},
  author = {Fan, H. and Fei, B. and Gentine, P. and Xiao, Yi and Chen, K. and Liu, Y. and Qu, Y. and Ling, F. and Bai, L.},
  journal = {Arxiv},
  year = {2025},
  doi = {https://doi.org/10.48550/arXiv.2502.02884}
}

@article{PedramStability,
  title = {{Hierarchical Implicit Neural Emulators}},
  author = {Jiang, R. and Zhang, X. and Jakhar, K. and Lu, P.Y. and Hassanzadeh, P. and Maire, M. and Willett, R.},
  journal = {Arxiv},
  year = {2025},
  doi = {https://doi.org/10.48550/arXiv.2506.04528}
}

@article{scikit-learn,
  title={Scikit-learn: Machine Learning in {P}ython},
  author={Pedregosa, F. and Varoquaux, G. and Gramfort, A. and Michel, V.
          and Thirion, B. and Grisel, O. and Blondel, M. and Prettenhofer, P.
          and Weiss, R. and Dubourg, V. and Vanderplas, J. and Passos, A. and
          Cournapeau, D. and Brucher, M. and Perrot, M. and Duchesnay, E.},
  journal={Journal of Machine Learning Research},
  volume={12},
  pages={2825--2830},
  year={2011}
}

@article{Lorenz, 
title={Deterministic nonperiodic ﬂow}, 
volume={20},
journal={J. Atmospheric Sci.}, 
publisher={Cambridge University Press}, 
author={Lorenz, E.N.}, 
year={1963}, 
pages={130–141}}

@book{Risken, 
    address={New York}, 
    title={The Fokker-Planck Equation – Methods of Solution and Applications},  
    publisher={Springer-Verlag}, 
    author={Risken, H.}, 
    year={1996},
    doi = {doi.org/10.1007/978-3-642-04898-2_150}
}

@article{adler2018deep,
  title={Deep bayesian inversion},
  author={Adler, Jonas and {\"O}ktem, Ozan},
  journal={arXiv preprint arXiv:1811.05910},
  volume={1},
  number={2},
  pages={7},
  year={2018},
  publisher={Eprint}
}

@article{kingma2014adam,
  title={Adam: A method for stochastic optimization},
  author={Kingma, Diederik P and Ba, Jimmy},
  journal={arXiv preprint arXiv:1412.6980},
  year={2014}
}

\end{document}